\documentclass[a4paper,11pt]{article}
\pdfoutput=1 

\usepackage{jheppub}
\usepackage{float}
\usepackage{epsfig}
\usepackage{amsmath}
\usepackage{graphicx}

\def\A0#1{\Pi_{\rm #1}(0)}
\def\AP0#1{\Pi'_{\rm #1}(0)}
\def\be{\begin{equation}}
\def\ee{\end{equation}}
\def\bea{\begin{array}}
\def\eea{\end{array}}
\def\beqa{\begin{eqnarray}}
\def\eeqa{\end{eqnarray}}
\def\beqas{\begin{eqnarray*}}
\def\eeqas{\end{eqnarray*}}

\def\bp{\begin{picture}}
\def\ep{\end{picture}}
\def\bc{\begin{center}}
\def\ec{\end{center}}
\def\bfig{\begin{figure}}
\def\efig{\end{figure}}

\def\bit{\begin{itemize}}
\def\eit{\end{itemize}}
\def\nn{\nonumber}
\def\f{\frac}

\def\[{\left[}
\def\]{\right]}
\def\({\left(}
\def\){\right)}

\def\..{\left.}
\def\.{\right.}
\def\tl{\tilde}
\def\ra{\rightarrow}
\def\la{\leftarrow}

\def\tm{\times}

\def\da{\dagger}

\def\la{\lambda}

\def\al{\alpha}
\def\bt{\beta}
\def\ka{\kappa}

\def\ep{\epsilon}

\def\Ga{\Gamma}
\def\ga{\gamma}

\def\pr{\prime}

\title{Flavor Structures Of Quarks and Leptons From Flipped SU(5) GUT with $A_4$ Modular Flavor Symmetry}
\author[a,b]{Xiao Kang Du,}
\author[a]{Fei Wang*.\note{*Corresponding author.}}
\affiliation[a]{School of Physics, Zhengzhou University, Zhengzhou 450000, P. R. China}
\affiliation[b]{Institute of Physics, Henan Academy of Sciences, Zhengzhou 450046, P. R. China}
\emailAdd{feiwang@zzu.edu.cn}
\abstract{We propose to generate the flavor structures of the Standard Model plus neutrinos from flipped SU(5) GUT with $A_4$ modular flavor symmetry. Possible way to assign different moduli values for quarks and leptons in modular GUT scheme is discussed. We propose to reduce the multiple modular symmetries to a single modular symmetry in the low energy effective theory with proper boundary conditions. We classify all possible scenarios in this scheme  according to the assignments of the modular $A_4$ representations for matter superfields and give the expressions of the quark and lepton mass matrices predicted by our scheme at the GUT scale. After properly selecting the modular weights for various superfields that can lead to better fitting, we can obtain the best-fit points with the corresponding $\chi^2$ values for the sample subscenarios. We find that the flavor structures of the Standard Model plus neutrinos can be fitted perfectly in such a $A_4$ modular flavor GUT scheme with single or two modulus fields. Especially, the $\chi^2_{total}$ of our fitting can be as low as $1.558$ for sample ${\bf IX^\pr}$ of scenario ${\bf III}$ even if only a single common modulus field for both quark and lepton sectors is adopted. The most predictive scenario ${\bf III}$, in which all superfields transform as triplets of $A_4$, can be fitted much better with two independent moduli fields $\tau_q,\tau_l$ for quark sector and lepton sector ( $\chi^2_{total}\approx 95$) than that with the single modulus case ($\chi^2_{total}\approx 282.4$).
}

\begin{document}
\maketitle \indent
\newpage
\section{\label{sec:int}Introduction}

Possible unification of strong and electroweak couplings of the Standard Model (SM) suggests the existence of Grand Unified Theory (GUT)~\cite{GUT,GUT1} at a very high energy, which can also possibly accommodate the unification of quarks and leptons. To tame the large quadratic divergences arising from the loop corrections for Higgs mass, low energy supersymmetry (SUSY) are always introduced in GUT, which is also preferable to realize genuine gauge coupling unification. The minimal SUSY SU(5) GUT~\cite{SUSYSU5} has been regarded as a very promising model of GUT. However, the rate for dimension-five operator induced proton decay in minimal SUSY SU(5) is very high, which put much pressure on such a  GUT model. Flipped SU(5) GUT~\cite{fSU5,fSU51,fSU52}, on the other hand, can naturally accommodate the economical missing-partner mechanism~\cite{missingpartner} to solve the notorious doublet-triplet (D-T) splitting problem and efficiently suppress the dimension-five proton decay rate. Besides, the flipped SU(5) GUT naturally include the right-handed (RH) neutrinos and can be further unified in SO(10) GUT. It can also be embedded in string theory and explain some cosmological puzzles. So, it is interesting to seek for the solutions of the SM problems (for example, the origin of free parameters and flavor structure, the origin of neutrino masses) in the framework of SUSY flipped SU(5) GUT.

The Yukawa-type couplings in the superpotential of SUSY GUT models, however, are not fixed by the GUT gauge symmetry. To address the flavor puzzle of SM from SUSY GUT, additional symmetry structures are in general needed to obtain certain flavor patterns at the GUT scale, for example, some non-Abelian discrete flavor symmetry such as $A_4$, $S_4$. 
As there is no hint of an exact flavor symmetry (neither in the quark sector nor in the lepton sector), only broken flavor symmetries have the chance of being realistic. To go beyond the unrealistic lowest-order predictions, a set of flavons with unknown coefficients are always necessary to break these flavor symmetries. The vacuum expectation values (VEVs) of flavons should be oriented along certain directions in flavor space, which require complicated vacuum alignment model buildings and are always not natural.

Recently, SUSY models with modular invariance had been proposed in~\cite{feruglio} to explain the SM flavor structure, which potentially make no use of any flavon fields other than the modulus field.
The invariance of the superpotential under the modular group requires the Yukawa couplings to be modular forms, which are holomorphic functions of the complex modulus that satisfy certain constraints. Besides, all higher-dimensional operators in the superpotential can be unambiguously determined in the limit of unbroken supersymmetry. Modular flavor models based on the inhomogeneous finite modular group of low levels, $S_3\simeq \Ga_2$~\cite{Kobayashi:2018vbk,Okada:2019xqk,Mishra:2020gxg,Kobayashi:2018wkl}, $A_4\simeq \Ga_3$~\cite{feruglio,Kobayashi:2018wkl,Criado:2018thu,Kobayashi:2018scp,Xing:2020ijf,Novichkov:2018yse,Okada:2018yrn,Ding:2019zxk,Okada:2019uoy,Nomura:2019jxj,Nomura:2019yft,Asaka:2019vev,Nomura:2019xsb,Zhang:2019ngf,Nomura:2019lnr,Kobayashi:2019gtp,Wang:2019xbo,Okada:2020rjb,Okada:2019mjf,King:2020qaj,Okada:2020brs,Asaka:2020tmo,Nomura:2020opk,Okada:2020dmb,Yao:2020qyy,Behera:2020sfe,Nomura:2020cog,Nagao:2020snm,Behera:2020lpd,King:2018fke,Abbas:2020qzc,Okada:2021qdf,Nomura:2021yjb,Kashav:2021zir,Okada:2021aoi,Kobayashi:2021ajl,Nagao:2021rio,deMedeirosVarzielas:2021pug,Abbas:2020vuy,Otsuka:2022rak,Nomura:2021pld,Nomura:2022hxs,Nomura:2022boj,Kashav:2022kpk,Aoki:2020eqf,Kang:2022psa,Gogoi:2022jwf,Gunji:2022xig,Tanimoto:2020lrl,Hutauruk:2020xtk,Kobayashi:2019mna,Kobayashi:2019xvz,Ding:2019gof}, $S_4\simeq \Ga_4$~\cite{Kobayashi:2019mna,Kobayashi:2019xvz,Ding:2019gof,Penedo:2018nmg,Novichkov:2018ovf,Novichkov:2019sqv,King:2019vhv,deMedeirosVarzielas:2019cyj,Wang:2019ovr,Okada:2019lzv,Wang:2020dbp,Qu:2021jdy,Nomura:2021ewm,Novichkov:2020ahb,Criado:2019tzk}, and $A_5\simeq \Ga_5$~\cite{Criado:2019tzk,Novichkov:2018nkm,deMedeirosVarzielas:2022ihu} with $\Ga_N\equiv \bar{\Ga}/\bar{\Ga}(N)$, had been proposed to explain the flavor structures of quarks or leptons in various papers. It is known that the fitting of the matter contents within proper GUT multiplets can already constrain stringently the low energy flavor structures. Combining the modular invariance with GUT, the Yukawa couplings for the quarks and leptons can be predicted with even less free parameters, which can improve the prediction power of the GUT theory.

 Some works had been done to explain the flavor puzzle of SM in the GUT framework with proper modular flavor symmetry, for example, SU(5) or SO(10) GUT  with $S_3$~\cite{Kobayashi:2019rzp,Du:2020ylx}, $A_4$~\cite{deAnda:2018ecu,Chen:2021zty,Ding:2021eva,Charalampous:2021gmf,Ding:2022bzs} and $S_4$~\cite{Ding:2021zbg,King:2021fhl,Kobayashi:2022,Zhao:2021jxg} modular symmetry, respectively. As the flipped SU(5) GUT can be advantageous in several aspects in comparison to ordinary SU(5), we propose to combine SUSY flipped SU(5) GUT with $A_4$ modular group (which is the smallest one that contains the triplet representation) to survey if the flavor structures of SM plus neutrino can be fitted more precisely in this modular flavor GUT framework.

 Most discussions in the framework of modular flavor GUT concentrate on the case with a single modulus field $\tau$, which corresponds to a single finite modular symmetry $\Ga_N$. In many circumstances, the fittings of the SM plus neutrino flavor structure are not good enough with a single modulus field, which on the other hand prefer multiple values of modulus Vacuum Expectation Values (VEVs) for various sectors. Such multiple values of modulus VEVs correspond to the existence of multiple independent moduli fields for a single finite modular group, which can origin from scenarios with multiple moduli fields transformed with multiple modular symmetry in the UV theory. We will present a new approach to reduce the UV theory with multiple moduli fields and multiple modular symmetry to IR theory with multiple moduli fields and a single modular symmetry.

This paper is organized as follows. In Sec~\ref{sec:ms} and Sec~\ref{sec:fSU5}, we briefly review the modular symmetry and the flipped SU(5) GUT. In Sec~\ref{sec:mm}, we discuss possible ways to generate multiple moduli values in GUT. In Sec~\ref{sec:cla}, we classify all possible scenarios in this scheme according to the assignments of the modular $A_4$ representations for matter superfields and give the analytical expressions of the predicted quark and lepton mass matrices. In Sec ~\ref{sec:num}, we carry out numerical fittings to survey if the predictions of such scenarios on SM plus neutrino flavor structures can be realistic. Sec~\ref{sec:con} contains our conclusions.
\section{\label{sec:ms}Modular Symmetry}
  The element in the modular group $\Ga\equiv SL(2, Z)$ can act on the upper half plane by
\begin{equation}
\ga:\tau \mapsto \frac{a\tau+b}{c\tau+d},~~~{\rm for}~\ga\equiv\begin{pmatrix}
a  ~&  b  \\
c  ~& d
\end{pmatrix}\in SL(2, Z),~~~\texttt{Im}(\tau)>0\,.
\end{equation}
The inhomogeneous modular group is $\bar{\Ga}=\Ga/\{I, -I\}$, while the infinite normal subgroups $\Ga(N)$ with positive integer $N =2,3,\cdots$ of $SL(2,Z)$ are given by
\begin{equation}
\Gamma(N)=\left\{\begin{pmatrix}
a  ~&  b \\
c  ~& d
\end{pmatrix}\in SL(2, Z)\left| \f{}{}\right.~\begin{pmatrix}
a  ~&  b \\
c  ~& d
\end{pmatrix} =\begin{pmatrix}
1  ~& 0 \\
0  ~& 1
\end{pmatrix}
(\texttt{mod}~N)\right\}\,.
\end{equation}
 The inhomogeneous finite modular group $\Gamma_N$ is defined by $\Gamma_N\cong \overline{\Gamma}/\overline{\Gamma}(N)$ with $\overline{\Gamma}(N)=\Gamma(N)$ for $N>2$ and $\overline{\Gamma}(N)=\Gamma(N)/\{I, -I\}$ for $N=2$.

Modular forms of weight $k$  and level $N$ are holomorphic functions $f(\tau)$ transforming under
the action of $\Ga(N)$ in the following way:
\beqa
f(\ga \tau)=(c\tau+d)^{k} f(\tau)~,~~\ga\in \Ga(N)~,
\eeqa
with $k$ an even and non-negative integer. It can be proved that  modular forms of weight $k$ and level $N$ form a linear space of finite dimension. So, after choosing proper basis in this linear space, the transformation of a set of modular forms $f_i(\tau)$ is described by a unitary representation $\rho$ of the finite modular group
\beqa
f_i(\ga\tau)=(c\tau+d)^{k}\rho(\ga)_{ij}f_j(\tau)~,~~\ga\in\bar{\Ga}.
\eeqa
See~\cite{feruglio} for detailed discussions.

In the framework of $N =1$ SUSY with typical modular symmetry, the superpotential is in general a function of the modulus field $\tau$ and superfields $\phi_i$. The superpotential should be invariant under the modular transformation while the Kahler potential should be invariant up to the addition of holomorphic and anti-holomorphic functions. The Kahler potential involving chiral matter fields $\phi_i$ with the modular weight $k_i$ is given by
\beqa
K\supseteq -h\ln(i\tau^*-i\tau)+\sum\limits_{\phi}\f{\phi^\da\phi}{(i\tau^*-i\tau)^{k_\phi}}~.
\eeqa

There are 4 inequivalent irreducible representations of $A_4$: three singlets $\mathbf{1}$, $\mathbf{1}'$, $\mathbf{1}''$ and a triplet $\mathbf{3}$.
The triplet representation $\mathbf{3}$ (in the basis where $T$ is diagonal) is given by
\begin{equation}
\label{eq:rep-gen}S=\frac{1}{3}\begin{pmatrix}
-1 ~& 2  ~& 2  \\
2  ~& -1 ~& 2 \\
2  ~& 2  ~& -1
\end{pmatrix}, ~\quad~
T=\begin{pmatrix}
1 ~&~ 0 ~&~ 0 \\
0 ~&~ \omega ~&~ 0 \\
0 ~&~ 0 ~&~ \omega^{2}
\end{pmatrix} \,.
\end{equation}
The finite modular group $A_4\cong \Ga_3$ can generated by
\beqa
S^2=(ST)^3=T^3=1~.
\eeqa
The decompositions of the direct product of $A_4$ representations are
\begin{eqnarray}
\nonumber&& \mathbf{1}'\otimes\mathbf{1}'=\mathbf{1}'',~~~ \mathbf{1}'\otimes\mathbf{1}''=\mathbf{1},~~~ \mathbf{1}''\otimes\mathbf{1}''=\mathbf{1}'\,,\\
&&\mathbf{3}\otimes \mathbf{3}= \mathbf{1}\oplus \mathbf{1'}\oplus \mathbf{1''}\oplus \mathbf{3}_S\oplus \mathbf{3}_A\,,
\label{A4:33product}
\end{eqnarray}
where $\mathbf{3}_{S(A)}$ denote the symmetric (antisymmetric) combinations, respectively.

Given two triplets $\alpha=(\alpha_1,\alpha_2,\alpha_3)$ and  $\beta=(\beta_1,\beta_2,\beta_3)$, the irreducible representations obtained from their product are:
\begin{eqnarray}
\nonumber &&\mathbf{1}=\alpha_1\beta_1+\alpha_2\beta_3+\alpha_3\beta_2\,, \\
\nonumber &&\mathbf{1}'=\alpha_3\beta_3+\alpha_1\beta_2+\alpha_2\beta_1\,, \\
\nonumber &&\mathbf{1}''=\alpha_2\beta_2+\alpha_1\beta_3+\alpha_3\beta_1\,, \\
\nonumber &&\mathbf{3}_S=(
2\alpha_1\beta_1-\alpha_2\beta_3-\alpha_3\beta_2,
2\alpha_3\beta_3-\alpha_1\beta_2-\alpha_2\beta_1,
2\alpha_2\beta_2-\alpha_1\beta_3-\alpha_3\beta_1)\,, \\
\label{eq:decomp-rules} &&\mathbf{3}_A=(
\alpha_2\beta_3-\alpha_3\beta_2,
\alpha_1\beta_2-\alpha_2\beta_1,
\alpha_3\beta_1-\alpha_1\beta_3)\,.
\end{eqnarray}

 The modular forms $Y_{\bf r}^{(k)}$ with level 3 and modular weight $k\leq 8$ under $A_4$ are collected in the appendix \ref{modular-weight}.
\section{\label{sec:fSU5}Flipped SU(5) GUT}
We briefly review the key ingredients in flipped SU(5) GUT, see~\cite{fSU5,fSU51,fSU52} for details. The gauge group for flipped $SU(5)$ GUT model is $SU(5)\times U(1)_{X}$ with the generator $U(1)_{Y^\pr}$ within $SU(5)$ defined by
\beqa
T_{\rm U(1)_{Y^\pr}}={\rm diag} \left(-{1\over 3}, -{1\over 3}, -{1\over 3},
 {1\over 2},  {1\over 2} \right).
\eeqa
The hypercharge can be given by
\beqa
Q_{Y} = {1\over 5} \left( Q_{X}-Q_{Y^\pr} \right).
\label{hypercharge}
\eeqa

The matter contents in each generation can be fitted into flipped $SU(5)$ by
\beqa
F_i(10, 1)=(Q_L, D_L^c , N_L^c),~~ \bar{f}_i(\bar{5},-3) =(L_L, U_L^c),~~ E_i(1,5)=E_L^c,
\eeqa
with the family index $i=1,2,3$. The Higgs sector contains
\beqa
H(10,1),~\overline{H}(\overline{10},-1),~h(5,-2),~\bar{h}(\bar{5},2),
\eeqa
to break the GUT and electroweak gauge symmetries.
The superpotential for Yukawa couplings is given as
\beqa
W\supseteq  y_{ij}^D F_i F_j h+ y_{ij}^U F_i \bar{f}_j \bar{h} + y_{ij}^E \bar{f}_i E_j h~,
\eeqa
while for the Higgs sector is
\beqa
W\supseteq H H h+\overline{H}\overline{H}\bar{h}+X(\overline{H}H-M_H^2)~,
\eeqa
to trigger the breaking of $SU(5)\tm U(1)_X$ into SM gauge group and solve the D-T splitting problem via missing partner mechanism.

In order to generate tiny neutrino masses via inverse seesaw mechanism, we need to introduce additional neutral superfield $S_i(1,0)$, whose fermionic components act as the new neutrino species in inverse seesaw mechanism. Relevant terms for neutrino masses are given as
\beqa
{\cal W}_{low} \supseteq Y^N_{ij} L_{L;i} N_{L;j}^c H_U+ Y_{ij}^S S_i N_{L;j}^c M_H +  \f{M_{SS;ij}}{2}S_i S_j~,
\eeqa
which can be embedded into the flipped SU(5) GUT model
\beqa
{\cal W}\supseteq Y^U_{ij} F_i \bar{f}_j \bar{h}+ Y^S_{ij} \overline{H}  S_i F_j +\f{M_{SS;ij}}{2} S_i S_j~.
\eeqa
Therefore, the neutrino mass matrix can be obtained to be
\beqa
{\cal M}=\(\bea{ccc}~0& m_D^T&0\\m_D&0&M_{SN}\\0&M_{SN}&M_{SS}\eea\)~,
\eeqa
 with $m_D\sim Y_2 v_u$ and $M_{SN}\sim Y_S M_H$ after electroweak symmetry breaking triggered by the VEVs of $h,\bar{h}$. Here $m_D,M_{SN},M_{SS}$ are all $3\times3$ matrices.
The effective neutrino mass matrix for the standard neutrinos can be approximately given by
\beqa
m_\nu\approx m_D^T M_{SN}^{-1} M_{SS} (M_{SN}^T)^{-1} m_D~,
\eeqa
when $M_{SS}\ll m_D\ll M_{SN}$.
\section{\label{sec:mm}Multiple Moduli Values in GUT}
The fitting of the low energy flavor structures can always be improved if quarks and leptons adopt different moduli values in single modular symmetry case, which may indicate the existence of multiple moduli fields and multiple modular symmetries in the UV theory, such as the string theory~\cite{mm:UV1,mm:UV2,mm:UV3}.

For any finite modular transformations ${\gamma_1, ..., \gamma_M}$ in $\Gamma_{N_1}^1\times \Gamma_{N_2}^2 \times \cdots \times \Gamma_{N_M}^M$, which is given by
 \begin{eqnarray}
&&\gamma_J: \tau_J \to \gamma_J \tau_J = \frac{a_J \tau_J + b_J}{c_J \tau_J + d_J} \,,
\end{eqnarray}
 the chiral superfield $\phi_i$ (as a function of $\tau_1$, ..., $\tau_M$ ) transforms as~\cite{deMedeirosVarzielas:2019cyj,King:2019vhv}
\begin{eqnarray}
 \phi_i(\tau_1, ...,\tau_M) &\to& \phi_i(\gamma_1\tau_1, ..., \gamma_M \tau_M) \nonumber\\
 &&= \prod_{J=1,...,M} (c_J\tau_J + d_J)^{-2k_{i,J}} \bigotimes_{J=1,...,M} \rho_{I_{i,J}}(\gamma_J) \phi_i(\tau_1, \tau_2, ...,\tau_M)\,,
 \label{eq:field_transformation2}
\end{eqnarray}
where $k_{i,J}$ and $I_{i,J}$ are the modular weight and representation of $\phi_i$ in $\Gamma_{N_J}^{J}$, respectively. The symbol $\bigotimes$ represents the direct product of the representation matrices for $\rho_{I_{i,1}}$, $\rho_{I_{i,2}}$, ..., $\rho_{I_{i,M}}$. Models with twin $S_4$ had been discussed in~\cite{King:2021fhl}, in which the breaking of the twin $S_4$ into their diagonal $S_4^D$ can be realized by bi-fundamental scalar VEVs.

In modular GUT framework, the superfields within a multiplet of GUT gauge group should transform identically with the same modulus field.  We know that some representation of the GUT group contains both quarks and leptons. So, it seems that the unification of matter contents will be spoiled if different values of modulus are assigned separately for quarks and leptons. We propose to reconcile such an inconsistency in the orbifold GUT scheme, for example, with a 5D ${\cal M}_4\tm S^1/Z_2$ orbifold. The generalization of 4D flipped SU(5) GUT to 5D flipped SU(5) orbifold GUT is straightforward.

Compactification on $S^1/Z_2$ is obtained by identifying the fifth coordinate $y$ under the two operations
\beqa
 Z : y \ra -y,~~~~T : y \ra y + 2\pi R.
\eeqa 
There are two inequivalent 3-branes located at $y=0$ and $y=\pi R$ which are denoted by $O$ and
$O^{\pr}$, respectively.

 The resulting 5D $N=1$ SUSY (corresponding to 4D $N=2$ SUSY) can also be reduced to 4D $N=1$ SUSY by proper boundary conditions (see~\cite{kawamura,at,hall,hebecker:2001wq,hebecker:2001jb} and examples in our previous works~\cite{fei2,fei3}). It is well known that the five-dimensional $N=1$ supersymmetric gauge theory has 8 real supercharges, corresponding to $N=2$ supersymmetry in four dimensions. The vector multiplet contains a vector boson $A_M$ where $M=0, 1, 2, 3, 5$, two Weyl gauginos $\lambda_{1,2}$, and a real scalar $\sigma$. In the ordinary four-dimensional $N=1$ language, it contains a vector multiplet $V(A_{\mu}, \lambda_1)$ and a chiral multiplet $\Sigma((\sigma+iA_5)/\sqrt 2, \lambda_2)$ that
transforms in the adjoint representation of the gauge group. On the other hand, the five-dimensional hypermultiplet has two physical complex scalars $\phi$ and $\phi^c$, a Dirac fermion $\Psi$, and can be decomposed into two 4-dimensional chiral multiplets $\Phi(\phi, \psi \equiv \Psi_R)$
and $\Phi^c(\phi^c, \psi^c \equiv \Psi_L)$, which transform as
each others conjugates under gauge transformations.

The general action \cite{nima2} for the gauge fields and the relevant
couplings to the bulk hypermultiplet $\Phi$ is
\begin{eqnarray}
S&=&\int{d^5x}\frac{1}{k g^2}
{\rm Tr}\left[\frac{1}{4}\int{d^2\theta} \left(W^\alpha W_\alpha
+{\rm h.c.}\right) \right.\nonumber\\
&&~~~~~~~~~~~~~~~~
\left.+\int{d^4\theta}\left((\sqrt{2}\partial_5+ {\bar \Sigma })
e^{-V}(-\sqrt{2}\partial_5+\Sigma )e^V+
\partial_5 e^{-V}\partial_5 e^V\right)\right] \nonumber\\
&+& \int{d^5x} \left[ \int{d^4\theta} \left( {\Phi}^c e^V {\bar \Phi}^c
+{\bar \Phi} e^{-V} \Phi \right)
+ \int{d^2\theta} \left( {\Phi}^c (\partial_5 -{1\over {\sqrt 2}} \Sigma)
\Phi + {\rm h. c.} \right)\right]~~
\label{VD-Lagrangian}
\end{eqnarray}
where $Tr(T^aT^b)=k\delta^{ab}$.

 Because the action is invariant under the parity operation $P\equiv Z$,
under this operation, the vector multiplet transforms as
\begin{eqnarray}
V(x^{\mu},y)&\to&  V(x^{\mu},-y) = P V(x^{\mu}, y) P^{-1}
~,~\,\nn\\
 \Sigma(x^{\mu},y) &\to &\Sigma(x^{\mu},-y) = - P \Sigma(x^{\mu},
y) P^{-1}~.~\,
\end{eqnarray}
If the hypermultiplet belongs to the fundamental or anti-fundamental
representations, since the parity satisfies $P=P^{-1}$, we have
\begin{eqnarray}
\Phi(x^{\mu},y)&\to& \Phi(x^{\mu}, -y)  = \eta_{\Phi} P
\Phi(x^{\mu},y) ~,~\,\nn\\
\Phi^c(x^{\mu},y) &\to& \Phi^c(x^{\mu}, -y)  = -\eta_{\Phi} P
\Phi^c(x^{\mu},y) ~.~\,
\end{eqnarray}
Alternatively, if the hypermultiplet belongs to the symmetric, anti-symmetric
or adjoint representations,  we have
\begin{eqnarray}
\Phi(x^{\mu},y)&\to& \Phi(x^{\mu}, -y)  = \eta_{\Phi} P
\Phi(x^{\mu},y) P ~,~\,\nn\\
\Phi^c(x^{\mu},y) &\to& \Phi^c(x^{\mu}, -y)  = -\eta_{\Phi} P
\Phi^c(x^{\mu},y) P ~,~\,
\end{eqnarray}
where $\eta_{\Phi} = \pm 1$. Similar results hold for the parity operation $P^\pr\equiv TZ$ at the fixed point $O^\pr (y=\pi R)$.

 The chiral superfields in ${\bf 10}_{\bf 1},\overline{\bf 5}_{\bf -3},{\bf 1}_{\bf -5}$ representations of flipped SU(5) within the corresponding hypermultiplets are placed in the 5D bulk. We impose the following boundary conditions (BCs) for each family in terms of $SU(3)_c\tm SU(2)_L\tm U(1)_{X}\tm U(1)_{Y^\pr}$ quantum numbers
\beqa
T_F^{a}({\bf 10}_{\bf 1})&=& Q_L^a({\bf 3},{\bf 2})^{(+,-)}_{\bf (1,1/6)}\oplus D_L^{c,a}(\bar{\bf 3},1)^{(+,-)}_{\bf (1,-2/3)}\oplus N_L^{c,a}(1,1)^{(+,+)}_{\bf (1,1)}~,\nn\\
T_F^{\pr,a}({\bf 10}_{\bf 1})&=& Q_L^a({\bf 3},{\bf 2})^{(+,-)}_{\bf (1,1/6)}\oplus D_L^{c,a}(\bar{\bf 3},1)^{(+,+)}_{\bf (1,-2/3)}\oplus N_L^{c,a}(1,1)^{(+,-)}_{\bf (1,1)}~,\nn\\
T_F^{\pr\pr,a}({\bf 10}_{\bf 1})&=& Q_L^a({\bf 3},{\bf 2})^{(+,+)}_{\bf (1,1/6)}\oplus D_L^{c,a}(\bar{\bf 3},1)^{(+,-)}_{\bf (1,-2/3)}\oplus N_L^{c,a}(1,1)^{+,-}_{\bf (1,1)}~,\nn\\
F_{\bar{f}}^{a}(\overline{\bf 5}_{\bf -3})&=& U_L^{c,a}(\bar{\bf 3},1)_{\bf (-3,1/3)}^{(+,+)}\oplus L_L^a({\bf 1},{\bf 2})_{\bf (-3,-1/2)}^{(+,-)}~,\nn\\
F_{\bar{f}}^{\pr,a}(\overline{\bf 5}_{\bf -3})&=&U_L^{c,a}(\bar{\bf 3},1)_{\bf (-3,1/3)}^{(+,-)}\oplus L_L^a({\bf 1},{\bf 2})_{\bf (-3,-1/2)}^{(+,+)}~,\nn\\
O_{E}^a({\bf 1}_{\bf -5})&=&E_L^{c,a}({\bf 1,1})^{(+,+)}_{\bf (-5,0)}~,~~~
O_S^a({\bf 1}_{\bf 0})= S^a({\bf 1,1})_{\bf (0,0)}^{(+,+)}~,
\label{orbifold-1}
\eeqa
with $a=1,2,3$ the indice for the three families and that of Higgs sector
\beqa
h({\bf 5}_{\bf -2})&=& H_T({\bf 3,1})_{(-2,-1/3)}^{(+,-)}\oplus H_D({\bf 1,2})_{(-2,1/2)}^{(+,+)}~,~\nn\\
\bar{h}({\bf \bar{5}}_{\bf 2})&=& H_T^\pr({\bf 3,1})_{(2,1/3)}^{(+,-)}\oplus H_U({\bf 1,2})^{(+,+)}_{(2,-1/2)}\nn~,\\
H({\bf {10}}_{\bf 1})&=& H_{TQ}({\bf 3},{\bf 2})^{(+,-)}_{\bf (1,1/6)}\oplus H_{TD}(\bar{\bf 3},1)^{(+,-)}_{\bf (1,-2/3)}\oplus H_N(1,1)^{(+,+)}_{\bf (1,1)}~,\nn\\
\overline{H}({\bf \overline{10}}_{\bf -1})&=& \overline{H}_{TQ}({\bf \bar{3}},{\bf \bar{2}})^{(+,-)}_{\bf (-1,-1/6)}\oplus \overline{H}_{TD}(\bar{\bf 3},1)^{(+,-)}_{\bf (1,-2/3)}\oplus \overline{H}_N(1,1)^{(+,+)}_{\bf (-1,-1)}~,\nn\\
O_X({\bf 1}_0)&=& X({\bf 1,1})_{\bf (0,0)}^{(+,+)}~.
\eeqa
Their conjugate superfields within the corresponding hypermultiplets are assigned with opposite parities, which are not written here explicitly.
 Such BCs can be realized by orbifold breaking via inner automorphism with the choice of parity
\beqa
P_{O(y=0)}&=&{\rm diag}\(~+1,~+1,~+1,~+1,~+1\),\nn\\
P_{O^\pr(y=\pi R)}&=&{\rm diag}\(~+1,~+1,~+1,~-1,~-1\),
\eeqa
and proper brane mass terms to change the boundary conditions from Neuman to Dirichilet. Further breaking\footnote{ We can also break the the GUT group directly into the SM gauge group via outer automorphism orbifold breaking to reduce the rank of the gauge group, for example, via charge conjugation~\cite{hebecker:2001wq,hebecker:2001jb} of combinations of $U(1)_{X}\tm U(1)_{Y^\pr}$ charges. }of $U(1)_{X}\tm U(1)_{Y^\pr}$ into $U(1)_Y$  can be triggered via proper Higgs field with the hypercharge given by eq.(\ref{hypercharge}), for example, the survived zero modes singlet components within  the $H(10,1)$ and $\overline{H}(\overline{10},-1)$ Higgses.
After orbifolding, the zero modes of bulk fields in eq.(\ref{orbifold-1}) reduce to the matter contents of SM and the corresponding RH neutrinos. We assume that the three families of $T_F^{\pr,a}({\bf 10}_{\bf 1}), T_F^{\pr\pr,a}({\bf 10}_{\bf 1})$ with $a=1,2,3$ transform under $A_4^Q\tm A_4^L$ as $(3,1)$ while $T_F^{a}({\bf 10}_{\bf 1})$ transform under $A_4^Q\tm A_4^L$ as $(1,3)$. Similarly, we require that, under $A_4^Q\tm A_4^L$, the down-type quark sector which lie within the three families of $F_{\bar{f}}^{a}(\overline{\bf 5}_{\bf -3})$ should transform under $(3,1)$ while the charged lepton sector $F_{\bar{f}}^{\pr,a}(\overline{\bf 5}_{\bf -3})$ ,$O_E^a({\bf 1}_{\bf -5})$ and the singlet sector  $O_S^a({\bf 1}_{\bf 0})$ transform as $(1,3)$. So, the zero modes for the quarks and leptons transform as $(3,1)$ and $(1,3)$ under $A_4^Q\tm A_4^L$, respectively.
 
The superpotential in the SU(5) preserving $O(y=0)$ brane can be written as
\beqa
{\cal L}&\supseteq& \delta(y)\int d^2\theta\left[ Y_{ab;23}^D T_{F}^{a;2} T_{F}^{b;3} h+ Y_{ab;12}^U T_{F}^{a;1} F_{\bar{f}}^{b;2} \bar{h} + Y_{ab;31}^U T_{F}^{a;3} F_{\bar{f}}^{b;1} \bar{h}+ Y_{ab;1}^E F_{\bar{f}}^{a;1} E^b h \right.\nn\\
 &+& \left.Y^S_{ab;1} \overline{H}  O_S^a T_{F}^{b;1}+\f{M_{SS;ab}}{2} O_S^a O_S^b+O_X(\overline{H}H-M_H^2)\right]~.
\label{5D:superpotential}
\eeqa
To simplify the expressions, we rewrite $T_F^{a},T_F^{\pr a},T_F^{\pr\pr a}$ as $T_F^{a;p}$ with $p=1,2,3$ and $F_{\bar{f}}^a,F_{\bar{f}}^{\pr,a}$ as $F_{\bar{f}}^{a;q}$ with $q=1,2$, respectively. Besides, we take the Yukawa couplings to be independent of the $p,q$ indices so that the low energy theory of 5D theory is identical to that of ordinary 4D flipped SU(5) GUT up to the RGE effects, that is, we choose $Y_{ab;pq}^D=Y_{ab}^D$,  $Y_{ab;pq}^U=Y_{ab}^U$, $Y_{ab;p}^E=Y_{ab}^E$ and $Y^{S}_{ab;p}=Y^S_{ab}$.

The product group $A_4^Q\tm A_4^L$ can be broken to diagonal $A_4^D$ by bi-fundamental superfields $\Phi(3,3)$ with VEVs of the form
\beqa
\langle\Phi\rangle_{i\al}=v_D \delta_{i\al}~,
\eeqa
in real basis where the singlet contraction is given by $(\tl{a}\tl{b})_{\bf 1}=\sum\limits_{i}\tl{a}_i\tl{b}_i$. Detailed discussions on the breaking of multiple modular symmetries by bi-fundamental Higgs fields can be found in~\cite{deMedeirosVarzielas:2019cyj}. We propose an alternative approach to break multiple modular symmetries via proper boundary conditions.

It is possible to break the multiple $A_4^Q\tm A_4^L$ by BCs to diagonal $A_4^D$, which is then identified to be the (single) modular $A_4$ symmetry in the low energy effective theory.
We can assign the following BCs for the bi-triplet ${\bf (3,3)}$ fields $\Phi_{i\al}$ of $A_4^Q\tm A_4^L$, with $i,\al$ the indices for $A_4^Q$ and $A_4^L$, respectively. 
To assign proper BCs that break the product group $A_4^Q\tm A_4^L$ to the diagonal $A_4^D$, we need to know the decomposition of the tensor indices in terms of survived diagonal subgroup $A_4^D$
\beqa
{\bf 3}\otimes{\bf 3}={\bf 1}\oplus{ {\bf 1}^\pr}\oplus{ {\bf 1}^{\pr\pr}}\oplus{\bf 3_s}\oplus{\bf 3_a},
\eeqa
with $\ga^Q\in A_4^Q$  and $\ga^L\in A_4^L$ being associated to the $\ga^D\in A_4^D$ by $\ga^Q=\ga^L=\ga^D$.
So, the $\Phi_{i\al}$ fields, which transform as bi-triplets of $A_4^Q\tm A_4^L$, will be reducible when both transformation parameters $\ga^Q,\ga^L$ are chosen to align to $\ga^D$ in $A_4^D$. The reducible tensor product can be decomposed in terms of the sum of irreducible representation $\Phi^{\bf r}$ of $A_4^D$
\beqa
\Phi_{\ka}^{\bf r}=\sum\limits_{i,\al} C_{i\al;\ka}^{\bf r} \Phi_{i\al}~,
\eeqa
with $\ka$ the indices for the irreducible representation of ${\bf r}$.
Proper Dirichlet or Neumann boundary conditions at the fix points can be assign to the fields that corresponds to various irreducible representation $\Phi_{\ka}^{\bf r}$ of survived $A_4^D$, which are proper combinations of the components $\Phi_{i\al}$~\footnote{General discussions on the BCs imposed for fields and combinations are given in~\cite{higgsless1,higgsless2,Csaki:2003dt}.}. For example, we can assign
\beqa
\Phi=(\Phi^{\bf 1})^{(++)}\oplus (\Phi^{\bf 1^\pr})^{(+-)}\oplus (\Phi^{\bf 1^{\pr\pr}})^{(+-)}\oplus (\Phi_{\ka}^{\bf 3_s})^{(+-)}\oplus (\Phi_{\ka}^{\bf 3_a})^{(+-)}~.
\label{BC2}
\eeqa
with
\beqa
\Phi^{\bf 1}&=&\f{1}{\sqrt{3}}\(\Phi_{11}+\Phi_{23}+\Phi_{32}\)~,\nn\\
\Phi^{\bf 1^\pr}&=&\f{1}{\sqrt{3}}\(\Phi_{12}+\Phi_{21}+\Phi_{33}\)~,\nn\\
\Phi^{\bf 1^{\pr\pr}}&=&\f{1}{\sqrt{3}}\(\Phi_{13}+\Phi_{22}+\Phi_{31}\)~,\nn\\
\Phi_{\ka}^{\bf 3_s}&=&\f{1}{\sqrt{6}}\(2\Phi_{11}-\Phi_{23}-\Phi_{32},
2\Phi_{33}-\Phi_{12}-\Phi_{21},2\Phi_{22}-\Phi_{13}-\Phi_{31} \)~,\nn\\
\Phi_{\ka}^{\bf 3_a}&=&\f{1}{\sqrt{2}}\(\Phi_{23}-\Phi_{32},
\Phi_{12}-\Phi_{21},\Phi_{13}-\Phi_{31}\)~,
\eeqa
so as that only the zero modes of the singlet (that is, $\Phi^{\bf 1}$) survives. BCs that lead to survived zero modes for any combination of the representations (but not all of them simultaneously) in eq.(\ref{BC2}) can be allowed to act as the BCs that break the $A_4^Q\tm A_4^L$ to $A_4^D$. Other choices of the combination of components $\Phi_{i\al}$ correspond to different symmetry breaking chains. For example, choices with $\Phi_{i\al}^{(++)}$ for fixed $'i'$ (or $'\al'$) corresponds to the breaking of $A_4^Q\tm A_4^L$ to $A_4^Q$ (or $A_4^L$), respectively. Survived zero modes for other combinations other than the previous BCs correspond to the fully breaking of $A_4^Q\tm A_4^L$.

The modular forms and superfield $\phi(\tau_Q,\tau_L)$ in $({\bf r_Q,r_L})$ representation of $A_4^Q\tm A_4^L$ with modular weights $(k_Q,k_L)$ will transform as
\begin{eqnarray}
\phi(\gamma_D\tau_Q,\gamma_D\tau_L ) &=& (c_D \tau_Q + d_D)^{-2k_Q}(c_D \tau_L + d_D)^{-2k_L} \rho_{\mathbf{r_Q}}(\gamma_D)\otimes\rho_{\mathbf{r_L}}(\gamma_D) \phi(\tau_Q,\tau_L)~,\nn\\
 Y_\alpha(\gamma_D\tau_{Q,L}) &=& (c_D \tau_{Q,L} + d_D)^{2k_\alpha} \rho_{\mathbf{\alpha}}(\gamma_D) Y_\alpha(\tau_{Q,L}) \,,
\end{eqnarray}
under the diagonal $A_4^D$, respectively. Here $\ga^Q=\ga^L=\ga^D$ after reduction of $A_4^Q\tm A_4^L$ to $A_4^D$ (for $\ga^Q\in A_4^Q$  and $\ga^L\in A_4^L$). As the UV theory involving multiple moduli fields is invariant under multiple modular transformations, the low energy effective theory with two (multiple) moduli fields is also obviously invariant under the single modular transformations of $A_4^D$.

\section{\label{sec:cla}Classification according to the choice of representation and modular weights}
According to the assignments of the modular $A_4$ representations for matter superfields and the values of modular weights, we can classify all the possible scenarios in this $A_4$ modular flavor flipped SU(5) GUT scheme. In our subsequent studies, the modular $A_4$ representations for matter superfields are given for single modulus scenarios. The classification of the scenarios according to the modular $A_4$ representations in the subsequent discussions can also be extended straightforwardly to the multiple modulus cases in five-dimensional theory, for example, with the corresponding replacements  
\beqas \rho_F\ra \rho_{T_F^{;p}},~\rho_{\bar{f}} \ra \rho_{F_{\bar{f}}^{;q}},~\rho_E\ra \rho_{O_E},~\rho_S\ra \rho_{O_S}.
\eeqas
It is also easy to extend the ordinary 4D superpotential for single modulus scenarios to the superpotential at the GUT symmetry preserving fixed point for multiple modulus scenarios by the replacements $F^a\ra T_F^{a, p}$, $\bar{f}^a\ra F_{\bar{f}}^{a, q}$, $E^a\ra O_E^a$ and $S^a\ra O_S^a$ similar to that in eq.(\ref{5D:superpotential}), within which the relevant coefficients of the Yukawa coupling terms are taken to be independent of such $p,q$ indices. We adopt the symbol conventions in~\cite{Chen:2021zty} with
\beqa
S_1^{(k)}(\tau)=Y_{\mathbf1}^{(k)}(\tau)\(\bea{ccc}1&0&0\\0&0&1\\0&1&0\eea\),~~
S^{(k)}_{\mathbf1'} =
Y_{\mathbf1'}^{(k)}(\tau)\left(
\begin{matrix}
 0 ~& 0 ~& 1 \\
 0 ~& 1 ~& 0 \\
 1 ~& 0 ~& 0 \\
\end{matrix}
\right)\,,~~
S^{(k)}_{\mathbf1''} =
Y_{\mathbf1''}^{(k)}(\tau)\left(
\begin{matrix}
 0 ~& 1 ~& 0 \\
 1 ~& 0 ~& 0 \\
 0 ~& 0 ~& 1 \\
\end{matrix}
\right)\,,\nn\\
\eeqa
and
\begin{eqnarray}
S^{(k)}_{{\mathbf3}}(\tau) &=&
\left(\begin{matrix}
 2Y^{(k)}_{{{\mathbf3}},1}(\tau) ~& -Y^{(k)}_{{{\mathbf3}},3}(\tau)  ~& -Y^{(k)}_{{{\mathbf3}},2}(\tau) \\
 -Y^{(k)}_{{{\mathbf3}},3}(\tau) ~& 2Y^{(k)}_{{{\mathbf3}},2}(\tau)  ~& -Y^{(k)}_{{{\mathbf3}},1}(\tau) \\
 -Y^{(k)}_{{{\mathbf3}},2}(\tau) ~& -Y^{(k)}_{{{\mathbf3}},1}(\tau)  ~& 2Y^{(k)}_{{{\mathbf3}},3}(\tau)
\end{matrix}
\right)\,,\nn\\
A^{(k)}_{{\mathbf3}}(\tau) &=&
\left(\begin{matrix}
 0 ~& Y^{(k)}_{{{\mathbf3}},3}(\tau)  ~&~ -Y^{(k)}_{{{\mathbf3}},2}(\tau) \\
 -Y^{(k)}_{{{\mathbf3}},3}(\tau) ~& 0  ~& Y^{(k)}_{{{\mathbf3}},1}(\tau) \\
 Y^{(k)}_{{{\mathbf3}},2}(\tau) ~& -Y^{(k)}_{{{\mathbf3}},1}(\tau)  ~& 0
\end{matrix}\right)\,.
\end{eqnarray}

\subsection{Up-type Quark Sector }

\bit
\item $\rho_{\bar{f}}={\bf 3},~\rho_{F}={\bf 3}$.

As noted in the previous paragraph, such choices of representations correspond to
\beqa
\rho_{T_F^{\pr\pr}}\equiv \rho_{T_F^{;3}} ={\bf 3},~~~~~\rho_{F_{\bar{f}^{\pr}}}\equiv\rho_{F_{\bar{f}^{;2}}} ={\bf 3},
\eeqa
for multiple modulus scenarios in five-dimensional cases. Similar replacements can be adopted for other choices of modular $A_4$ representations and the assignments of the modular weights.

According to the production expansions of irreducible representation {\bf 3}s of $A_4$ in~eq.(\ref{A4:33product}), we can get the Yukawa couplings for the up-type quarks when the corresponding modular weights $k_{\phi_i}$ for various fields $\phi_i$ are given 
\beqa
&&k_{\bar{f}}+k_{{F}}=0; ~~\(y_{U}\)_{ij}= \beta_1 S_\mathbf{1}^0(\tau)~,\\
&&k_{\bar{f}}+k_{{F}}=2;~~\(y_{U}\)_{ij}=\beta_1 S_\mathbf{3}^{(2)}(\tau)+\beta_2 A_\mathbf{3}^{(2)}(\tau) ~,\nn\\
&&k_{\bar{f}}+k_{{F}}=4;~~\(y_{U}\)_{ij}=\beta_1\,S^{(4)}_{\mathbf3} + \beta_2 A^{(4)}_{\mathbf3} + \beta_3 S^{(4)}_{\mathbf1} +\beta_4  S^{(4)}_{\mathbf1'}\,,\nn\\
&&k_{\bar{f}}+k_{{F}}=6;~~\(y_{U}\)_{ij}=\beta_1\,S^{(6)}_{\mathbf{3}I} +\beta_2 A^{(6)}_{\mathbf{3}I} + \beta_3 S^{(6)}_{\mathbf{3}II} +\beta_4 A^{(6)}_{\mathbf{3}II} +\beta_5 S^{(6)}_{\mathbf1}\,,\nn\\
&&k_{\bar{f}}+k_{{F}}=8;\nn\\
&&\(y_{U}\)_{ij}= \beta_1 \,S^{(8)}_{\mathbf{3}I} +\beta_2 A^{(8)}_{\mathbf{3}I} +\beta_3 S^{(8)}_{\mathbf{3}II} +\beta_4 A^{(8)}_{\mathbf{3}II} + \beta_5 S^{(8)}_{\mathbf1} + \beta_6 S^{(8)}_{\mathbf1'} +\beta_7 S^{(8)}_{\mathbf1''}\,,\nn
\eeqa
Note that $S_1^{(2)}(\tau)=0$.

Here we adopt the symbol convention in~\cite{Chen:2021zty} with
\beqa
S_1^{(k)}(\tau)=Y_{\mathbf1}^{(k)}(\tau)\(\bea{ccc}1&0&0\\0&0&1\\0&1&0\eea\),
~~~S^{(k)}_{\mathbf1'} =
Y_{\mathbf1'}^{(k)}(\tau)\left(
\begin{matrix}
 0 ~& 0 ~& 1 \\
 0 ~& 1 ~& 0 \\
 1 ~& 0 ~& 0 \\
\end{matrix}
\right)\,,~~~~~~~~~~~~~~~~~~~~~
\eeqa
\begin{eqnarray}S^{(k)}_{\mathbf1''} =
Y_{\mathbf1''}^{(k)}(\tau)\left(
\begin{matrix}
 0 ~& 1 ~& 0 \\
 1 ~& 0 ~& 0 \\
 0 ~& 0 ~& 1 \\
\end{matrix}
\right)\,,
S^{(k)}_{{\mathbf3}}(\tau) =
\left(\begin{matrix}
 2Y^{(k)}_{{{\mathbf3}},1}(\tau) ~& -Y^{(k)}_{{{\mathbf3}},3}(\tau)  ~& -Y^{(k)}_{{{\mathbf3}},2}(\tau) \\
 -Y^{(k)}_{{{\mathbf3}},3}(\tau) ~& 2Y^{(k)}_{{{\mathbf3}},2}(\tau)  ~& -Y^{(k)}_{{{\mathbf3}},1}(\tau) \\
 -Y^{(k)}_{{{\mathbf3}},2}(\tau) ~& -Y^{(k)}_{{{\mathbf3}},1}(\tau)  ~& 2Y^{(k)}_{{{\mathbf3}},3}(\tau)
\end{matrix}
\right)\,,
\end{eqnarray}
and
\begin{eqnarray}
A^{(k)}_{{\mathbf3}}(\tau) =
\left(\begin{matrix}
 0 ~& Y^{(k)}_{{{\mathbf3}},3}(\tau)  ~&~ -Y^{(k)}_{{{\mathbf3}},2}(\tau) \\
 -Y^{(k)}_{{{\mathbf3}},3}(\tau) ~& 0  ~& Y^{(k)}_{{{\mathbf3}},1}(\tau) \\
 Y^{(k)}_{{{\mathbf3}},2}(\tau) ~& -Y^{(k)}_{{{\mathbf3}},1}(\tau)  ~& 0
\end{matrix}\right)\,.
\end{eqnarray}

\item $\rho_{\bar{f}}={\bf 3},~\rho_{F}={\bf 1,1^\pr,1^{\pr\pr}}$.

Given the modular weights $k_{\phi_i}$ for various fields $\phi_i$, we have the Yukawa couplings
\beqa
k\equiv k_{\bar{f}}+k_{{F}_i}=2,4;~~~~~~~~~~~~~\(y_{U}\)_{i*}=\beta_1\left(\bea{ccc}Y^{(k)}_{{\mathbf3},1}& Y^{(k)}_{{\mathbf3},3}& Y^{(k)}_{{\mathbf3},2}\\
Y^{(k)}_{{\mathbf3},3}& Y^{(k)}_{{\mathbf3},2}&Y^{(k)}_{{\mathbf3},1}\\Y^{(k)}_{{\mathbf3},2}&Y^{(k)}_{{\mathbf3},1}&Y^{(k)}_{{\mathbf3},3}\\
\eea\right)~,\eeqa
Here the row matrix $i=1$ corresponds to $\rho_{F}={\bf 1}$; $i=2$ corresponds to $\rho_{F}={\bf 1^\pr}$; $i=3$ corresponds to $\rho_{F}={\bf 1^{\pr\pr}}$.
\beqa
&&k\equiv k_{\bar{f}}+k_{F}=6,8;~\nn\\
&&~\(y_{U}\)_{i*}=\beta_1 \left(\bea{ccc}Y^{(k)}_{{\mathbf 3I},1}& Y^{(k)}_{{\mathbf3I},3}& Y^{(k)}_{{\mathbf3I},2}\\Y^{(k)}_{{\mathbf3I},3}& Y^{(k)}_{{\mathbf3I},2}&Y^{(k)}_{{\mathbf3I},1}\\
Y^{(k)}_{{\mathbf3I},2}&Y^{(k)}_{{\mathbf3I},1}&Y^{(k)}_{{\mathbf3I},3}\eea\right)
+\beta_2\left(\bea{ccc}Y^{(k)}_{{\mathbf3II},1}& Y^{(k)}_{{\mathbf3II},3}& Y^{(k)}_{{\mathbf3II},2}\\Y^{(k)}_{{\mathbf3II},3}& Y^{(k)}_{{\mathbf3II},2}&Y^{(k)}_{{\mathbf3II},1}\\
Y^{(k)}_{{\mathbf3II},2}&Y^{(k)}_{{\mathbf3II},1}&Y^{(k)}_{{\mathbf3II},3}
\eea\right)~.
\eeqa
Here the row matrix $i=1$ corresponds to $\rho_{F}={\bf 1}$; $i=2$ corresponds to $\rho_{F}={\bf 1^\pr}$; $i=3$ corresponds to $\rho_{F}={\bf 1^{\pr\pr}}$.

\item  $\rho_{\bar{f}}={\bf 1,1^\pr,1^{\pr\pr}},~\rho_{F}={\bf 3}$.

Given the modular weights, we have the Yukawa couplings
\beqa
k\equiv k_{{F}}+k_{\bar{f}_j}=2,4;~~~~~~~~\(y_{U}\)_{*j}=\beta_1\left(\bea{ccc}Y^{(k)}_{{\mathbf3},1}& Y^{(k)}_{{\mathbf3},3}& Y^{(k)}_{{\mathbf3},2}\\
Y^{(k)}_{{\mathbf3},3}& Y^{(k)}_{{\mathbf3},2}&Y^{(k)}_{{\mathbf3},1}\\Y^{(k)}_{{\mathbf3},2}&Y^{(k)}_{{\mathbf3},1}&Y^{(k)}_{{\mathbf3},3}\\
\eea\right)~.\eeqa
Here the column matrix $j=1$ corresponds to $\rho_{\bar{f}}={\bf 1}$; $j=2$ corresponds to $\rho_{\bar{f}}={\bf 1^\pr}$; $j=3$ corresponds to $\rho_{\bar{f}}={\bf 1^{\pr\pr}}$.
\beqa
&&k\equiv k_{\bar{f}}+k_{F}=6,8;\nn\\
&&\(y_{U}\)_{*j}=\beta_1 \left(\bea{ccc}Y^{(k)}_{{\mathbf 3I},1}& Y^{(k)}_{{\mathbf3I},3}& Y^{(k)}_{{\mathbf3I},2}\\Y^{(k)}_{{\mathbf3I},3}& Y^{(k)}_{{\mathbf3I},2}&Y^{(k)}_{{\mathbf3I},1}\\
Y^{(k)}_{{\mathbf3I},2}&Y^{(k)}_{{\mathbf3I},1}&Y^{(k)}_{{\mathbf3I},3}\eea\right)
+\beta_2\left(\bea{ccc}Y^{(k)}_{{\mathbf3II},1}& Y^{(k)}_{{\mathbf3II},3}& Y^{(k)}_{{\mathbf3II},2}\\Y^{(k)}_{{\mathbf3II},3}& Y^{(k)}_{{\mathbf3II},2}&Y^{(k)}_{{\mathbf3II},1}\\
Y^{(k)}_{{\mathbf3II},2}&Y^{(k)}_{{\mathbf3II},1}&Y^{(k)}_{{\mathbf3II},3}
\eea\right)~.
\eeqa
Here the column matrix $j=1$ corresponds to $\rho_{F}={\bf 1}$; $j=2$ corresponds to $\rho_{F}={\bf 1^\pr}$; $j=3$ corresponds to $\rho_{F}={\bf 1^{\pr\pr}}$.

\item $\rho_{\bar{f}}={\bf 1,1^\pr,1^{\pr\pr}},~\rho_{F}={\bf 1,1^\pr,1^{\pr\pr}}$.

Given the modular weights, we have the Yukawa couplings
\beqa
&& k_{\bar{f}_i}+k_{F_j}=0; ~~~\(y_{U}\)_{ij}=\beta_1\left\{\bea{c}1,~~~~~~\rho_{\bar{f}_i}\otimes \rho_{F_j}={\bf 1}\\0,~~~~~~\rho_{\bar{f}_i}\otimes \rho_{F_j}\neq {\bf 1}\eea\right.\nn\\
&& k_{\bar{f}_i}+k_{F_j}=2; ~~~\(y_{U}\)_{ij}= 0~,\nn\\
&& k_{\bar{f}_i}+k_{F_j}=4; ~~~\(y_{U}\)_{ij}=\beta_1\left\{\bea{c}Y_{\bf 1}^{(4)},~~~~~~\rho_{\bar{f}_i}\otimes \rho_{F_j}={\bf 1}\\ Y_{\bf 1^\pr}^{(4)},~~~~~~~\rho_{\bar{f}_i}\otimes \rho_{F_j}= {\bf 1^{\pr\pr}}\\
0~~,~~~~~~~~~~~otherwise\eea\right.\nn\\
&& k_{\bar{f}_i}+k_{F_j}=6; ~~~\(y_{U}\)_{ij}=\beta_1\left\{\bea{c}Y_{\bf 1}^{(6)},~~~~~~\rho_{\bar{f}_i}\otimes \rho_{F_j}={\bf 1}\\~~~~0~,~~~~~~~\rho_{\bar{f}_i}\otimes \rho_{F_j}\neq {\bf 1}\eea\right.\nn\\
&& k_{\bar{f}_i}+k_{F_j}=8; ~~~\(y_{U}\)_{ij}=\beta_1\left\{\bea{c}Y_{\bf 1}^{(8)},~~~~~~\rho_{\bar{f}_i}\otimes \rho_{F_j}={\bf 1}\\ Y_{\bf 1^\pr}^{(8)},~~~~~~~\rho_{\bar{f}_i}\otimes \rho_{F_j}= {\bf 1^{\pr\pr}}\\
Y_{\bf 1^{\pr\pr}}^{(8)},~~~~~~~\rho_{\bar{f}_i}\otimes \rho_{F_j}= {\bf 1^{\pr}}\\
0~~,~~~~~~~~~~~otherwise\eea\right.
\eeqa

\eit

The Dirac neutrino mass terms take the same form as $\(y_{U}\)_{ij}$.

\subsection{Down-type Quark Sector }
\bit
\item $\rho_{F}={\bf 3}$.

As noted in the previous paragraphs, such choices of representations correspond to
$\rho_{T_F^{\pr}}\equiv\rho_{T_F^{;2}} ={\bf 3}$, $\rho_{T_F^{;3}} ={\bf 3}$ for multiple modulus scenarios in five-dimensional cases. Similar replacements can be adopted for other choices of modular $A_4$ representations and the assignments of the modular weights.

With the following assignments of modular weights, we have the form of the Yukawa couplings
\beqa
&&2 k_{{F}}=0; ~~~\(y_{D}\)_{ij}=\al_1 S_\mathbf{1}^0(\tau)~,\nn\\
&&2 k_{{F}}=2; ~~~\(y_{D}\)_{ij}=\al_1 S_\mathbf{3}^{(2)}(\tau) ~,\nn\\
&&2k_{{F}}=4; ~~~\(y_{D}\)_{ij}=\al_1\,S^{(4)}_{\mathbf3}  +\al_2 S^{(4)}_{\mathbf1} + \al_3 S^{(4)}_{\mathbf1'}\,,\nn\\
&&2k_{{F}}=6; ~~~\(y_{D}\)_{ij}=\al_1\,S^{(6)}_{\mathbf{3}I} +\al_2 S^{(6)}_{\mathbf{3}II} + \al_3 S^{(6)}_{\mathbf1}\,,\nn\\
&&2k_{{F}}=8; ~~~\(y_{D}\)_{ij}=\al_1 S^{(8)}_{\mathbf{3}I} +\al_2 S^{(8)}_{\mathbf{3}II} +\al_3 S^{(8)}_{\mathbf1}+\al_4 S^{(8)}_{\mathbf1'} +\al_5 S^{(8)}_{\mathbf1''}\,,
\eeqa
\item  $\rho_{F}={\bf 1,1^\pr,1^{\pr\pr}}$.

Similarly, given the modular weights, the Yukawa couplings take the form
\beqa
&& k_{{F}_i}+k_{{F}_j}=0: ~~~\(y_{D}\)_{ij}=\al_1\left\{\bea{c}1,~~~~~~\rho_{F_i}\otimes \rho_{F_j}={\bf 1}\\0,~~~~~~\rho_{F_i}\otimes \rho_{F_j}\neq {\bf 1}\eea\right.\nn\\
&& k_{{F}_i}+k_{{F}_j}=2: ~~~\(y_{D}\)_{ij}= 0~,\nn\\
&& k_{{F}_i}+k_{{F}_j}=4: ~~~\(y_{D}\)_{ij}=\al_1\left\{\bea{c}Y_{\bf 1}^{(4)},~~~~~\rho_{F_i}\otimes \rho_{F_j}={\bf 1}\\ Y_{\bf 1^\pr}^{(4)},~~~~~~~\rho_{F_i}\otimes \rho_{F_j}= {\bf 1^{\pr\pr}}\\
0~~,~~~~~~~~~~~otherwise\eea\right.~\nn\\
&& k_{{F}_i}+k_{{F}_j}=6: ~~~\(y_{D}\)_{ij}=\al_1\left\{\bea{c}Y_{\bf 1}^{(6)},~~~~~~\rho_{F_i}\otimes \rho_{F_j}={\bf 1}\\~~~~0~,~~~~~~~\rho_{F_i}\otimes \rho_{F_j}\neq {\bf 1}\eea\right.\nn\\
&& k_{{F}_i}+k_{{F}_j}=8: ~~~\(y_{D}\)_{ij}=\al_1\left\{\bea{c}Y_{\bf 1}^{(8)},~~~~~~\rho_{\bar{f}_i}\otimes \rho_{F_j}={\bf 1}\\ Y_{\bf 1^\pr}^{(8)},~~~~~~~\rho_{\bar{f}_i}\otimes \rho_{F_j}= {\bf 1^{\pr\pr}}\\
Y_{\bf 1^{\pr\pr}}^{(8)},~~~~~~~\rho_{\bar{f}_i}\otimes \rho_{F_j}= {\bf 1^{\pr}}\\
0~~,~~~~~~~~~~~otherwise\eea\right.
\eeqa

\eit
\subsection{Charged lepton Sector }
\bit
\item $\rho_{\bar{f}}={\bf 3},\rho_{E}={\bf 3}$.

Such choices of representations correspond to
$\rho_{F_{\bar{f}}^{\pr}}\equiv\rho_{F_{\bar{f}}^{;2}} ={\bf 3}$, $\rho_{O_E} ={\bf 3}$ for multiple modulus scenarios in five-dimensional cases. Similar replacements can be adopted for other choices of modular $A_4$ representations.

Given the modular weights $k_{\phi_i}$ for various fields $\phi_i$, we have the Yukawa couplings
\beqa
&&k_{\bar{f}}+k_{{E}}=0; ~~\(y_{E}\)_{ij}=\ga_1 S_\mathbf{1}^0(\tau)~,\nn\\
&&k_{\bar{f}}+k_{{E}}=2; ~~\(y_{E}\)_{ij}=\ga_1 S_\mathbf{3}^{(2)}(\tau)+\ga_2 A_\mathbf{3}^{(2)}(\tau) ~,\nn\\
&&k_{\bar{f}}+k_{{E}}=4; ~~\(y_{E}\)_{ij}=\ga_1 \,S^{(4)}_{\mathbf3} +\ga_2  A^{(4)}_{\mathbf3} + \ga_3 S^{(4)}_{\mathbf1} +\ga_4 S^{(4)}_{\mathbf1'}\,,\nn\\
&&k_{\bar{f}}+k_{{E}}=6; ~~\(y_{E}\)_{ij}=\ga_1\,S^{(6)}_{\mathbf{3}I} +\ga_2 A^{(6)}_{\mathbf{3}I} +\ga_3 S^{(6)}_{\mathbf{3}II} +\ga_4 A^{(6)}_{\mathbf{3}II} +\ga_5 S^{(6)}_{\mathbf1}\,,\nn\\
&&k_{\bar{f}}+k_{{E}}=8;~~\\
&&\(y_{E}\)_{ij}=\ga_1 \,S^{(8)}_{\mathbf{3}I} +\ga_2 A^{(8)}_{\mathbf{3}I} +\ga_3 S^{(8)}_{\mathbf{3}II} +\ga_4 A^{(8)}_{\mathbf{3}II} +\ga_5 S^{(8)}_{\mathbf1} +\ga_6 S^{(8)}_{\mathbf1'} +\ga_7 S^{(8)}_{\mathbf1''}\,,\nn
\eeqa
\item $\rho_{\bar{f}}={\bf 3},\rho_E={\bf 1,1^\pr,1^{\pr\pr}}$.

Given the modular weights, the Yukawa couplings take the form
\beqa
k\equiv k_{\bar{f}}+k_{E_j}=2,4;~~~~~\(y_{E}\)_{*j}=\ga_1\left(\bea{ccc}Y^{(k)}_{{\mathbf3},1}& Y^{(k)}_{{\mathbf3},3}& Y^{(k)}_{{\mathbf3},2}\\
Y^{(k)}_{{\mathbf3},3}&Y^{(k)}_{{\mathbf3},2}&Y^{(k)}_{{\mathbf3},1}\\Y^{(k)}_{{\mathbf3},2}& Y^{(k)}_{{\mathbf3},1}&Y^{(k)}_{{\mathbf3},3}
\eea\right)~,\label{yukawaE1}\eeqa
Here the column matrix $j=1$ corresponds to $\rho_{E}={\bf 1}$; $j=2$ corresponds to $\rho_{E}={\bf 1^\pr}$; $j=3$ corresponds to $\rho_{E}={\bf 1^{\pr\pr}}$.
\beqa
&&k\equiv k_{\bar{f}}+k_{E}=6,8;\\
&&\(y_{E}\)_{*j}=\ga_1 \left(\bea{ccc}Y^{(k)}_{{\mathbf 3I},1}& Y^{(k)}_{{\mathbf3I},3}& Y^{(k)}_{{\mathbf3I},2}\\Y^{(k)}_{{\mathbf3I},3}& Y^{(k)}_{{\mathbf3I},2}&Y^{(k)}_{{\mathbf3I},1}\\
Y^{(k)}_{{\mathbf3I},2}&Y^{(k)}_{{\mathbf3I},1}&Y^{(k)}_{{\mathbf3I},3}\eea\right)
+\ga_2\left(\bea{ccc}Y^{(k)}_{{\mathbf3II},1}& Y^{(k)}_{{\mathbf3II},3}& Y^{(k)}_{{\mathbf3II},2}\\Y^{(k)}_{{\mathbf3II},3}& Y^{(k)}_{{\mathbf3II},2}&Y^{(k)}_{{\mathbf3II},1}\\
Y^{(k)}_{{\mathbf3II},2}&Y^{(k)}_{{\mathbf3II},1}&Y^{(k)}_{{\mathbf3II},3}
\eea\right)~.\nn\label{yukawaE2}
\eeqa
Here the column matrix $j=1$ corresponds to $\rho_{E}={\bf 1}$; $j=2$ corresponds to $\rho_{E}={\bf 1^\pr}$; $j=3$ corresponds to $\rho_{E}={\bf 1^{\pr\pr}}$.

\item $\rho_{\bar{f}}={\bf 1,1^\pr,1^{\pr\pr}},\rho_E={\bf 3}$.

\bit
\item For $k_{\bar{f}_i}+k_{E}=2,4$, the expression of $\(y_{E}\)_{i*}$ is the same as eq.(\ref{yukawaE1}).
\item For $k_{\bar{f}_i}+k_{E}=6,8$, the expression of $\(y_{E}\)_{i*}$ is the same as eq.(\ref{yukawaE2}).
 \eit
Here the row matrix $i=1$ corresponds to $\rho_{\bar{f}}={\bf 1}$; $j=2$ corresponds to $\rho_{\bar{f}}={\bf 1^\pr}$; $j=3$ corresponds to $\rho_{\bar{f}}={\bf 1^{\pr\pr}}$.

\item  $\rho_{\bar{f}}={\bf 1,1^\pr,1^{\pr\pr}},\rho_E={\bf 1,1^\pr,1^{\pr\pr}}$.

Given the modular weights, the Yukawa couplings take the form
\beqa
&& k_{\bar{f}_i}+k_{{E}_j}=0; ~~~\(y_{E}\)_{ij}=\ga_1\left\{\bea{c}1,~~~~~~\rho_{\bar{f}_i}\otimes \rho_{{E}_j}={\bf 1}\\0,~~~~~~\rho_{\bar{f}_i}\otimes \rho_{{E}_j}\neq {\bf 1}\eea\right.\nn\\
&& k_{\bar{f}_i}+k_{{E}_j}=2; ~~~\(y_{E}\)_{ij}= 0~,\nn\\
&& k_{\bar{f}_i}+k_{{E}_j}=4; ~~~\(y_{E}\)_{ij}=\ga_1\left\{\bea{c}Y_{\bf 1}^{(4)},~~~~~~\rho_{\bar{f}_i}\otimes \rho_{{E}_j}={\bf 1}\\ Y_{\bf 1^\pr}^{(4)},~~~~~~~\rho_{\bar{f}_i}\otimes \rho_{{E}_j}= {\bf 1^{\pr\pr}}\\
0~~,~~~~~~~~~~~otherwise\eea\right.\nn\eeqa\beqa
&& k_{\bar{f}_i}+k_{{E}_j}=6; ~~~\(y_{E}\)_{ij}=\ga_1\left\{\bea{c}Y_{\bf 1}^{(6)},~~~~~~\rho_{\bar{f}_i}\otimes \rho_{{E}_j}={\bf 1}\\~~~~0~,~~~~~~~\rho_{\bar{f}_i}\otimes \rho_{{E}_j}\neq {\bf 1}\eea\right.\nn\\
&& k_{\bar{f}_i}+k_{{E}_j}=8; ~~~\(y_{E}\)_{ij}=\ga_1\left\{\bea{c}Y_{\bf 1}^{(8)},~~~~~~\rho_{\bar{f}_i}\otimes \rho_{F_j}={\bf 1}\\ Y_{\bf 1^\pr}^{(8)},~~~~~~~\rho_{\bar{f}_i}\otimes \rho_{F_j}= {\bf 1^{\pr\pr}}\\
Y_{\bf 1^{\pr\pr}}^{(8)},~~~~~~~\rho_{\bar{f}_i}\otimes \rho_{F_j}= {\bf 1^{\pr}}\\
0~~,~~~~~~~~~~~otherwise\eea\right.
\eeqa
\eit
\subsection{Neutrino sector}
\bit
\item $\rho_{F}={\bf 3},\rho_{S}={\bf 3}$.

Such choices of representations correspond to
$$\rho_{T_F}\equiv\rho_{T_F^{;1}} ={\bf 3}, ~\rho_{F_{\bar{f}}^{;2}} ={\bf 3}, ~\rho_{O_S} ={\bf 3}$$
 for multiple modulus scenarios in five-dimensional cases. Similar replacements can be adopted for other choices of modular $A_4$ representations.

Given the modular weights $k_{\phi_i}$, the $S-N$ mixing matrix takes the form
\beqa
&&k_{F}+k_{S}=0; ~~~{\cal M}^{SN}_{ij}=\la_1\Lambda_1 S_\mathbf{1}^0(\tau)~,\nn\\
&&k_{F}+k_{S}=2; ~~~{\cal M}^{SN}_{ij}=\la_1\Lambda_1 S_\mathbf{3}^{(2)}(\tau)+\la_2\Lambda_1 A_\mathbf{3}^{(2)}(\tau) ~,\nn\\
&&k_{F}+k_{S}=4; ~~~{\cal M}^{SN}_{ij}=\la_1\,\Lambda_1 S^{(4)}_{\mathbf3} +\la_2 \Lambda_1 A^{(4)}_{\mathbf3} +\la_3\Lambda_1 S^{(4)}_{\mathbf1} + \la_4\Lambda_1 S^{(4)}_{\mathbf1'}\,,\nn\\
&&k_{F}+k_{S}=6;\nn\\
 &&{\cal M}^{SN}_{ij}=\la_1\,\Lambda_1 S^{(6)}_{\mathbf{3}I} +\la_2 \Lambda_1 A^{(6)}_{\mathbf{3}I} +\la_1 \Lambda_3 S^{(6)}_{\mathbf{3}II} +\la_4\Lambda_1 A^{(6)}_{\mathbf{3}II} +\la_5\Lambda_1 S^{(6)}_{\mathbf1}\,,\nn\\
&&k_{F}+k_{S}=8;~~~\nn\\
&&{\cal M}^{SN}_{ij}= \la_1\Lambda_1\,S^{(8)}_{\mathbf{3}I} +\la_2 \Lambda_1 A^{(8)}_{\mathbf{3}I} + \la_3\Lambda_1 S^{(8)}_{\mathbf{3}II} +\la_4\Lambda_1  A^{(8)}_{\mathbf{3}II} + \la_5\Lambda_1 S^{(8)}_{\mathbf1}\nn\\
&&~~~~~~~ +\la_6 \Lambda_1 S^{(8)}_{\mathbf1'} +\la_7\Lambda_1 S^{(8)}_{\mathbf1''}\,,
\eeqa
with $\Lambda_1$ the typical mass scale for $S-N$ mixing.

The mass matrix for ${\cal M}^{SS}$ takes the form
\beqa
&& 2k_{S}=0;~~~~{\cal M}^{SS}_{ij}=\ka_1\Lambda_2 S_1^0(\tau)~,\\
&&2 k_{S}=2; ~~~~{\cal M}^{SS}_{ij}=\ka_1\Lambda_2 S_3^{(2)}(\tau) ~,\nn\\
&&2k_{S}=4; ~~~~{\cal M}^{SS}_{ij}=\ka_1\Lambda_2 \,S^{(4)}_{\mathbf3}  +\ka_2\Lambda_2 S^{(4)}_{\mathbf1} +\ka_3\Lambda_2 S^{(4)}_{\mathbf1'}\,,\nn\\
&&2k_{S}=6; ~~~~{\cal M}^{SS}_{ij}=\ka_1\Lambda_2\,S^{(6)}_{\mathbf{3}I} +\ka_2\Lambda_2 S^{(6)}_{\mathbf{3}II} +\ka_3\Lambda_2 S^{(6)}_{\mathbf1}\,,\nn\\
&&2k_{S}=8;\nn\\
&&{\cal M}^{SS}_{ij}=\ka_1 \Lambda_2\,S^{(8)}_{\mathbf{3}I}  +\ka_2 \Lambda_2 S^{(8)}_{\mathbf{3}II} +\ka_3 \Lambda_2 S^{(8)}_{\mathbf1} +\ka_4 \Lambda_2 S^{(8)}_{\mathbf1'} +\ka_5\Lambda_2 S^{(8)}_{\mathbf1''}\,,\nn
\eeqa
with various values of the modular weights $k_{S}$. Here $\Lambda_2$ denotes the small mass scale for new neutrinos $S_i$, which is a small lepton number violating parameter that is responsible for the smallness of the light neutrinos.

\item $\rho_{F}={\bf 3},\rho_S={\bf 1,1^\pr,1^{\pr\pr}}$.

The $S-N$ mixing matrices take the forms
\beqa
k\equiv k_{F}+k_{S_j}=2,4;~~~~~{\cal M}^{SN}_{*j}=\la_1\Lambda_1 \left(\bea{ccc}Y^{(k)}_{{\mathbf3},1}& Y^{(k)}_{{\mathbf3},3}& Y^{(k)}_{{\mathbf3},2}\\Y^{(k)}_{{\mathbf3},3}& Y^{(k)}_{{\mathbf3},2}&Y^{(k)}_{{\mathbf3},1}\\
Y^{(k)}_{{\mathbf3},2}&Y^{(k)}_{{\mathbf3},1}&Y^{(k)}_{{\mathbf3},3}\\
\eea\right)~.\label{SN1}\eeqa\\
While for the modular weight choices $k\equiv k_{F}+k_{S}=6,8$, the $S-N$ mixing matrices takes the forms
\beqa
&& \nn\\&&{\cal M}^{SN}_{*j}=\la_1\Lambda_1 \alpha \left(\bea{ccc}Y^{(k)}_{{\mathbf 3I},1}& Y^{(k)}_{{\mathbf3I},3}& Y^{(k)}_{{\mathbf3I},2}\\Y^{(k)}_{{\mathbf3I},3}& Y^{(k)}_{{\mathbf3I},2}&Y^{(k)}_{{\mathbf3I},1}\\
Y^{(k)}_{{\mathbf3I},2}&Y^{(k)}_{{\mathbf3I},1}&Y^{(k)}_{{\mathbf3I},3}\eea\right)
+\la_2\Lambda_1  \beta\left(\bea{ccc}Y^{(k)}_{{\mathbf3II},1}& Y^{(k)}_{{\mathbf3II},3}& Y^{(k)}_{{\mathbf3II},2}\\Y^{(k)}_{{\mathbf3II},3}& Y^{(k)}_{{\mathbf3II},2}&Y^{(k)}_{{\mathbf3II},1}
\\Y^{(k)}_{{\mathbf3II},2}&Y^{(k)}_{{\mathbf3II},1}&Y^{(k)}_{{\mathbf3II},3}
\eea\right).\label{SN2}
\eeqa
Here the column matrix $j=1$ corresponds to $\rho_{S}={\bf 1}$; $j=2$ corresponds to $\rho_{S}={\bf 1^\pr}$; $j=3$ corresponds to $\rho_{S}={\bf 1^{\pr\pr}}$.

The mass matrix for ${\cal M}^{SS}$ is given by
\beqa
&& 2k_{S}=0;~~~~{\cal M}^{SS}_{ij}=\ka_1\Lambda_2\left\{\bea{c}1,~~~~~~\rho_{S_i}\otimes \rho_{S_j}={\bf 1}\\0,~~~~~~~\rho_{S_i}\otimes \rho_{S_j}\neq {\bf 1}\eea\right.\nn\\
&&2 k_{S}=2; ~~~~{\cal M}^{SS}_{ij}=0 ~,\nn\\
&&2k_{S}=4; ~~~~{\cal M}^{SS}_{ij}=\ka_1\Lambda_2\left\{\bea{c}Y_{\bf 1}^{(4)},~~~~~~\rho_{S_i}\otimes \rho_{S_j}={\bf 1}\\ Y_{\bf 1^\pr}^{(4)},~~~~~~~\rho_{S_i}\otimes \rho_{S_j}= {\bf 1^{\pr\pr}}\\
0~~,~~~~~~~~~~~otherwise\eea\right.\nn\\
&&2k_{S}=6; ~~~~{\cal M}^{SS}_{ij}=\ka_1\Lambda_2\left\{\bea{c}Y_{\bf 1}^{(6)},~~~~~~\rho_{S_i}\otimes \rho_{S_j}={\bf 1}\\~~~0~~,~~~~~~~\rho_{S_i}\otimes \rho_{S_j}\neq {\bf 1}\eea\right.\nn\\
&&2k_{S}=8; ~~~~{\cal M}^{SS}_{ij}=\ka_1\Lambda_2\left\{\bea{c}Y_{\bf 1}^{(8)},~~~~~~\rho_{S_i}\otimes \rho_{S_j}={\bf 1}\\ Y_{\bf 1^\pr}^{(8)},~~~~~~~\rho_{S_i}\otimes \rho_{S_j}= {\bf 1^{\pr\pr}}\\
Y_{\bf 1^{\pr\pr}}^{(8)},~~~~~~~\rho_{S_i}\otimes \rho_{S_j}= {\bf 1^{\pr}}\\
0~~,~~~~~~~~~~~otherwise\eea\right.
\label{ISS:mass}
\eeqa
with various values of the modular weights $k_{S}$.
\item $\rho_{F}={\bf 1,1^\pr,1^{\pr\pr}}, \rho_S={\bf 3}$.

For the modular weights $k_{F_i}+k_{S}=2,4$, the expression of ${\cal M}^{SN}_{i*}$ takes the same form as eq.(\ref{SN1}). While for $k_{F_i}+k_{S}=6,8$, the expression of ${\cal M}^{SN}_{i*}$ takes the same form as eq.(\ref{SN2}).
Here the column matrix $i=1$ corresponds to $\rho_{F}={\bf 1}$; $i=2$ corresponds to $\rho_{F}={\bf 1^\pr}$; $i=3$ corresponds to $\rho_{F}={\bf 1^{\pr\pr}}$.

\item $\rho_{F}={\bf 1,1^\pr,1^{\pr\pr}}, \rho_S={\bf 1,1^\pr,1^{\pr\pr}}$.

Similarly, given the values of the modular weights $k_{F_i},k_{S_j}$, we can obtain the $S-N$ mixing matrix
\beqa
&& k_{F_i}+k_{S_j}=0; ~~~{\cal M}^{SN}_{ij}=\la_1\Lambda_1\left\{\bea{c}1,~~~~~~\rho_{F_i}\otimes \rho_{S_j}={\bf 1}\\0,~~~~~~\rho_{F_i}\otimes \rho_{S_j}\neq {\bf 1}\eea\right.\nn\\
&& k_{F_i}+k_{S_j}=2; ~~~{\cal M}^{SN}_{ij}= 0~,\nn\\
&& k_{F_i}+k_{S_j}=4; ~~~{\cal M}^{SN}_{ij}=\la_1\Lambda_1\left\{\bea{c}Y_{\bf 1}^{(4)},~~~~~~\rho_{F_i}\otimes \rho_{S_j}={\bf 1}\\ Y_{\bf 1^\pr}^{(4)},~~~~~~~\rho_{F_i}\otimes \rho_{S_j}= {\bf 1^{\pr\pr}}\\
0~~,~~~~~~~~~~~otherwise\eea\right.\nn\\
&& k_{F_i}+k_{S_j}=6; ~~~{\cal M}^{SN}_{ij}=\la_1\Lambda_1\left\{\bea{c}Y_{\bf 1}^{(6)},~~~~~~\rho_{F_i}\otimes \rho_{S_j}={\bf 1}\\~~~~0~,~~~~~~~\rho_{F_i}\otimes \rho_{S_j}\neq {\bf 1}\eea\right.\nn\eeqa\beqa
&& k_{F_i}+k_{S_j}=8; ~~~
~{\cal M}^{SN}_{ij}=\la_1\Lambda_1\left\{\bea{c}Y_{\bf 1}^{(8)},~~~~~~\rho_{F_i}\otimes \rho_{S_j}={\bf 1}\\ Y_{\bf 1^\pr}^{(8)},~~~~~~~\rho_{F_i}\otimes \rho_{S_j}= {\bf 1^{\pr\pr}}\\
Y_{\bf 1^{\pr\pr}}^{(8)},~~~~~~~\rho_{F_i}\otimes \rho_{S_j}= {\bf 1^{\pr}}\\
0~~.~~~~~~~~~~~otherwise\eea\right.
\eeqa

The mass matrix for ${\cal M}^{SS}_{ij}$ take the same form as eq.(\ref{ISS:mass}).
\eit
\subsection{Possible Scenarios}
We will not survey the scenarios in which all matter fields
are assigned to $A_4$ singlets, since the Yukawa couplings in the superpotentials would be less constrained by modular symmetry and  more free parameters would be involved in general.
We neglect these cases in which one column or one row of the fermion mass matrix is vanishing, since at least one fermion would be massless. If any two generations of fermion fields are assigned to the same singlet representation of $A_4$, we require their modular weights to be different to eliminate unwanted degeneracies.

\subsubsection{\label{All:3}~~~$\rho_{\bar{f}}={\bf 3}$, $\rho_{F}={\bf 3}$, $\rho_{E}={\bf 3}$, $\rho_{S}={\bf 3}$.}

As noted in the previous section, such assignments of modular representations for four-dimensional single modulus cases in modular flavor GUT correspond to the assignments
\beqas \rho_{T_F^{;p}} ={\bf 3}~~~(p=1,2,3)~,~\rho_{F_{\bar{f}}^{;q}}={\bf 3}~~~(q=1,2)~,~\rho_{O_E}={\bf 3}~,~\rho_{O_S}={\bf 3}~, \eeqas
for five-dimensional multiple modulus cases. The superpotential given for each subscenarios in single modulus cases can easily be extended to multiple modulus cases (at the GUT symmetry preserving fix point brane) via the replacements
\beqa
Y^D F F h &\ra&  Y_{;23}^D T_{F}^{;2} T_{F}^{;3} h~,\nn\\
Y^U F \bar{f} \bar{h} &\ra& Y_{;12}^U T_{F}^{;1} F_{\bar{f}}^{b;2} \bar{h} + Y_{;31}^U T_{F}^{;3} F_{\bar{f}}^{;1} \bar{h}~,\nn\\
Y^E \bar{f} E h &\ra& Y_{;1}^E F_{\bar{f}}^{;1} E h~,\nn\\
Y^S \overline{H} S F &\ra& Y^S_{;1} \overline{H}  O_S T_{F}^{;1}~,\nn\\
\f{M_{SS;}}{2} S S &\ra& \f{M_{SS;}}{2} O_S O_S~,
\eeqa
where the Yukawa couplings $Y^K_{;pq}$ in multiple modulus cases (or $Y^K$ in single modulus cases) can be decomposed into the product of some free coefficients $\al_{;pq}$ (or $\al$) and typical modular forms $Y_{\bf r}^{2k}$, respectively. We will not repeat again the replacements for each subscenario in our subsequent discussions. The free coefficients $\al_{;pq}$ of the Yukawa coupling terms\footnote{For example, the coefficient $\al_{1;pq}$ (here it is $\al_{1;23}$) in the Yukawa term $$W\supseteq \al_{1;23} Y_{\bf 3}^{(4)}(T_F^{a;2} T_F^{a;3})_{{\bf 3}S} h~,$$ are taken to be $\al_{1;23}=\al_1$, which is independent of the $p,q$ indices.} in multiple modulus cases are taken to be independent of the $p,q$ indices of $T_F^{;p}$ and $F_{\bar{f}}^{;q}$ and the modular weights are also taken to be universal for each type of matter representation in our 5D multiple modulus cases. That is, we choose $k_{T_F^{;p}}=k_F$, $k_{F_{\bar{f}}^{;q}}=k_{\bar{f}}$ and also $k_{O_E}=k_E$, $k_{O_S}=k_S$ for multiple modulus cases. With such choices, the mass matrices for matter contents in both single and multiple modulus scenarios can take almost identical forms (except for possible different choices of modulus values). Consequently, the form of the Yukawa couplings take the same form in both single and multiple modulus cases except that the quarks and leptons can adopt different values of modulus fields in the multiple modulus cases. We should note again that, under $A_4^Q\tm A_4^L$, the superfields $ T_F^{;2},T_F^{;3},F_{\bar{f}}^{;1}$ can transform non-trivially under $A_4^Q$ with the corresponding modular weights $k_F,k_F,k_{\bar{f}}$, respectively. The superfields $T_F^{;1}, F_{\bar{f}}^{;2},O_E, O_S$ can transform non-trivially under $A_4^L$ with the corresponding modular weights $k_F,k_{\bar{f}},k_E,k_S$, respectively.

We neglect sub-scenarios which would obviously lead to non-hierarchical structures of up, down quark mass matrices and lepton mass matrices. We concentrate on the sub-scenarios with the following choices of modular weights
\bit
\item $(k_{\bar{f}},k_{F},k_{E},k_{S})=(0,2,0,0)$.

The mass matrices for fermions are given as
\beqa
{\cal M}_U/v_u\equiv \(y_{U}\)_{ij}&=&\beta_1 S_{\mathbf3}^{(2)}(\tau)+ \beta_2 A_{\mathbf3}^{(2)}(\tau)~,\nn\\
{\cal M}_D/v_d\equiv\(y_{D}\)_{ij}&=& \al_1 S^{(4)}_{\mathbf3}  + \al_2 S^{(4)}_{\mathbf1} +\al_3 S^{(4)}_{\mathbf1'}\,,\nn\\
{\cal M}_E/v_d\equiv\(y_{E}\)_{ij}&=&\ga_1 S_{\mathbf1}^0(\tau)~,\nn\\
{\cal M}_N^{Dirac}/v_u\equiv  \(y_{N}\)^{Dirac}_{ij}&=&\(y_{U}\)^T~,\nn\\
{\cal M}^{SN}_{ij}&=&\la_1 \Lambda_1 S_{\mathbf3}^{(2)}(\tau)+\la_2\Lambda_1 A_{\mathbf3}^{(2)}(\tau) ~,\nn\\
{\cal M}^{SS}_{ij}&=&\ka_1 \Lambda_2 S_{\mathbf1}^0(\tau)~,
\eeqa
with the superpotential
\beqa
W&\supseteq& \[\al_1 Y_{\bf 3}^{(4)}(F F)_{{\bf 3}S} h +\al_2 Y_{\bf 1}^{(4)}(F F)_{\bf 1} h + \al_3 Y_{\bf 1^\pr}^{(4)} (F F)_{\bf 1^{\pr\pr}} h\]\nn\\
 &+&\[\beta_1 Y_{\bf 3}^{(2)} (F \bar{f})_{{\bf 3}S} \bar{h} +\beta_2 Y_{\bf 3}^{(2)} (F \bar{f})_{{\bf 3}A} \bar{h}\]\nn\\
  &+&\ga_1 (\bar{f} E)_{\bf 1} h
+\[\lambda_1 Y_{\bf 3}^{(2)}  (S F)_{{\bf 3}S} \overline{H}+\lambda_2 Y_{\bf 3}^{(2)}  (S F)_{{\bf 3}A} \overline{H}  \]+\f{\ka_1}{2} \Lambda_2 (S S)_{\bf 1}~.
\eeqa
The parameter $\al_1, \beta_1,\ga_1,\la_1,\ka_1$ can be taken to be real while others $\al_2,\al_3,\beta_2,\la_2$ are in general complex. Following the replacements discussed in the previous paragraphs, the extension to the multiple modulus superpotential at the GUT symmetry preserving fix point brane is straightforward. As an example, the superpotential in the fix point $O(y=0)$ should take the form
\beqa
{\cal L}&\supseteq&\delta(y)\int d^2\theta\left\{\f{}{} \[\al_1 Y_{\bf 3}^{(4)}(T_{F}^{;2} T_{F}^{;3})_{{\bf 3}S} h +\al_2 Y_{\bf 1}^{(4)}
(T_{F}^{;2} T_{F}^{;3})_{\bf 1} h + \al_3 Y_{\bf 1^\pr}^{(4)} (T_{F}^{;2} T_{F}^{;3})_{\bf 1^{\pr\pr}} h\]\right.\nn\\
 &+&\(\beta_1 Y_{\bf 3}^{(2)}\[ (T_{F}^{;1} F_{\bar{f}}^{b;2})_{{\bf 3}S}+(T_{F}^{;3} F_{\bar{f}}^{;1})_{{\bf 3}S}\] \bar{h} +\beta_2 Y_{\bf 3}^{(2)}\[ (T_{F}^{;1} F_{\bar{f}}^{b;2})_{{\bf 3}A}+(T_{F}^{;3} F_{\bar{f}}^{;1})_{{\bf 3}A} \]\bar{h}\)\nn\\
  &+&\ga_1 (F_{\bar{f}}^{;1} E)_{\bf 1} h
+\[\lambda_1 Y_{\bf 3}^{(2)}  (O_S T_{F}^{;1})_{{\bf 3}S} \overline{H}
+\lambda_2 Y_{\bf 3}^{(2)}  (O_S T_{F}^{;1})_{{\bf 3}A} \overline{H} \]
\nn\\&+&\left.\f{\ka_1}{2} \Lambda_2 (O_S O_S)_{\bf 1}\f{}{}\right\}~.
\eeqa
The coefficients $\al_i, \beta_i, \cdots$ in the Yukawa couplings are taken to be independent of the $p,q$ indices of $T_F^{;p}$ and $F_{\bar{f}}^{;q}$.

\item $(k_{\bar{f}},k_{F},k_{E},k_{S})=(2,2,0,0)$.

The mass matrices for fermions are given as
\beqa
{\cal M}_U/v_u\equiv \(y_{U}\)_{ij}&=&\beta_1\,S^{(4)}_{\mathbf3} + \beta_2 A^{(4)}_{\mathbf3} +\beta_3 S^{(4)}_{\mathbf1} +\beta_4 S^{(4)}_{\mathbf1'}\,,\nn\\
{\cal M}_D/v_d\equiv \(y_{D}\)_{ij}&=& \al_1 S^{(4)}_{\mathbf3}  + \al_2 S^{(4)}_{\mathbf1} +\al_3 S^{(4)}_{\mathbf1'}\,,\nn\\
{\cal M}_E/v_d\equiv \(y_{E}\)_{ij}&=&\ga_1 S_{\mathbf3}^{(2)}(\tau)+\ga_2 A_{\mathbf3}^{(2)}(\tau)~,\nn\\
{\cal M}_N^{Dirac}/v_u\equiv\(y_{N}\)^{Dirac}_{ij}&=&\(y_{U}\)^T~,\nn\\
{\cal M}^{SN}_{ij}&=&\la_1 \Lambda_1 S_{\mathbf3}^{(2)}(\tau)+\la_2\Lambda_1 A_{\mathbf3}^{(2)}(\tau) ~,\nn\\
{\cal M}^{SS}_{ij}&=&\ka_1 \Lambda_2 S_{\mathbf1}^0(\tau)~,
\eeqa
with the superpotential
\beqa
W&\supseteq& \(\al_1 Y_{\bf 3}^{(4)}(F F)_{{\bf 3}S} h +\al_2 Y_{\bf 1}^{(4)}(F F)_{\bf 1} h + \al_3 Y_{\bf 1^\pr}^{(4)} (F F)_{\bf 1^{\pr\pr}} h\)\nn\\
 &+&\(\beta_1 Y_{\bf 3}^{(4)} (F \bar{f})_{{\bf 3}S} \bar{h} +\beta_2 Y_{\bf 3}^{(4)} (F \bar{f})_{{\bf 3}A} \bar{h}+ \beta_3 Y_{\bf 1}^{(4)} (F \bar{f})_{\bf 1} \bar{h} +\beta_4 Y_{\bf 1^\pr}^{(4)} (F \bar{f})_{\bf 1^{\pr\pr}} \bar{h}  \)\nn\\
  &+&\(\ga_1 Y_{\bf 3}^{(2)} (\bar{f} E)_{{\bf 3}S} h+\ga_2 Y_{\bf 3}^{(2)} (\bar{f} E)_{{\bf 3}A} h   \)
+\(\lambda_1 Y_{\bf 3}^{(2)}  (S F)_{{\bf 3}S} \overline{H}+\lambda_2 Y_{\bf 3}^{(2)}  (S F)_{{\bf 3}A} \overline{H}\)\nn\\
&+&\f{\ka_1}{2} \Lambda_2 (S S)_{\bf 1}~.
\eeqa
The parameter $\al_1, \beta_1,\ga_1,\la_1,\ka_1$ can be taken to be real while others $\al_2,\al_3,\beta_2,\beta_3,\beta_4,\ga_2,\la_2$ are in general complex.

\item $(k_{\bar{f}},k_{F},k_{E},k_{S})=(4,2,0,0)$.

The mass matrices for fermions are given as
\beqa
{\cal M}_U/v_u\equiv\(y_{U}\)_{ij}&=&\beta_1\,S^{(6)}_{\mathbf{3}I} + \beta_2 A^{(6)}_{\mathbf{3}I} +\beta_3 S^{(6)}_{\mathbf{3}II} +\beta_4 A^{(6)}_{\mathbf{3}II} +\beta_5 S^{(6)}_{\mathbf1}\,,\nn\\
{\cal M}_D/v_d\equiv\(y_{D}\)_{ij}&=& \al_1 S^{(4)}_{\mathbf3}  + \al_2 S^{(4)}_{\mathbf1} +\al_3 S^{(4)}_{\mathbf1'}\,,\nn\\
{\cal M}_E/v_d\equiv\(y_{E}\)_{ij}&=&\ga_1\,S^{(4)}_{\mathbf3} +\ga_2 A^{(4)}_{\mathbf3} +\ga_3 S^{(4)}_{\mathbf1} + \ga_4 S^{(4)}_{\mathbf1'}\,,\nn\eeqa\beqa
{\cal M}_N^{Dirac}/v_u\equiv\(y_{N}\)^{Dirac}_{ij}&=&\(y_{U}\)^T~,\nn\\
{\cal M}^{SN}_{ij}&=&\la_1 \Lambda_1 S_{\mathbf3}^{(2)}(\tau)+\la_2\Lambda_1 A_{\mathbf3}^{(2)}(\tau) ~,\nn\\
{\cal M}^{SS}_{ij}&=&\ka_1 \Lambda_2 S_{\mathbf1}^0(\tau)~,\nn
\eeqa
with the superpotential
\beqa
W&\supseteq& \(\al_1 Y_{\bf 3}^{(4)}(F F)_{{\bf 3}S} h +\al_2 Y_{\bf 1}^{(4)}(F F)_{\bf 1} h + \al_3 Y_{\bf 1^\pr}^{(4)} (F F)_{\bf 1^{\pr\pr}} h\)\nn\\
 &+&\(\beta_1 Y_{\bf 3}^{(6)} (F \bar{f})_{{\bf 3}I,S} \bar{h}+\beta_2 Y_{\bf 3}^{(6)} (F \bar{f})_{{\bf 3}I,A} \bar{h}+ \beta_3 Y_{\bf 1}^{(6)} (F \bar{f})_{{\bf 3}II,S} \bar{h}\right.\nn\\
  &+&\left.\beta_4 Y_{\bf 3}^{(6)} (F \bar{f})_{{\bf 3}II,A} \bar{h}+ \beta_5 Y_{\bf 1}^{(6)} (F \bar{f})_{\bf 1} \bar{h}  \)\nn\\
  &+&\(\ga_1 Y_{\bf 3}^{(4)} (\bar{f} E)_{{\bf 3}S} h+\ga_2 Y_{\bf 3}^{(4)} (\bar{f} E)_{{\bf 3}A} h  +\ga_3 Y_{\bf 1}^{(4)} (\bar{f} E)_{\bf 1} h+\ga_4 Y_{\bf 1^{\pr}}^{(4)} (\bar{f} E)_{\bf 1^{\pr\pr}} h  \)\nn\\
&+&\(\lambda_1 Y_{\bf 3}^{(2)}  (S F)_{{\bf 3}S} \overline{H}+\lambda_2 Y_{\bf 3}^{(2)}  (S F)_{{\bf 3}A} \overline{H}\)+\f{\ka_1}{2} \Lambda_2 (S S)_{\bf 1}~.
\eeqa
The parameter $\al_1, \beta_1,\ga_1,\la_1,\ka_1$ can be taken to be real while others $\al_2,\al_3,\beta_2$,$\beta_3$,$\beta_4,\beta_5$, $\ga_2,\ga_3$,$\ga_4,\la_2$ are in general complex.

\item $(k_{\bar{f}},k_{F},k_{E},k_{S})=(4,2,2,0)$.

The mass matrices for fermions are given as
\beqa
{\cal M}_U/v_u\equiv \(y_{U}\)_{ij}&=&\beta_1\,S^{(6)}_{\mathbf{3}I} + \beta_2 A^{(6)}_{\mathbf{3}I} +\beta_3 S^{(6)}_{\mathbf{3}II} +\beta_4 A^{(6)}_{\mathbf{3}II} +\beta_5 S^{(6)}_{\mathbf1}\,,\nn\\
{\cal M}_D/v_d\equiv\(y_{D}\)_{ij}&=& \al_1 S^{(4)}_{\mathbf3}  + \al_2 S^{(4)}_{\mathbf1} +\al_3 S^{(4)}_{\mathbf1'}\,,\nn\\
{\cal M}_E/v_d\equiv\(y_{E}\)_{ij}&=&\ga_1\,S^{(6)}_{\mathbf{3}I} +\ga_2 A^{(6)}_{\mathbf{3}I} +\ga_3 S^{(6)}_{\mathbf{3}II} +\ga_4 A^{(6)}_{\mathbf{3}II} +\ga_5 S^{(6)}_{\mathbf1}\,,\nn\\
{\cal M}_N^{Dirac}/v_u\equiv\(y_{N}\)^{Dirac}_{ij}&=&\(y_{U}\)^T~,\nn\\
{\cal M}^{SN}_{ij}&=&\la_1 \Lambda_1 S_3^{(2)}(\tau)+\la_2\Lambda_1 A_3^{(2)}(\tau) ~,\nn\\
{\cal M}^{SS}_{ij}&=&\ka_1 \Lambda_2 S_1^0(\tau)~,
\eeqa
with the superpotential
\beqa
W&\supseteq& \(\al_1 Y_{\bf 3}^{(4)}(F F)_{{\bf 3}S} h +\al_2 Y_{\bf 1}^{(4)}(F F)_{\bf 1} h + \al_3 Y_{\bf 1^\pr}^{(4)} (F F)_{\bf 1^{\pr\pr}} h\)\nn\\
 &+&\(\beta_1 Y_{\bf 3}^{(6)} (F \bar{f})_{{\bf 3}I,S} \bar{h}+\beta_2 Y_{\bf 3}^{(6)} (F \bar{f})_{{\bf 3}I,A} \bar{h}+ \beta_3 Y_{\bf 1}^{(6)} (F \bar{f})_{{\bf 3}II,S} \bar{h} +\beta_4 Y_{\bf 3}^{(6)} (F \bar{f})_{{\bf 3}II, A} \bar{h}\right.\nn\\
 &+& \left.\beta_5 Y_{\bf 1}^{(6)} (F \bar{f})_{\bf 1} \bar{h}\)
 +\(\ga_1 Y_{\bf 3}^{(6)} (\bar{f} E)_{{\bf 3}I,S} h+\ga_2 Y_{\bf 3}^{(6)} (\bar{f} E)_{{\bf 3}I, A} h  +\ga_3 Y_{\bf 1}^{(6)} (\bar{f} E)_{{\bf 3}II,S} h~\right.\nn\\
 &+&\left.\ga_4 Y_{\bf 1^{\pr}}^{(6)} (\bar{f} E)_{{\bf 3}II,A} h +\ga_5 Y_{\bf 1}^{(6)} (\bar{f} E)_{\bf 1} h\)\nn\\
&+&\(\lambda_1 Y_{\bf 3}^{(2)}  (S F)_{{\bf 3}S} \overline{H}+\lambda_2 Y_{\bf 3}^{(2)}  (S F)_{{\bf 3}A} \overline{H}\)+\f{\ka_1}{2} \Lambda_2 (S S)_{\bf 1}~.
\eeqa
The parameter $\al_1, \beta_1,\ga_1,\la_1,\ka_1$ can be taken to be real while other parameters $\al_2$,$\al_3$,$\beta_2$,$\beta_3$,$\beta_4$,$\beta_5$,$\ga_2$,$\ga_3$,$\ga_4$,$\ga_5$,$\la_2$ are in general complex.

\item $(k_{\bar{f}},k_{F},k_{E},k_{S})=(4,2,2,2)$.

The mass matrices for fermions are given as
\beqa
{\cal M}_U/v_u\equiv \(y_{U}\)_{ij}&=&\beta_1\,S^{(6)}_{\mathbf{3}I} + \beta_2 A^{(6)}_{\mathbf{3}I} +\beta_3 S^{(6)}_{\mathbf{3}II} +\beta_4 A^{(6)}_{\mathbf{3}II} +\beta_5 S^{(6)}_{\mathbf1}\,,\nn\eeqa\beqa
{\cal M}_D/v_d\equiv\(y_{D}\)_{ij}&=& \al_1 S^{(4)}_{\mathbf3}  + \al_2 S^{(4)}_{\mathbf1} +\al_3 S^{(4)}_{\mathbf1'}\,,\nn\\
{\cal M}_E/v_d\equiv\(y_{E}\)_{ij}&=&\ga_1\,S^{(6)}_{\mathbf{3}I} +\ga_2 A^{(6)}_{\mathbf{3}I} +\ga_3 S^{(6)}_{\mathbf{3}II} +\ga_4 A^{(6)}_{\mathbf{3}II} +\ga_5 S^{(6)}_{\mathbf1}\,,\nn\\
{\cal M}_N^{Dirac}/v_u\equiv\(y_{N}\)^{Dirac}_{ij}&=&\(y_{U}\)^T~,\nn\\
{\cal M}^{SN}_{ij}&=&\la_1\,\Lambda_1 S^{(4)}_{\mathbf3} +\la_2 \Lambda_1 A^{(4)}_{\mathbf3} +\la_3\Lambda_1 S^{(4)}_{\mathbf1} +\la_4 \Lambda_1 S^{(4)}_{\mathbf1'}\,,\nn\\
{\cal M}^{SS}_{ij}&=&\ka_1\Lambda_2 \,S^{(4)}_{\mathbf3}  +\ka_2\Lambda_2 S^{(4)}_{\mathbf1} +\ka_3 \Lambda_2 S^{(4)}_{\mathbf1'}\,,
\eeqa
with the superpotential
\beqa
W&\supseteq& \(\al_1 Y_{{\bf 3}}^{(4)}(F F)_{{\bf 3}S} h +\al_2 Y_{\bf 1}^{(4)}(F F)_{\bf 1} h + \al_3 Y_{\bf 1^\pr}^{(4)} (F F)_{\bf 1^{\pr\pr}} h\)\nn\\
 &+&\(\beta_1 Y_{{\bf 3}}^{(6)} (F \bar{f})_{{\bf 3} I,S} \bar{h}+\beta_2 Y_{{\bf 3}}^{(6)} (F \bar{f})_{{\bf 3}I,A} \bar{h}+ \beta_3 Y_{\bf 1}^{(6)} (F \bar{f})_{{\bf 3}II,S} \bar{h} +\beta_4 Y_{{\bf 3}}^{(6)} (F \bar{f})_{{\bf 3}II, A} \bar{h}\right.\nn\\
 &+& \left.\beta_5 Y_{\bf 1}^{(6)} (F \bar{f})_{\bf 1} \bar{h}  \)
  +\(\ga_1 Y_{{\bf 3}}^{(6)} (\bar{f} E)_{{\bf 3}I,S} h+\ga_2 Y_{{\bf 3}}^{(6)} (\bar{f} E)_{{\bf 3}I, A} h  \right.\nn\\
  &+&\left.\ga_3 Y_{\bf 1}^{(6)} (\bar{f} E)_{{\bf 3}II,S} h+\ga_4 Y_{\bf 1^{\pr}}^{(6)} (\bar{f} E)_{{\bf 3}II,A} h +\ga_5 Y_{\bf 1}^{(6)} (\bar{f} E)_{\bf 1} h\)\nn \\
&+&\(\lambda_1 Y_{{\bf 3}}^{(4)}  (S F)_{{\bf 3}S} \overline{H}+\lambda_2 Y_{{\bf 3}}^{(4)}  (S F)_{{\bf 3}A} \overline{H}+\lambda_3 Y_{\bf 1}^{(4)}  (S F)_{\bf 1} \overline{H}+\lambda_4 Y_{\bf 1^\pr}^{(4)}  (S F)_{\bf 1^{\pr\pr}} \overline{H}\)\nn\\
&+&\(\f{\ka_1}{2}\Lambda_2 Y_{{\bf 3}}^{(4)} (S S)_{{\bf 3}S}+\f{\ka_2}{2}\Lambda_2 Y_{\bf 1}^{(4)} (S S)_{\bf 1}+ \f{\ka_2}{2} \Lambda_2 Y_{\bf 1^\pr}^{(4)}(S S)_{\bf 1^{\pr\pr}} \)~.
\eeqa
The parameter $\al_1, \beta_1,\ga_1,\la_1,\ka_1$ can be taken to be real while others $\al_2,\al_3,\beta_2,\cdots$ are in general complex.

\item $(k_{\bar{f}},k_{F},k_{E},k_{S})=(4,4,2,2)$.

The mass matrices for fermions are given as
\beqa
{\cal M}_U/v_u\equiv \(y_{U}\)_{ij}&=&\beta_1 \,S^{(8)}_{\mathbf{3}I} +\beta_2 A^{(8)}_{\mathbf{3}I} +\beta_3 S^{(8)}_{\mathbf{3}II} +\beta_4 A^{(8)}_{\mathbf{3}II} + \beta_5 S^{(8)}_{\mathbf1} + \beta_6 S^{(8)}_{\mathbf1'} +\beta_7 S^{(8)}_{\mathbf1''}\,,\nn\\
{\cal M}_D/v_d\equiv\(y_{D}\)_{ij}&=& \al_1 S^{(8)}_{\mathbf{3}I} +\al_2 S^{(8)}_{\mathbf{3}II} +\al_3 S^{(8)}_{\mathbf1}+\al_4 S^{(8)}_{\mathbf1'} +\al_5 S^{(8)}_{\mathbf1''}\,,\nn\,\\
{\cal M}_E/v_d\equiv\(y_{E}\)_{ij}&=&\ga_1\,S^{(6)}_{\mathbf{3}I} +\ga_2 A^{(6)}_{\mathbf{3}I} +\ga_3 S^{(6)}_{\mathbf{3}II} +\ga_4 A^{(6)}_{\mathbf{3}II} +\ga_5 S^{(6)}_{\mathbf1}\,,\nn\\
{\cal M}_N^{Dirac}/v_u\equiv\(y_{N}\)^{Dirac}_{ij}&=&\(y_{U}\)^T~,\nn\\
{\cal M}^{SN}_{ij}&=&\la_1\,\Lambda_1 S^{(6)}_{\mathbf{3}I} + \la_2\Lambda_1 A^{(6)}_{\mathbf{3}I} +\la_3 \Lambda_1 S^{(6)}_{\mathbf{3}II} +\la_4\Lambda_1 A^{(6)}_{\mathbf{3}II} +\la_5\Lambda_1 S^{(6)}_{\mathbf1}\,,\nn\\
{\cal M}^{SS}_{ij}&=&\ka_1\Lambda_2 \,S^{(4)}_{\mathbf3}  +\ka_2\Lambda_2 S^{(4)}_{\mathbf1} +\ka_3 \Lambda_2 S^{(4)}_{\mathbf1'}\,,
\eeqa
with the superpotential
\beqa
W&\supseteq& \(\al_1 Y_{{\bf 3}I}^{(8)}(F F)_{{\bf 3}S} h +\al_2 Y_{{\bf 3}II}^{(8)}(F F)_{{\bf 3}S} h
+\al_3 Y_{\bf 1}^{(8)}(F F)_{\bf 1} h
+\al_4 Y_{\bf 1^\pr}^{(8)}(F F)_{\bf 1^{\pr\pr}} h \right.\nn\\
&+&\left. \al_5 Y_{\bf 1^{\pr\pr}}^{(8)} (F F)_{\bf 1^{\pr}} h\)
 +\(\beta_1 Y_{{\bf 3}I}^{(8)} (F \bar{f})_{{\bf 3}S} \bar{h}+\beta_2 Y_{{\bf 3}I}^{(8)} (F \bar{f})_{{\bf 3}A} \bar{h}+\beta_3 Y_{{\bf 3}II}^{(8)} (F \bar{f})_{{\bf 3}S} \bar{h}\right.\nn\\
 &+& \beta_4 Y_{{\bf 3}II}^{(8)} (F \bar{f})_{{\bf 3}A} \bar{h}+
    \beta_5 Y_{\bf 1}^{(8)} (F \bar{f})_{\bf 1} \bar{h}
    +\left.\beta_6 Y_{\bf 1^\pr}^{(8)} (F \bar{f})_{\bf 1^{\pr\pr}} \bar{h}+ \beta_7 Y_{\bf 1^{\pr\pr}}^{(8)} (F \bar{f})_{\bf 1^\pr} \bar{h}  \)\nn\\
  &+&\(\ga_1 Y_{{\bf 3}I}^{(6)} (\bar{f} E)_{{\bf 3}S} h+\ga_2 Y_{{\bf 3}I}^{(6)} (\bar{f} E)_{{\bf 3}A} h  +\ga_3 Y_{{\bf 3}II}^{(6)} (\bar{f} E)_{{\bf 3}S} h+\ga_4 Y_{{\bf 3}II}^{(6)} (\bar{f} E)_{{\bf 3}A} h \right.\nn\\
&+&\left. \ga_5 Y_{\bf 1}^{(6)} (\bar{f} E)_{\bf 1} h\)+\(\lambda_1 Y_{{\bf 3}I}^{(6)}  (S F)_{{\bf 3}S} \overline{H}+\lambda_2 Y_{{\bf 3}I}^{(6)}  (S F)_{{\bf 3}A} \overline{H}+\lambda_3 Y_{{\bf 3}II}^{(6)}  (S F)_{{\bf 3}S} \overline{H}\right.\nn\\
&+&\left.\lambda_4 Y_{{\bf 3}II}^{(6)}  (S F)_{{\bf 3}A} \overline{H}+\lambda_5 Y_{\bf 1}^{(6)}  (S F)_{\bf 1} \overline{H}\)\nn\\
&+&\(\f{\ka_1}{2}\Lambda_2 Y_{{\bf 3}}^{(4)} (S S)_{{\bf 3}S}+\f{\ka_2}{2}\Lambda_2 Y_{\bf 1}^{(4)} (S S)_{\bf 1}+ \f{\ka_2}{2} \Lambda_2 Y_{\bf 1^\pr}^{(4)}(S S)_{\bf 1^{\pr\pr}} \)~.
\eeqa
The $5$ parameter $\al_1, \beta_1,\ga_1,\la_1,\ka_1$ can be taken to be real while others $20$ parameters $\al_2,\al_3,\beta_2,\cdots$ are in general complex.
\item $(k_{\bar{f}},k_{F},k_{E},k_{S})=(4,4,4,2)$.

This scenario will have $5$ real and $22$ complex free parameters, which is less constrained. So we will discard such scenarios.
\item $(k_{\bar{f}},k_{F},k_{E},k_{S})=(4,4,4,4)$.

This scenario will have $5$ real and $26$ complex free parameters, which is less constrained. So we will discard such scenarios.

\eit
\subsubsection{ $\rho_{\bar{f}_i}={\bf 1,1^\pr,1^{\pr\pr}},~\rho_{F}={\bf 3},~\rho_{E}={\bf 3}, \rho_{S}={\bf 3}$}
We have ten combinations for three generation $\bar{f}_i$, namely
\beqa
&&({\bf 1,1,1})_{\bar{f}_I},~ ({\bf 1^{\pr},1,1})_{\bar{f}_{II}}, ~({\bf 1^{\pr\pr},1,1})_{\bar{f}_{III}},~({\bf 1^{\pr},1^{\pr},1})_{\bar{f}_{IV}},~({\bf 1^{\pr},1^{\pr},1^{\pr}})_{\bar{f}_{V}},\nn\\
 &&({\bf 1^{\pr\pr},1^{\pr\pr},1})_{\bar{f}_{VI}},({\bf 1^{\pr\pr},1^{\pr\pr},1^{\pr\pr}})_{\bar{f}_{VII}},({\bf 1^{\pr}, 1^{\pr},1^{\pr\pr}})_{\bar{f}_{VIII}},({\bf 1^{\pr}, 1^{\pr\pr},1^{\pr\pr}})_{\bar{f}_{IX}},
 ({\bf 1, 1^{\pr},1^{\pr\pr}})_{\bar{f}_{X}}~,
\label{combination} \eeqa
 and modular weight choice $$(k_{\bar{f}},k_{F},k_{E},k_{S})=((k_{1},k_{2},k_3),k_{F},k_{E},k_{S}).$$

The general form of the superpotential can be written as
\beqa
W&\supseteq& \(\sum\limits_{M}\al_{1,M} Y_{{\bf 3}}^{(2k_F)}(F F)_{{\bf 3}S} h +\al_2 Y_{\bf 1}^{(2k_F)}(F F)_{\bf 1} h + \al_3 Y_{\bf 1^\pr}^{(2k_F)} (F F)_{\bf 1^{\pr\pr}} h
\right.\\
&+&\left.\al_4 Y_{\bf 1^{\pr\pr}}^{(2k_F)} (F F)_{\bf 1^{\pr}} h \)+
\(\sum\limits_{i,M}\beta_{i,M} Y_{{\bf 3},M}^{(\tl{k}_i)} (F \bar{f}_i)_{{\bf 3}} \bar{h} \)
  +\(\sum\limits_{i,M}\ga_{i,M} Y_{{\bf 3},M}^{(\hat{k}_i)} (\bar{f}_i E)_{{\bf 3}} h\)\nn\\
&+&\(\sum\limits_{M}\lambda_{1,M} Y_{{\bf 3}}^{(k_S+k_F)}  (S F)_{{\bf 3}S} \overline{H}+\sum\limits_{M}\lambda_{2,M} Y_{{\bf 3}}^{(k_S+k_F)}  (S F)_{{\bf 3}A} \overline{H}    +\lambda_{3} Y_{\bf 1}^{(k_S+k_F)}  (S F)_{\bf 1} \overline{H}\right.\nn\\
&+&~\left.\la_4 Y_{\bf 1^\pr}^{(k_S+k_F)}  (S F)_{\bf 1^{\pr\pr}} \overline{H}
+\la_5 Y_{\bf 1^{\pr\pr}}^{(k_S+k_F)}  (S F)_{\bf 1^{\pr}} \overline{H}\)\nn\\
&+&\f{\Lambda_2}{2}  \( \sum\limits_{M}\ka_{1,M} Y_{{\bf 3}}^{(2k_S)} (S S)_{{\bf 3}S}
+\ka_2 Y_{\bf 1}^{(2k_S)}(SS)_{\bf 1} + \ka_3 Y_{\bf 1^\pr}^{(2k_S)} (S S)_{\bf 1^{\pr\pr}}
+\ka_4 Y_{\bf 1^{\pr\pr}}^{(2k_S)} (SS)_{\bf 1^{\pr}}\)~,\nn
\eeqa
with the index $'M'$ taken values in $\emph{2,4,6I,6II,8I,8II}$, depending on the values of the modular weights. Some of the coefficients $\al_{i,M},\al_i,\beta_{i,M},\ga_{i,M},\cdots $ vanish, which also depend on the values of the modular weights.

We can define
\beqa
U[a,b]&\equiv& \bt_a Y^{(\tl{k}_a)}_{{\mathbf3},b}+\tl{\bt}_a Y^{(6)}_{{\mathbf3I},b}+\tl{\bt}^\pr_a Y^{(6)}_{{\mathbf3II},b}+\hat{\bt}_a Y^{(8)}_{{\mathbf3I},b}+\hat{\bt}^\pr_a Y^{(8)}_{{\mathbf3II},b}~,\nn\\
L[a,b]&\equiv& \ga_a Y^{(\hat{k}_a)}_{{\mathbf3},b}+\tl{\ga}_a Y^{(6)}_{{\mathbf3I},b}+\tl{\ga}^\pr_a Y^{(6)}_{{\mathbf3II},b}+\hat{\ga}_a Y^{(8)}_{{\mathbf3I},b}+\hat{\ga}^\pr_a Y^{(8)}_{{\mathbf3II},b}~.
\eeqa
for $\tl{k}_a,\hat{k}_a\neq 6,8$.
The up quark mass matrices are given as
\beqa
\(y_{U}\)_{\bar{f}_{I}}&=&\left(\bea{ccc}
U[1,1],&U[2,1],&U[3,1],\\
U[1,3],&U[2,3],&U[3,3],\\
U[1,2],&U[2,2],&U[3,2],
\eea\right),~
\(y_{U}\)_{\bar{f}_{II}}=\left(\bea{ccc}
U[1,3],&U[2,1],&U[3,1],\\
U[1,2],&U[2,3],&U[3,3],\\
U[1,1],&U[2,2],&U[3,2],
\eea\right),~\nn\\
\(y_{U}\)_{\bar{f}_{III}}&=&\left(\bea{ccc}
U[1,2],&U[2,1],&U[3,1],\\
U[1,1],&U[2,3],&U[3,3],\\
U[1,3],&U[2,2],&U[3,2],
\eea\right),
\(y_{U}\)_{\bar{f}_{IV}}=\left(\bea{ccc}
U[1,3],&U[2,3],&U[3,1],\\
U[1,2],&U[2,2],&U[3,3],\\
U[1,1],&U[2,1],&U[3,2],
\eea\right),~\nn\\
\(y_{U}\)_{\bar{f}_{V}}&=&\left(\bea{ccc}
U[1,3],&U[2,3],&U[3,3],\\
U[1,2],&U[2,2],&U[3,2],\\
U[1,1],&U[2,1],&U[3,1],
\eea\right),
\(y_{U}\)_{\bar{f}_{VI}}=\left(\bea{ccc}
U[1,2],&U[2,2],&U[3,1],\\
U[1,1],&U[2,1],&U[3,3],\\
U[1,3],&U[2,3],&U[3,2],
\eea\right),~\nn\\
\(y_{U}\)_{\bar{f}_{VII}}&=&\left(\bea{ccc}
U[1,2],&U[2,2],&U[3,2],\\
U[1,1],&U[2,1],&U[3,1],\\
U[1,3],&U[2,3],&U[3,3],
\eea\right),
\(y_{U}\)_{\bar{f}_{VIII}}=\left(\bea{ccc}
U[1,3],&U[2,3],&U[3,2],\\
U[1,2],&U[2,2],&U[3,1],\\
U[1,1],&U[2,1],&U[3,3],
\eea\right),~\nn\eeqa\beqa
\(y_{U}\)_{\bar{f}_{IX}}&=&\left(\bea{ccc}
U[1,3],&U[2,2],&U[3,2],\\
U[1,2],&U[2,1],&U[3,1],\\
U[1,1],&U[2,3],&U[3,3],
\eea\right),
\(y_{U}\)_{\bar{f}_{IX}}=\left(\bea{ccc}
U[1,1],&U[2,3],&U[3,2],\\
U[1,3],&U[2,2],&U[3,1],\\
U[1,2],&U[2,1],&U[3,3],
\eea\right).\nn\\\label{up:massv1}
\eeqa

The lepton mass matrices are given as the transpose of eq.(\ref{up:massv1}) after replacing $U[a,b]$ with $L[a,b]$
\beqa
\(y_{E}\)_{\bar{f}_{I}}&=&\left(\bea{ccc}
L[1,1],&L[1,3],&L[1,2],\\
L[2,1],&L[2,3],&L[2,2],\\
L[3,1],&L[3,3],&L[3,2],
\eea\right),
\(y_{E}\)_{\bar{f}_{II}}=\left(\bea{ccc}
L[1,3],&L[1,2],&L[1,1],\\
L[2,1],&L[2,3],&L[2,2],\\
L[3,1],&L[3,3],&L[3,2],
\eea\right),~\nn\\
\(y_{E}\)_{\bar{f}_{III}}&=&\left(\bea{ccc}
L[1,2],&L[1,1],&L[1,3],\\
L[2,1],&L[2,3],&L[2,2],\\
L[3,1],&L[3,3],&L[3,2],
\eea\right),
\(y_{E}\)_{\bar{f}_{IV}}=\left(\bea{ccc}
L[1,3],&L[1,2],&L[1,1],\\
L[2,3],&L[2,2],&L[2,1],\\
L[3,1],&L[3,3],&L[3,2],
\eea\right),~\nn\\
\(y_{E}\)_{\bar{f}_{V}}&=&\left(\bea{ccc}
L[1,3],&L[1,2],&L[1,1],\\
L[2,3],&L[2,2],&L[2,1],\\
L[3,3],&L[3,2],&L[3,1],
\eea\right),~
\(y_{E}\)_{\bar{f}_{VI}}=\left(\bea{ccc}
L[1,2],&L[1,1],&L[1,3],\\
L[2,2],&L[2,1],&L[2,3],\\
L[3,1],&L[3,3],&L[3,2],
\eea\right),~\nn\\
\(y_{E}\)_{\bar{f}_{VII}}&=&\left(\bea{ccc}
L[1,2],&L[1,1],&L[1,3],\\
L[2,2],&L[2,1],&L[2,3],\\
L[3,2],&L[3,1],&L[3,3],
\eea\right),~
\(y_{E}\)_{\bar{f}_{VIII}}=\left(\bea{ccc}
L[1,3],&L[1,2],&L[1,1],\\
L[2,3],&L[2,2],&L[2,1],\\
L[3,2],&L[3,1],&L[3,3],
\eea\right),~\nn\eeqa\beqa
\(y_{E}\)_{\bar{f}_{IX}}&=&\left(\bea{ccc}
L[1,3],&L[1,2],&L[1,1],\\
L[2,2],&L[2,1],&L[2,3],\\
L[3,2],&L[3,1],&L[3,3],
\eea\right),~
\(y_{E}\)_{\bar{f}_{X}}=\left(\bea{ccc}
L[1,1],&L[1,3],&L[1,2],\\
L[2,3],&L[2,2],&L[2,1],\\
L[3,2],&L[3,1],&L[3,3],
\eea\right),~\nn\\
\label{lepton:massv1}
\eeqa
The mass matrices for down type quark and neutrinos are given as
\beqa
\(y_{D}\)_{ij}&=& \al_{1,M}S^{(2k_F)}_{\mathbf3M}+\al_2 S^{(2k_F)}_{\mathbf1}
+\al_3 S^{(2k_F)}_{\mathbf1'}+\al_4 S^{(2k_F)}_{\mathbf1''},\nn\\
{\cal M}^{SN}_{ij}&=&\la_{1,M} S^{(k_S+k_F)}_{\mathbf3M} + \la_{2,M} A^{(k_S+k_F)}_{\mathbf3M}
+\la_3 S^{(k_S+k_F)}_{\mathbf1}
+\la_4 S^{(k_S+k_F)}_{\mathbf1'}+\la_5 S^{(k_S+k_F)}_{\mathbf1''} ~,\nn\\
{\cal M}^{SS}_{ij}&=&\Lambda_2\[\ka_{1,M}  S^{(2k_S)}_{\mathbf3M}+\ka_2  S^{(2k_S)}_{\mathbf1}
+\ka_3 S^{(2k_S)}_{\mathbf1'}+\ka_4 S^{(2k_S)}_{\mathbf1''}\] ~,\nn\\
~~~\(y_{N}\)^{Dirac}_{ij}&=&\(y_{U}\)^T~,
\eeqa
with the index $'M'$ taken values in $\emph{2,4,6I,6II,8I,8II}$, depending on the values of the modular weights.
We should note that coefficients for those modular weights without certain representations should be set to vanish. For example, the coefficient $\al_4=0$ when $2k_F=4$; and the coefficient $\al_3=\al_4=0$ when $2k_F=6$, as the $Y^{(4)}_{1^{\pr\pr}},Y^{(6)}_{1^{\pr}},Y^{(6)}_{1^{\pr\pr}}$ are not independent modular forms.

\bit
\item For $k_F=2,k_S=0$:

\bit
\item  $\tl{k}_i\equiv k_i+k_F=2,4$ and $\hat{k}_i\equiv k_i+k_E=2,4$:

The forms of the up-type quark and lepton mass matrices take the form in (\ref{up:massv1}) and (\ref{lepton:massv1}) with $\tl{\bt}_a=\tl{\bt}^\pr_a=\hat{\bt}_a=\hat{\bt}^\pr_a=0$ and $\tl{\ga}_a=\tl{\ga}^\pr_a=\hat{\ga}_a=\hat{\ga}^\pr_a=0$.
\beqa
\(y_{U}\)_{\bar{f}_I}&=&
\left(\bea{ccc}
\bt_1 Y^{(\tl{k}_1)}_{{\mathbf3},1}&\bt_2 Y^{(\tl{k}_2)}_{{\mathbf3},1}& \bt_3 Y^{(\tl{k}_3)}_{{\mathbf3},1}\\
\bt_1 Y^{(\tl{k}_1)}_{{\mathbf3},3}&\bt_2 Y^{(\tl{k}_2)}_{{\mathbf3},3}& \bt_3 Y^{(\tl{k}_3)}_{{\mathbf3},3}\\
\bt_1 Y^{(\tl{k}_1)}_{{\mathbf3},2}&\bt_2 Y^{(\tl{k}_2)}_{{\mathbf3},2}& \bt_3 Y^{(\tl{k}_3)}_{{\mathbf3},2}
\eea\right),~
\(y_{E}\)_{\bar{f}_I}=\left(\bea{ccc}
\ga_1 Y^{(\hat{k}_1)}_{{\mathbf3},1}&\ga_1 Y^{(\hat{k}_1)}_{{\mathbf3},3}& \ga_1 Y^{(\hat{k}_1)}_{{\mathbf3},2}\\
\ga_2Y^{(\hat{k}_2)}_{{\mathbf3},1}&\ga_2 Y^{(\hat{k}_2)}_{{\mathbf3},3}&\ga_2 Y^{(\hat{k}_2)}_{{\mathbf3},2}\\
\ga_3Y^{(\hat{k}_3)}_{{\mathbf3},1}&\ga_3 Y^{(\hat{k}_3)}_{{\mathbf3},3}&\ga_3 Y^{(\hat{k}_3)}_{{\mathbf3},2}\eea\right),\nn\\
\eeqa

\beqa
\(y_{U}\)_{\bar{f}_{II}}&=&
\left(\bea{ccc}
\bt_1 Y^{(\tl{k}_1)}_{{\mathbf3},3}&\bt_2 Y^{(\tl{k}_2)}_{{\mathbf3},1}& \bt_3 Y^{(\tl{k}_3)}_{{\mathbf3},1}\\
\bt_1 Y^{(\tl{k}_1)}_{{\mathbf3},2}&\bt_2 Y^{(\tl{k}_2)}_{{\mathbf3},3}& \bt_3 Y^{(\tl{k}_3)}_{{\mathbf3},3}\\
\bt_1 Y^{(\tl{k}_1)}_{{\mathbf3},1}&\bt_2 Y^{(\tl{k}_2)}_{{\mathbf3},2}& \bt_3 Y^{(\tl{k}_3)}_{{\mathbf3},2}
\eea\right),~
\(y_{E}\)_{\bar{f}_{II}}=\left(\bea{ccc}
\ga_1 Y^{(\hat{k}_1)}_{{\mathbf3},3}&\ga_1 Y^{(\hat{k}_1)}_{{\mathbf3},2}& \ga_1 Y^{(\hat{k}_1)}_{{\mathbf3},1}\\
\ga_2Y^{(\hat{k}_2)}_{{\mathbf3},1}&\ga_2 Y^{(\hat{k}_2)}_{{\mathbf3},3}&\ga_2 Y^{(\hat{k}_2)}_{{\mathbf3},2}\\
\ga_3Y^{(\hat{k}_3)}_{{\mathbf3},1}&\ga_3 Y^{(\hat{k}_3)}_{{\mathbf3},3}&\ga_3 Y^{(\hat{k}_3)}_{{\mathbf3},2}\eea\right),\nn\\
\eeqa

\beqa
\(y_{U}\)_{\bar{f}_{III}}&=&
\left(\bea{ccc}
\bt_1 Y^{(\tl{k}_1)}_{{\mathbf3},2}&\bt_2 Y^{(\tl{k}_2)}_{{\mathbf3},1}& \bt_3 Y^{(\tl{k}_3)}_{{\mathbf3},1}\\
\bt_1 Y^{(\tl{k}_1)}_{{\mathbf3},1}&\bt_2 Y^{(\tl{k}_2)}_{{\mathbf3},3}& \bt_3 Y^{(\tl{k}_3)}_{{\mathbf3},3}\\
\bt_1 Y^{(\tl{k}_1)}_{{\mathbf3},3}&\bt_2 Y^{(\tl{k}_2)}_{{\mathbf3},2}& \bt_3 Y^{(\tl{k}_3)}_{{\mathbf3},2}
\eea\right),~
\(y_{E}\)_{\bar{f}_{III}}=\left(\bea{ccc}
\ga_1 Y^{(\hat{k}_1)}_{{\mathbf3},2}&\ga_1 Y^{(\hat{k}_1)}_{{\mathbf3},1}& \ga_1 Y^{(\hat{k}_1)}_{{\mathbf3},3}\\
\ga_2Y^{(\hat{k}_2)}_{{\mathbf3},1}&\ga_2 Y^{(\hat{k}_2)}_{{\mathbf3},3}&\ga_2 Y^{(\hat{k}_2)}_{{\mathbf3},2}\\
\ga_3Y^{(\hat{k}_3)}_{{\mathbf3},1}&\ga_3 Y^{(\hat{k}_3)}_{{\mathbf3},3}&\ga_3 Y^{(\hat{k}_3)}_{{\mathbf3},2}\eea\right),\nn\\
\eeqa

\beqa
\(y_{U}\)_{\bar{f}_{IV}}&=&
\left(\bea{ccc}
\bt_1 Y^{(\tl{k}_1)}_{{\mathbf3},3}&\bt_2 Y^{(\tl{k}_2)}_{{\mathbf3},3}& \bt_3 Y^{(\tl{k}_3)}_{{\mathbf3},1}\\
\bt_1 Y^{(\tl{k}_1)}_{{\mathbf3},2}&\bt_2 Y^{(\tl{k}_2)}_{{\mathbf3},2}& \bt_3 Y^{(\tl{k}_3)}_{{\mathbf3},3}\\
\bt_1 Y^{(\tl{k}_1)}_{{\mathbf3},1}&\bt_2 Y^{(\tl{k}_2)}_{{\mathbf3},1}& \bt_3 Y^{(\tl{k}_3)}_{{\mathbf3},2}
\eea\right),~
\(y_{E}\)_{\bar{f}_{IV}}=\left(\bea{ccc}
\ga_1 Y^{(\hat{k}_1)}_{{\mathbf3},3}&\ga_1 Y^{(\hat{k}_1)}_{{\mathbf3},2}& \ga_1 Y^{(\hat{k}_1)}_{{\mathbf3},1}\\
\ga_2Y^{(\hat{k}_2)}_{{\mathbf3},3}&\ga_2 Y^{(\hat{k}_2)}_{{\mathbf3},2}&\ga_2 Y^{(\hat{k}_2)}_{{\mathbf3},1}\\
\ga_3Y^{(\hat{k}_3)}_{{\mathbf3},1}&\ga_3 Y^{(\hat{k}_3)}_{{\mathbf3},3}&\ga_3 Y^{(\hat{k}_3)}_{{\mathbf3},2}\eea\right),\nn\\
\eeqa

\beqa
\(y_{U}\)_{\bar{f}_{V}}&=&
\left(\bea{ccc}
\bt_1 Y^{(\tl{k}_1)}_{{\mathbf3},3}&\bt_2 Y^{(\tl{k}_2)}_{{\mathbf3},3}& \bt_3 Y^{(\tl{k}_3)}_{{\mathbf3},3}\\
\bt_1 Y^{(\tl{k}_1)}_{{\mathbf3},2}&\bt_2 Y^{(\tl{k}_2)}_{{\mathbf3},2}& \bt_3 Y^{(\tl{k}_3)}_{{\mathbf3},2}\\
\bt_1 Y^{(\tl{k}_1)}_{{\mathbf3},1}&\bt_2 Y^{(\tl{k}_2)}_{{\mathbf3},1}& \bt_3 Y^{(\tl{k}_3)}_{{\mathbf3},1}
\eea\right),~
\(y_{E}\)_{\bar{f}_{V}}=\left(\bea{ccc}
\ga_1 Y^{(\hat{k}_1)}_{{\mathbf3},3}&\ga_1 Y^{(\hat{k}_1)}_{{\mathbf3},2}& \ga_1 Y^{(\hat{k}_1)}_{{\mathbf3},1}\\
\ga_2Y^{(\hat{k}_2)}_{{\mathbf3},3}&\ga_2 Y^{(\hat{k}_2)}_{{\mathbf3},2}&\ga_2 Y^{(\hat{k}_2)}_{{\mathbf3},1}\\
\ga_3Y^{(\hat{k}_3)}_{{\mathbf3},3}&\ga_3 Y^{(\hat{k}_3)}_{{\mathbf3},2}&\ga_3 Y^{(\hat{k}_3)}_{{\mathbf3},1}
\eea\right),\nn\\
\eeqa

\beqa
\(y_{U}\)_{\bar{f}_{VI}}&=&
\left(\bea{ccc}
\bt_1 Y^{(\tl{k}_1)}_{{\mathbf3},2}&\bt_2 Y^{(\tl{k}_2)}_{{\mathbf3},2}& \bt_3 Y^{(\tl{k}_3)}_{{\mathbf3},1}\\
\bt_1 Y^{(\tl{k}_1)}_{{\mathbf3},1}&\bt_2 Y^{(\tl{k}_2)}_{{\mathbf3},1}& \bt_3 Y^{(\tl{k}_3)}_{{\mathbf3},3}\\
\bt_1 Y^{(\tl{k}_1)}_{{\mathbf3},3}&\bt_2 Y^{(\tl{k}_2)}_{{\mathbf3},3}& \bt_3 Y^{(\tl{k}_3)}_{{\mathbf3},2}
\eea\right),~
\(y_{E}\)_{\bar{f}_{VI}}=\left(\bea{ccc}
\ga_1 Y^{(\hat{k}_1)}_{{\mathbf3},2}&\ga_1 Y^{(\hat{k}_1)}_{{\mathbf3},1}& \ga_1 Y^{(\hat{k}_1)}_{{\mathbf3},3}\\
\ga_2Y^{(\hat{k}_2)}_{{\mathbf3},2}&\ga_2 Y^{(\hat{k}_2)}_{{\mathbf3},1}&\ga_2 Y^{(\hat{k}_2)}_{{\mathbf3},3}\\
\ga_3Y^{(\hat{k}_3)}_{{\mathbf3},1}&\ga_3 Y^{(\hat{k}_3)}_{{\mathbf3},3}&\ga_3 Y^{(\hat{k}_3)}_{{\mathbf3},2}
\eea\right),\nn\\
\eeqa

\beqa
\(y_{U}\)_{\bar{f}_{VII}}&=&
\left(\bea{ccc}
\bt_1 Y^{(\tl{k}_1)}_{{\mathbf3},2}&\bt_2 Y^{(\tl{k}_2)}_{{\mathbf3},2}& \bt_3 Y^{(\tl{k}_3)}_{{\mathbf3},2}\\
\bt_1 Y^{(\tl{k}_1)}_{{\mathbf3},1}&\bt_2 Y^{(\tl{k}_2)}_{{\mathbf3},1}& \bt_3 Y^{(\tl{k}_3)}_{{\mathbf3},1}\\
\bt_1 Y^{(\tl{k}_1)}_{{\mathbf3},3}&\bt_2 Y^{(\tl{k}_2)}_{{\mathbf3},3}& \bt_3 Y^{(\tl{k}_3)}_{{\mathbf3},3}
\eea\right),~
\(y_{E}\)_{\bar{f}_{VII}}=\left(\bea{ccc}
\ga_1 Y^{(\hat{k}_1)}_{{\mathbf3},2}&\ga_1 Y^{(\hat{k}_1)}_{{\mathbf3},1}& \ga_1 Y^{(\hat{k}_1)}_{{\mathbf3},3}\\
\ga_2Y^{(\hat{k}_2)}_{{\mathbf3},2}&\ga_2 Y^{(\hat{k}_2)}_{{\mathbf3},1}&\ga_2 Y^{(\hat{k}_2)}_{{\mathbf3},3}\\
\ga_3Y^{(\hat{k}_3)}_{{\mathbf3},2}&\ga_3 Y^{(\hat{k}_3)}_{{\mathbf3},1}&\ga_3 Y^{(\hat{k}_3)}_{{\mathbf3},3}
\eea\right),\nn\\
\eeqa

\beqa
\(y_{U}\)_{\bar{f}_{VIII}}&=&
\left(\bea{ccc}
\bt_1 Y^{(\tl{k}_1)}_{{\mathbf3},3}&\bt_2 Y^{(\tl{k}_2)}_{{\mathbf3},3}& \bt_3 Y^{(\tl{k}_3)}_{{\mathbf3},2}\\
\bt_1 Y^{(\tl{k}_1)}_{{\mathbf3},2}&\bt_2 Y^{(\tl{k}_2)}_{{\mathbf3},2}& \bt_3 Y^{(\tl{k}_3)}_{{\mathbf3},1}\\
\bt_1 Y^{(\tl{k}_1)}_{{\mathbf3},1}&\bt_2 Y^{(\tl{k}_2)}_{{\mathbf3},1}& \bt_3 Y^{(\tl{k}_3)}_{{\mathbf3},3}
\eea\right),~
\(y_{E}\)_{\bar{f}_{VIII}}=\left(\bea{ccc}
\ga_1 Y^{(\hat{k}_1)}_{{\mathbf3},3}&\ga_1 Y^{(\hat{k}_1)}_{{\mathbf3},2}& \ga_1 Y^{(\hat{k}_1)}_{{\mathbf3},1}\\
\ga_2Y^{(\hat{k}_2)}_{{\mathbf3},3}&\ga_2 Y^{(\hat{k}_2)}_{{\mathbf3},2}&\ga_2 Y^{(\hat{k}_2)}_{{\mathbf3},1}\\
\ga_3Y^{(\hat{k}_3)}_{{\mathbf3},2}&\ga_3 Y^{(\hat{k}_3)}_{{\mathbf3},1}&\ga_3 Y^{(\hat{k}_3)}_{{\mathbf3},3}
\eea\right),\nn\\
\eeqa

\beqa
\(y_{U}\)_{\bar{f}_{IX}}&=&
\left(\bea{ccc}
\bt_1 Y^{(\tl{k}_1)}_{{\mathbf3},3}&\bt_2 Y^{(\tl{k}_2)}_{{\mathbf3},2}& \bt_3 Y^{(\tl{k}_3)}_{{\mathbf3},2}\\
\bt_1 Y^{(\tl{k}_1)}_{{\mathbf3},2}&\bt_2 Y^{(\tl{k}_2)}_{{\mathbf3},1}& \bt_3 Y^{(\tl{k}_3)}_{{\mathbf3},1}\\
\bt_1 Y^{(\tl{k}_1)}_{{\mathbf3},1}&\bt_2 Y^{(\tl{k}_2)}_{{\mathbf3},3}& \bt_3 Y^{(\tl{k}_3)}_{{\mathbf3},3}
\eea\right),~
\(y_{E}\)_{\bar{f}_{IX}}=\left(\bea{ccc}
\ga_1 Y^{(\hat{k}_1)}_{{\mathbf3},3}&\ga_1 Y^{(\hat{k}_1)}_{{\mathbf3},2}& \ga_1 Y^{(\hat{k}_1)}_{{\mathbf3},1}\\
\ga_2Y^{(\hat{k}_2)}_{{\mathbf3},2}&\ga_2 Y^{(\hat{k}_2)}_{{\mathbf3},1}&\ga_2 Y^{(\hat{k}_2)}_{{\mathbf3},3}\\
\ga_3Y^{(\hat{k}_3)}_{{\mathbf3},2}&\ga_3 Y^{(\hat{k}_3)}_{{\mathbf3},1}&\ga_3 Y^{(\hat{k}_3)}_{{\mathbf3},3}
\eea\right),\nn\\
\eeqa

\beqa
\(y_{U}\)_{\bar{f}_{X}}&=&
\left(\bea{ccc}
\bt_1 Y^{(\tl{k}_1)}_{{\mathbf3},1}&\bt_2 Y^{(\tl{k}_2)}_{{\mathbf3},3}& \bt_3 Y^{(\tl{k}_3)}_{{\mathbf3},2}\\
\bt_1 Y^{(\tl{k}_1)}_{{\mathbf3},3}&\bt_2 Y^{(\tl{k}_2)}_{{\mathbf3},2}& \bt_3 Y^{(\tl{k}_3)}_{{\mathbf3},1}\\
\bt_1 Y^{(\tl{k}_1)}_{{\mathbf3},2}&\bt_2 Y^{(\tl{k}_2)}_{{\mathbf3},1}& \bt_3 Y^{(\tl{k}_3)}_{{\mathbf3},3}
\eea\right),~
\(y_{E}\)_{\bar{f}_{X}}=\left(\bea{ccc}
\ga_1 Y^{(\hat{k}_1)}_{{\mathbf3},1}&\ga_1 Y^{(\hat{k}_1)}_{{\mathbf3},3}& \ga_1 Y^{(\hat{k}_1)}_{{\mathbf3},2}\\
\ga_2Y^{(\hat{k}_2)}_{{\mathbf3},3}&\ga_2 Y^{(\hat{k}_2)}_{{\mathbf3},2}&\ga_2 Y^{(\hat{k}_2)}_{{\mathbf3},1}\\
\ga_3Y^{(\hat{k}_3)}_{{\mathbf3},2}&\ga_3 Y^{(\hat{k}_3)}_{{\mathbf3},1}&\ga_3 Y^{(\hat{k}_3)}_{{\mathbf3},3}
\eea\right),\nn\\
\eeqa

\item  $\tl{k}_i\equiv k_i+k_F=6$ and $\hat{k}_i\equiv k_i+k_E=6$:

The superpotential can be written as
\beqa
W&\supseteq& \(\al_1 Y_{{\bf 3}}^{(4)}(F F)_{{\bf 3}S} h +\al_2 Y_{\bf 1}^{(4)}(F F)_{\bf 1} h + \al_3 Y_{\bf 1^\pr}^{(4)} (F F)_{\bf 1^{\pr\pr}} h\)\nn\\
 &+&\(\sum\limits_{i}\tl{\beta}_i Y_{{\bf 3}I}^{(6)} (F \bar{f}_i)_{{\bf 3}} \bar{h}
 +\sum\limits_{i}\tl{\beta}^\pr_i Y_{{\bf 3}II}^{(6)} (F \bar{f}_i)_{{\bf 3}} \bar{h} \)\nn\\
 &+&\(\sum\limits_{i}\tl{\ga}_i Y_{{\bf 3}I}^{(\hat{k}_i)} (\bar{f}_i E)_{{\bf 3}} h
  +\sum\limits_{i}\tl{\ga}_i^\pr Y_{{\bf 3}II}^{(\hat{k}_i)} (\bar{f}_i E)_{{\bf 3}} h\)\nn\\
&+&\(\lambda_1 Y_{{\bf 3}}^{(2)}  (S F)_{{\bf 3}S} \overline{H}+\lambda_2 Y_{{\bf 3}}^{(2)}  (S F)_{{\bf 3}A} \overline{H}    \)+\f{\ka_1}{2} \Lambda_2 (S S)_{\bf 1}~.
\eeqa
The parameter $\al_1, \tl{\beta}_i,\tl{\ga}_i,\la_1,\ka_1$ can be taken to be real while others $\al_2,\al_3,\tl{\beta}_i^\pr,\tl{\ga}_i^\pr,\la_2$ are in general complex.
The form of the up-type quark and lepton mass matrices take the form in eq.(\ref{up:massv1}) and eq.(\ref{lepton:massv1}), with ${\bt}_a=\hat{\bt}_a=\hat{\bt}^\pr=0$ and ${\ga}_a=\hat{\ga}_a=\hat{\ga}_a^\pr=0$.

\item  $\tl{k}_i\equiv k_i+k_F=6$ and $\hat{k}_i\equiv k_i+k_E=2,4$ for any $i$:

Similar discussions can be given here.
The form of the up-type quark and lepton mass matrices take the form in eq.(\ref{up:massv1}) and eq.(\ref{lepton:massv1}), with ${\bt}_a=\hat{\bt}_a=\hat{\bt}^\pr=0$ and
$\tl{\ga}_a=\tl{\ga}^\pr_a=\hat{\ga}_a=\hat{\ga}_a^\pr=0$.

For $\tl{k}_i=8$, we need just repeat the replacement $\tl{\bt}_a\ra\hat{\bt}_a$ and
$\tl{\bt}^\pr_a\ra \hat{\bt}_a^\pr$.
\item  $\tl{k}_i\equiv k_i+k_F=2,4$ and $\hat{k}_i\equiv k_i+k_E=6$ for any $i$:

With similar discussions, we can obtain that the form of the up-type quark and lepton mass matrices, which take the form in eq.(\ref{up:massv1}) and eq.(\ref{lepton:massv1}), with $\tl{\bt}_a=\tl{\bt}^\pr_a=\hat{\bt}_a=\hat{\bt}^\pr=0$  and ${\ga}_a=\hat{\ga}_a=\hat{\ga}_a^\pr=0$.

For $\hat{k}_i=8$, we need just repeat the replacement
$\tl{\ga}_a\ra\hat{\ga}_a$ and
$\tl{\ga}^\pr_a\ra \hat{\ga}_a^\pr$.
\item  Some of the $\tl{k}_i\equiv k_i+k_F=2,4$ and some of the $\tl{k}_j=6,8$; some of the  $\hat{k}_i\equiv k_i+k_E=6,8$  and some of the $\hat{k}_j=2,4$:

 Similar discussions can be given here to obtain the mass matrices.

\eit

\item For $k_F=2,k_S=2$:

Only the expressions for ${\cal M}^{SN}$ and ${\cal M}^{SS}$ change with respect to the $k_F=2,k_S=0$ case, which are given as
\beqa
{\cal M}^{SN}_{ij}&=&
\la_1 \Lambda_1 S_{\bf 3}^{(4)}(\tau)+\la_2\Lambda_1 A_{\bf 3}^{(4)}(\tau)+
\la_3\Lambda_1 S_{\bf 1}^{(4)}(\tau)+\la_4\Lambda_1 S_{{\bf 1}^\pr}^{(4)}(\tau) ~\nn\\
{\cal M}^{SS}_{ij}&=&\ka_1 \Lambda_2 S_{\bf 3}^{(4)}(\tau)
+\ka_2\Lambda_2 S_{\bf 1}^{(4)}(\tau)+\ka_3\Lambda_2 S_{{\bf 1}^\pr}^{(4)}(\tau)~,
\eeqa
with the corresponding superpotential
\beqa
W&\supseteq& \(\lambda_1 Y_{{\bf 3}}^{(4)}  (S F)_{{\bf 3}S} \overline{H}+\lambda_2 Y_{{\bf 3}}^{(4)}  (S F)_{{\bf 3}A} \overline{H}+\lambda_3 Y_{\bf 1}^{(4)}  (S F)_{\bf 1} \overline{H}+\lambda_1 Y_{\bf 1^\pr}^{(4)}  (S F)_{\bf 1^{\pr\pr}} \overline{H}\)~,\nn\\
&+&\[\f{\ka_1}{2} \Lambda_2 Y_{{\bf 3}}^{(4)} (S S)_{{\bf 3}S}
+\f{\ka_2}{2} \Lambda_2 Y_{\bf 1}^{(4)} (S S)_{\bf 1}+\f{\ka_3}{2} \Lambda_2 Y_{\bf 1^\pr}^{(4)} (S S)_{\bf 1^{\pr\pr}}\]~.
\eeqa
The parameters $\la_1,\ka_1$ can be taken to be real while the others are complex.

\item For $k_F=4,k_S=0$:

The superpotential can be written as
\beqa
W&\supseteq& \(\al_1 Y_{{\bf 3}I}^{(8)}(F F)_{{\bf 3}S} h+\al_2 Y_{{\bf 3}II}^{(8)}(F F)_{{\bf 3}S} h +\al_3 Y_{\bf 1}^{(8)}(F F)_{\bf 1} h + \al_4 Y_{\bf 1^\pr}^{(8)} (F F)_{\bf 1^{\pr\pr}} h \right.\nn\\&+&\left.
\al_5 Y_{\bf 1^{\pr\pr}}^{(8)} (F F)_{\bf 1^{\pr}} h\)
 +\(\sum\limits_{i,A}\beta_{i;A} Y_{{\bf 3};A}^{(\tl{k}_i)} (F \bar{f}_i)_{{\bf 3}} \bar{h} \)
  +\(\sum\limits_{i,A}\ga_{i;A} Y_{{\bf 3};A}^{(\hat{k}_i)} (\bar{f}_i E)_{{\bf 3}} h\)\nn\\
&+&\(\lambda_1 Y_{{\bf 3}}^{(4)}  (S F)_{{\bf 3}S} \overline{H}+\lambda_2 Y_{{\bf 3}}^{(4)}  (S F)_{{\bf 3}A} \overline{H}+\lambda_3 Y_{\bf 1}^{(4)}  (S F)_{\bf 1} \overline{H}+\lambda_1 Y_{\bf 1^\pr}^{(4)}  (S F)_{\bf 1^{\pr\pr}} \overline{H}\)\nn\\
&+&\f{\ka_1}{2} \Lambda_2 (S S)_{\bf 1}~,
\eeqa
with the indices $'A'$ taken values in $\emph{2,4,6I,6II,8I,8II}$, depending on the values of $k_F+k_{\bar{f}_i}$.
The mass matrices for the down type quarks and neutrinos are
\beqa
\(y_{D}\)_{ij}&=& \al_1 S^{(8)}_{\mathbf3I}  +\al_2 S^{(8)}_{\mathbf3II}+ \al_3 S^{(8)}_{\mathbf1} +\al_4 S^{(8)}_{\mathbf1'}+\al_5 S^{(8)}_{\mathbf1^{\pr\pr}}\,,\nn\\
{\cal M}^{SN}_{ij}&=&\la_1 \Lambda_1 S_{\bf 3}^{(4)}(\tau)+\la_2\Lambda_1 A_{\bf 3}^{(4)}(\tau)+
\la_3\Lambda_1 S_{\bf 1}^{(4)}(\tau)+\la_4\Lambda_1 S_{{\bf 1}^\pr}^{(4)}(\tau) ~,\nn\\
{\cal M}^{SS}_{ij}&=&\ka_1 \Lambda_2 S_{\bf 1}^0(\tau)~,~~~\(y_{N}\)^{Dirac}_{ij}=\(y_{U}\)^T~.
\eeqa

The up quark mass matrices and lepton quark matrices take the form eq.(\ref{up:massv1}) and eq.(\ref{lepton:massv1}).
For $\tl{k}_a\equiv k_F+k_{\bar{f}_i}=2,4$, the coefficients $\tl{\beta}_a=\tl{\beta}^\pr_a=\hat{\beta}_a=\hat{\beta}_a^\pr=0$; for $\tl{k}_a=6$, ${\beta}_a=\hat{\beta}_a=\hat{\beta}_a^\pr=0$; for $\tl{k}_a=8$, $\tl{\beta}_a=\tl{\beta}^\pr_a={\beta}_a=0$. Similar expressions hold for lepton mass matrices.


\item For $k_F=4,k_S=2$:

Only the expressions for ${\cal M}^{SN}$ and ${\cal M}^{SS}$ change with respect to the $k_F=4,k_S=0$ case, which are given as
\beqa
{\cal M}^{SN}_{ij}&=&
 \la_1 \Lambda_1 S_{{\bf 3}I}^{(6)}(\tau)+\la_2\Lambda_1 A_{{\bf 3}I}^{(6)}(\tau)
+\la_3 \Lambda_1 S_{{\bf 3}II}^{(6)}(\tau)+\la_4\Lambda_1 A_{{\bf 3}II}^{(6)}(\tau)
+\la_5\Lambda_1 S_{\bf 1}^{(6)}(\tau)~,\nn\\
{\cal M}^{SS}_{ij}&=&\ka_1 \Lambda_2 S_{\bf 3}^{(4)}(\tau)
+\ka_2\Lambda_2 S_{\bf 1}^{(4)}(\tau)+\ka_3\Lambda_2 S_{{\bf 1}^\pr}^{(4)}(\tau)~,
\eeqa
with the corresponding superpotential
\beqa
W&\supseteq&
\(\lambda_1 Y_{{\bf 3}I}^{(6)}  (S F)_{{\bf 3}S} \overline{H}+\lambda_2 Y_{{\bf 3}I}^{(6)}  (S F)_{{\bf 3}A} \overline{H}+ \lambda_3 Y_{{\bf 3}II}^{(6)}  (S F)_{{\bf 3}S} \overline{H}+\lambda_4 Y_{{\bf 3}II}^{(6)}  (S F)_{{\bf 3}A} \overline{H}\right.\nn\\&+&\left. \lambda_5 Y_{\bf 1}^{(6)}  (S F)_{\bf 1} \overline{H} \)+\[\f{\ka_1}{2} \Lambda_2 Y_{{\bf 3}}^{(4)} (S S)_{{\bf 3}S}
+\f{\ka_2}{2} \Lambda_2 Y_{\bf 1}^{(4)} (S S)_{\bf 1}+\f{\ka_3}{2} \Lambda_2 Y_{\bf 1^\pr}^{(4)} (S S)_{\bf 1^{\pr\pr}}\].\nn\\
\eeqa

\eit

\subsubsection{ $\rho_{\bar{f}}={\bf 3},~\rho_{F}={\bf 1,1^\pr,1^{\pr\pr}},~\rho_{E}={\bf 3}, \rho_{S}={\bf 3}$}
We have ten combinations for three generation $\rho_{F_i}$, namely
\beqa
 &&({\bf 1,1,1})_{F_I},~ ({\bf 1^{\pr},1,1})_{F_{II}}, ~({\bf 1^{\pr\pr},1,1})_{F_{III}},~({\bf 1^{\pr},1^{\pr},1})_{F_{IV}},~({\bf 1^{\pr},1^{\pr},1^{\pr}})_{F_{V}},\nn\\
 &&({\bf 1^{\pr\pr},1^{\pr\pr},1})_{F_{VI}},({\bf 1^{\pr\pr},1^{\pr\pr},1^{\pr\pr}})_{F_{VII}},({\bf 1^{\pr}, 1^{\pr},1^{\pr\pr}})_{F_{VIII}},({\bf 1^{\pr}, 1^{\pr\pr},1^{\pr\pr}})_{F_{IX}},
 ({\bf 1, 1^{\pr},1^{\pr\pr}})_{F_{X}}\nn
 \eeqa
 and modular weight choice $(k_{\bar{f}},k_{F},k_{E},k_{S})=(k_{\bar{f}},(k_{1},k_{2},k_3),k_{E},k_{S})$.

The possible form of the superpotential can be written as
\beqa
W&\supseteq& \(\al_1 Y_{\bf 1}^{(k_{F_i}+k_{F_j})}(F_i F_j)_{\bf 1} h +\al_2 Y_{\bf 1^\pr}^{(k_{F_i}+k_{F_j})}(F_i F_j)_{\bf 1^{\pr\pr}} h + \al_3 Y_{\bf 1^{\pr\pr}}^{(k_{F_i}+k_{F_j})} (F_i F_j)_{\bf 1^{\pr}} h\)\nn\\
 &+&\(\sum\limits_{i,M}\beta_i Y_{{\bf 3}; M}^{(k_i+k_{\bar{f}})} (F_i \bar{f})_{{\bf 3}} \bar{h}\) \nn\\
 & +&\(\sum\limits_{i,M}\ga_1 Y_{{\bf 3}; M}^{(\hat{k}_i)} (\bar{f} E)_{{\bf 3}S} h+\sum\limits_{i,M}\ga_2 Y_{{\bf 3}; M}^{(\hat{k}_i)} (\bar{f} E)_{{\bf 3}A} h
  +\sum\limits_{i={\bf 1},{\bf 1}^{\pr\pr},{\bf 1}^\pr}{\ga}_i Y_{\bf 1,1^\pr,1^{\pr\pr}}^{(\hat{k}_i)} (\bar{f} E)_{\bf 1,1^{\pr\pr},1^\pr} h\)\nn\\
&+&\(
\sum\limits_{i,M} \la_i Y_{{\bf 3}; M}^{(k_S+k_{F_i})} (S F_i)_{{\bf 3}} \overline{H}\)\nn\\
&+&\f{1}{2} \Lambda_2\( \sum\limits_{M}\ka_1 Y_{{\bf 3}; M}^{(2k_S)} (S S)_{{\bf 3}S} +
+\sum\limits_{i={\bf 1},{\bf 1}^{\pr\pr},{\bf 1}^\pr}{\ka}_i Y_{\bf 1,1^\pr,1^{\pr\pr}}^{(2k_S)} (S S)_{\bf 1,{\bf 1}^{\pr\pr},{\bf 1}^\pr}\)~,
\eeqa
with the index $'M'$ taken values in $\emph{2,4,6I,6II,8I,8II}$, depending on the values of the modular weights. In general, $\al_{1,2,3}$ should be replaced by $\al_{1,2,3;ij}$. To keep the simply and predictive power of the model, we keep only three free parameters $\al_{1,2,3}$. The parameter $\al_1, \beta_i,\ga_1,\la_i,\ka_1$ can be taken to be real while others are in general complex.

Define $Y_{{\bf 1}}^{k_{F_i}+k_{F_j}}=U(i,j)$,
$~Y_{{\bf 1}^\pr}^{k_{F_i}+k_{F_j}}=V(i,j)$,$~Y_{{\bf 1}^{\pr\pr}}^{k_{F_i}+k_{F_j}}=W(i,j)$
and
\beqa
\label{neutrino:simplication}
U[i,j]&=&\bt_a Y^{(\tl{k}_a)}_{{\mathbf3},b}+\tl{\bt}_a Y^{(6)}_{{\mathbf3I},b}+\tl{\bt}^\pr_a Y^{(6)}_{{\mathbf3II},b}+\hat{\bt}_a Y^{(8)}_{{\mathbf3I},b}+\hat{\bt}^\pr_a Y^{(8)}_{{\mathbf3II},b}~,\nn\\
S[i,j]&=&\la_a Y^{(\hat{k}_a)}_{{\mathbf3},b}+\tl{\la}_a Y^{(6)}_{{\mathbf3I},b}+\tl{\la}^\pr_a Y^{(6)}_{{\mathbf3II},b}+\hat{\la}_a Y^{(8)}_{{\mathbf3I},b}+\hat{\la}^\pr_a Y^{(8)}_{{\mathbf3II},b}~,
\eeqa
 with $\tl{k}_a\equiv k_{F_i}+k_{\bar{f}}$ and $\hat{k}_a\equiv k_{F_i}+k_{S}$.

The mass matrices for down-type quark are given by
\beqa
\label{example:F111}
\(y_{D}\)_{I}&=&\al_1\(\bea{ccc}U(1,1)&U(1,2)&U(1,3)\\U(2,1)&U(2,2)&U(2,3)\\U(3,1)&U(3,2)&U(3,3) \eea\),
\(y_{D}\)_{II}=\(\bea{ccc}\al_2 V(1,1)&\al_3 W(1,2)&\al_3 W(1,3) \\\al_3 W(2,1)&\al_1U(2,2)&\al_1U(2,3)\\\al_3 W(3,1)&\al_1 U(3,2)&\al_1 U(3,3) \eea\),\nn\\
\(y_{D}\)_{III}&=&\(\bea{ccc}\al_3 W(1,1)&\al_2 V(1,2)&\al_2V(1,3) \\\al_2 V(2,1)&\al_1U(2,2)&\al_1U(2,3)\\\al_2 V(3,1)&\al_1 U(3,2)&\al_1 U(3,3) \eea\),
\(y_{D}\)_{IV}=\(\bea{ccc}\al_2 V(1,1)&\al_2 V(1,2)&\al_3 W(1,3) \\\al_2 V(2,1)&\al_2 V(2,2)&\al_3 W(2,3)\\\al_3W(3,1)&\al_3W(3,2)&\al_1 U(3,3) \eea\),\nn\eeqa\beqa
\(y_{D}\)_{V}&=&\al_2\(\bea{ccc} V(1,1)& V(1,2)&V(1,3) \\ V(2,1)&V(2,2)& V(2,3)\\V(3,1)&V(3,2)&V(3,3) \eea\),
\(y_{D}\)_{VI}=\(\bea{ccc} \al_3 W(1,1)&  \al_3 W(1,2)&\al_2 V(1,3) \\  \al_3 W(2,1)& \al_3 W(2,2)& \al_2 V(2,3)\\\al_2 V(3,1)&\al_2 V(3,2)&\al_1 U(3,3) \eea\),\nn\\
\(y_{D}\)_{VII}&=&\al_3\(\bea{ccc} W(1,1)&  W(1,2)& W(1,3) \\   W(2,1)& W(2,2)& W(2,3)\\ W(3,1)&W(3,2)&W(3,3) \eea\),
\(y_{D}\)_{VIII}=\(\bea{ccc}\al_2 V(1,1)&\al_2 V(1,2)&\al_1 U(1,3) \\\al_2 V(2,1)&\al_2 V(2,2)&\al_1 U(2,3)\\\al_1 U(3,1)&\al_1 U(3,2)&\al_3 W(3,3) \eea\),\nn\eeqa\beqa
\(y_{D}\)_{IX}&=&\(\bea{ccc}\al_2 V(1,1)&\al_1 U(1,2)&\al_1 U(1,3) \\\al_1 U(2,1)&\al_3 W(2,2)&\al_3 W(2,3)\\\al_1 U(3,1)&\al_3 W(3,2)&\al_3 W(3,3) \eea\),
\(y_{D}\)_{X}=\(\bea{ccc}\al_1 U(1,1)&\al_3 W(1,2)&\al_2 V(1,3) \\\al_3 W(2,1)&\al_2 V(2,2)&\al_1 U(2,3)\\\al_2 V(3,1)&\al_1 U(3,2)&\al_3 W(3,3) \eea\),\nn
\eeqa
The mass matrices for up-type quark take the forms as the {\bf transpose} of eq.(\ref{up:massv1}) while ${\cal M}_{SN}$ takes the same form as eq.(\ref{up:massv1}) after replacing $U[a,b]$ with $S[a,b]$. For example,
\beqa
\(y_{U}\)_{\bar{f}_{I}}&=&\left(\bea{ccc}
U[1,1],&U[1,3],&U[1,2],\\
U[2,1],&U[2,3],&U[2,2],\\
U[3,1],&U[3,3],&U[3,2],
\eea\right),~{\cal M}_{SN;I}=\left(\bea{ccc}
S[1,1],&S[2,1],&S[3,1],\\
S[1,3],&S[2,3],&S[3,3],\\
S[1,2],&S[2,2],&S[3,2],
\eea\right),~
\eeqa
Similar to the previous cases, when $\tl{k}_a\equiv k_{F_a}+k_{\bar{f}}=2,4$, the coefficients $\tl{\beta}_a=\tl{\beta}^\pr_a=\hat{\beta}_a=\hat{\beta}_a^\pr=0$; for $\tl{k}_a=6$, ${\beta}_a=\hat{\beta}_a=\hat{\beta}_a^\pr=0$; for $\tl{k}_a=8$, $\tl{\beta}_a=\tl{\beta}^\pr_a={\beta}_a=0$. Similar discussions hold for ${\cal M}_{SN}$ with different choices of $\hat{k}_a\equiv k_{S}+k_{F_a}$.

The expressions for ${\cal M}_{SS}$ are also similar to the discussions in subsection (\ref{All:3}).

\bit
\item $k_{\bar{f}}+k_E=0$, which means $(k_{\bar{f}} ,k_E)=(0,0)$.

The charged lepton mass matrices are given by
\beqa
\(y_{E}\)_{ij}&=&\ga_1 S_{\bf 1}^0(\tau)~.
\eeqa
\item $k_{\bar{f}}+k_E=2$, which means $(k_{\bar{f}} ,k_E)=(2,0),(0,2)$

The lepton mass matrices are given by
\beqa
\(y_{L}\)&=& \ga_1 S_{\mathbf3}^{(2)}(\tau)+\ga_2 A_{\mathbf3}^{(2)}(\tau)~.
\eeqa

\item $k_{\bar{f}}+k_E=4$, which means $(k_{\bar{f}} ,k_E )=(2,2),(4,0),(0,4)$.

The charged lepton mass matrices are given by
\beqa
\(y_{E}\)_{ij}&=&\ga_1\,S^{(4)}_{\mathbf3} +\ga_2 A^{(4)}_{\mathbf3} +\ga_3 S^{(4)}_{\mathbf1} + \ga_4 S^{(4)}_{\mathbf1'}\,.
\eeqa

\item $k_{\bar{f}}+k_E=6$, which means $(k_{\bar{f}},k_E)=(6,0),(4,2),(2,4),(6,0)$.

The charged lepton mass matrices are given by
\beqa
\(y_{E}\)_{ij}&=&\ga_1\,S^{(6)}_{\mathbf{3}I} +\ga_2 A^{(6)}_{\mathbf{3}I} +\ga_3 S^{(6)}_{\mathbf{3}II} +\ga_4 A^{(6)}_{\mathbf{3}II} +\ga_5 S^{(6)}_{\mathbf1}\,.
\eeqa

\eit
\subsubsection{ $\rho_{\bar{f}}={\bf 3},~\rho_{F}={\bf 3},~\rho_{E}={\bf 1,1^\pr,1^{\pr\pr}}, \rho_{S}={\bf 3}$}

We will adopt the notations for the combinations for ${\bf 1,1^\pr,1^{\pr\pr}}$ in eq.(\ref{combination}). In this case, only the lepton mass matrices will change.

 \beqa
W&\supseteq& \sum\limits_{i}\ga_i Y_{\bf 3}^{(\hat{k}_i)} (\bar{f} E_i)_{\bf 3} h~,~~
~\hat{k}_i\equiv k_{\bar{f}}+k_{E_i}~.
 \eeqa
The lepton mass matrices take the form of {\bf transpose} of eq.(\ref{lepton:massv1}).

\subsubsection{ $\rho_{\bar{f}}={\bf 3},~\rho_{F}={\bf 3},~\rho_{E}={\bf 3}, \rho_{S}={\bf 1,1^\pr,1^{\pr\pr}}$}

We will adopt the notations for the combinations for ${\bf 1,1^\pr,1^{\pr\pr}}$ in eq.(\ref{combination}). In this case, only the mass matrices in the neutrino sector will change.

The relevant superpotential for neutrino sector is given by
 \beqa
W&\supseteq& \(
\sum\limits_{i,M} \la_i Y_{{\bf 3}; M}^{(k_{S_i}+k_{F})} (S_i F)_{\bf 3} \overline{H}\)\nn\\
&+&\f{\Lambda_2}{2} \(\ka_1 Y_{\bf 1}^{(k_{S_i}+k_{S_j})}(S_i S_j)_{\bf 1} +\ka_2 Y_{\bf 1^\pr}^{(k_{S_i}+k_{S_j})}(S_i S_j)_{\bf 1^{\pr\pr}} + \ka_3 Y_{\bf 1^{\pr\pr}}^{(k_{S_i}+k_{S_j})} (S_i S_j)_{\bf 1^{\pr}}\).
 \eeqa
The matrices ${\cal M}_{SN}$ takes the form as the transpose of eq.(\ref{up:massv1}) after replacing $U[a,b]$ with $S[a,b]$, adopting the notation in eq.(\ref{neutrino:simplication}). The discussions are also similar to the case presented below eq.(\ref{example:F111}), depending on the values of $k_S+k_{F_i}$.

We can define
$$Y_{1}^{k_{S_i}+k_{S_j}}=NU(i,j),
~Y_{1^\pr}^{k_{S_i}+k_{S_j}}=NV(i,j), ~Y_{1^{\pr\pr}}^{k_{S_i}+k_{S_j}}=NW(i,j).$$

The mass matrices ${\cal M}_{SS}/\Lambda_2$ takes the same form as eq.(\ref{example:F111}) after the repalcement
 \beqa
 U(i,j)\ra NU(i,j)~, V(i,j)\ra NV(i,j)~, W(i,j)\ra NW(i,j)~, \al_i\ra \ka_i~.
  \eeqa

 The value $\ka_1$ can be chosen to be real while $\ka_2,\ka_3$ are complex.
\section{\label{sec:num}Numerical Fitting}
 We numerically scan the parameter spaces for various scenarios of the flipped SU(5) GUT with $A_4$ modular flavor symmetry to fit our theoretical prediction to the experimental data on the flavor structures for SM plus neutrinos. To keep the predictive power of our modular GUT scheme, we concentrate on the scenarios in which at least one of the $\bar{f},F,E,S$ superfields transforms as the triplet of $A_4$ modular group. The VEVs of the complex modulus fields are taken to be free parameters, which in principle can be determined by some modulus stability mechanism, for example, the KKLT-type settings. The VEV of the modulus field $\tau$ can be chosen to lie in the fundamental domain $\mathcal{D}=\left\{\tau| \texttt{Im}(\tau)>0, |\texttt{Re}(\tau)|\leq\frac{1}{2}, |\tau|\geq1\right\}$.

  The GUT-scale flavor structures of quarks and leptons predicted by our models need to be evolved to the EW scale with the renormalization group equation (RGE) before the implement of the $\chi^2$ fit to the experimental data of SM and neutrino flavor structures. In our realistic numerical calculations, we use two-loop RGE to evolve our predicted flavor structures to the SUSY scale and fit our results to the low energy flavor parameters in~\cite{value1:rge2013} obtained after evolution and matching from $M_Z$ to the SUSY scale with $M_{SUSY}=1~{\rm TeV},\tan\beta=5$ and SM input values from PDG review~\cite{{PDG:2020}}. Our two-loop RGE evolution program, which is based on the SARAH~\cite{SARAH1,SARAH2,SARAH3} package, takes into account the interactions in neutrino seesaw mechanism. In obtaining such low energy flavor parameters, minimal supersymmetric extension of the Standard Model (MSSM) is assumed with the $\tan\beta$ enhanced one-loop SUSY threshold corrections~\cite{value1:rge2013}. The mass matrices at the SUSY scale predicted by our model after RGE evolutions  can be diagonalized to extract the corresponding lepton, quark masses and mixing matrices.

   To optimize our fitting procedure, we choose to best-fit the quark sector and lepton sector separately. We require that the $\chi^2$ values for both sectors should to be less than 100, respectively. The forms of $\chi^2$ functions are given as
 \beqa
\chi^2_{q,l}\equiv \sum\limits_{i_{q,l}}\chi^2_{i_{q,l}}~,~~\chi^2_{i_{q,l}}&=&\left(\frac{{obs}_{i_{q,l}} - \left<{obs}_{i_{q,l}}\right>}{\sigma_{i_{q,l}}}\right)^2~,
 \eeqa
with $'obs_{i_{q,l}}'$ the mass and flavor mixing parameters predicted by our model at the SUSY scale. As noted in the previous paragraph, the central values $'\left<{obs}_{i_{q,l}}\right>'$ of flavor parameters for quarks and charged leptons adopt the corresponding numerical results in~\cite{value1:rge2013}. Neutrino data for Normal Ordering (NO) neutrino masses~\cite{Esteban:2020cvm} are adopted in our fitting, which is slightly preferred over the inverted ordering (IO) masses by the present data~\cite{Esteban:2020cvm}. The $'\sigma_{i_{q,l}}'$ are the $1\sigma$ deviations of the corresponding observables (see Table~\ref{tab:central-data}).
\begin{table}[t!]
\centering
\begin{tabular}{|c|c|} \hline
Parameters & Value \\ \hline
$y_u/10^{-6}$ & $6.3^{+1.3}_{-2.6}$ \\
$y_c/10^{-3}$ & $2.776\pm0.095$ \\
$y_t$ & $0.8685^{+0.0090}_{-0.0084}$ \\
$y_d/10^{-5}$ & $1.364^{+0.198}_{-0.087}$ \\
$y_s/10^{-4}$ & $2.70^{+0.14}_{-0.15}$ \\
$y_b/10^{-2}$ & $1.388^{+0.013}_{-0.014}$ \\
$\theta^q_{12}$ & $0.22736^{+0.00072}_{-0.00071}$ \\
$\theta^q_{13}/10^{-3}$ & $3.72 \pm 0.13$ \\
$\theta^q_{23}/10^{-2}$ & $4.296^{+0.064}_{-0.065}$ \\
$\delta^q_{CP}$ & $1.208\pm0.054$ \\
\hline
$y_e/10^{-6}$ & $2.8482^{+0.0022}_{-0.0021}$ \\
$y_{\mu}/10^{-4}$ & $6.0127^{+0.0047}_{-0.0044}$ \\
$y_{\tau}/10^{-2}$ & $1.208^{+0.00078}_{-0.00077}$ \\
$\frac{\Delta m_{21}^2}{10^{-5}{\rm eV}^2}$ & $7.42^{+0.21}_{-0.20}$ \\
$\frac{\Delta m_{31}^2}{10^{-3}{\rm eV}^2}$ & $2.517^{+0.026}_{-0.028}$ \\
$\sin^2\theta_{12}^l$ & $0.304\pm 0.012$ \\
$\sin^2\theta_{13}^l/10^{-2}$ & $2.221^{+0.068}_{-0.062}$ \\
$\sin^2\theta_{23}^l$ & $0.570^{+0.018}_{-0.024}$ \\
\hline
\end{tabular}
\caption{\label{tab:central-data}
The central values and the corresponding $1\sigma$ deviations of various flavor parameters for quarks and leptons, collected from \cite{value1:rge2013} and \cite{Esteban:2020cvm} (NO for neutrino masses are adopted). }
\end{table}

The lepton sector $\chi^2_l$ function can be constructed with the predicted mass ratios $m_e/m_\mu$, $m_\mu/m_\tau$, $\Delta m_{21}^2/\Delta m_{31}^2$ and the lepton mixing parameters $\sin^2\theta_{12}^l$, $\sin^2\theta_{13}^l$, $\sin^2\theta_{23}^l$, $\delta_{CP}^l$.
Similarly, the $\chi^2_q$ function for the quark sector can be constructed from the predicted quark mass ratios $m_u/m_c$, $m_c/m_t$, $m_d/m_s$, $m_s/m_b$ and the quark mixing parameters $\theta_{12}^q$, $\theta_{13}^q$, $\theta_{23}^q$, $\delta_{CP}^q$.
 The experimental values of the neutrino mixing parameters are taken from NuFIT v5.0 with Super-Kamiokanda atmospheric data~\cite{Esteban:2020cvm}. Dimensionless input parameters, such as the values of the moduli fields and the ratios of the coupling constants, can determine the mixing angles, CP violation phases and the fermion mass ratios. As noted in our previous sections, some phases of the input Yukawa-type coefficients can be removed by field redefinitions while the remaining ones are in general complex.

 To obtain the best-fit parameters in our fitting, we scan randomly the allowed parameter regions to find good seeds for further MCMC scanning. In practice, we try to find the best-fit points for the quark sector first, and then perform the numerical fitting for the lepton sector with the best-fit value of $\tau$ in the quark sector. However, it is sometimes difficult to obtain good fittings with a common $\tau$ for both sectors. Multiple $\tau$ values (for quark and lepton sectors, respectively) are then used in the fitting if the single modulus scenarios do not work well.

 After numerical calculations and fittings, we can obtain the best-fit points for the following scenarios
 \bit
 \item {\bf I:}~ $\rho_{\bar{f}}={\bf 3},~\rho_{F}={\bf 3},~\rho_{E}={\bf 3},~\rho_{S}={\bf 3}$.

This scenario, in which all ${\bar{f}}, F, E, S$ superfields transform as triplets of $A_4$ modular flavor symmetry, is the most predictive scenario in the modular flipped SU(5) GUT scheme. We carry out numerical fittings for each models in this scenario with different modular weights. Our numerical results indicate that all cases with modular weight $k_F=2$ can not lead to good fittings to experimental data while some case with modular weight $k_F=4$ can work well.

  The values for the best-fit point with modular weight $k_{\bar{f}}=k_F=4,k_E=k_S=2$ are given in Table.~\ref{table:fFES3}. We can see that, when the quark sector and lepton sector share the same modulus value $\tau$, we get $\chi_q^2=46.122$ and a much larger $\chi_l^2=236.259$. In order to improve the fitting accuracy for the lepton sector, we can adopt multiple moduli fields (with a single modular $A_4$ symmetry) in our modular GUT scheme by introducing the other independent modulus value $\tau_l$ for leptons, which can reduce the $\chi_l^2$ of lepton sector to $48.806$. It should be noted that, after introducing additional  $\tau_l$, the best-fit parameters for the quark sector are almost unchanged, while those for the lepton sector do not change much. The changes in the lepton sector will have only tiny effects on the fitting for the quark sector (by affecting the RGE evolutions of flavor parameters in the quark sector) and can be ignored. See more discussions in the next section.

 \item {\bf II:} $\rho_{\bar{f}}\in \{{\bf 1},{\bf {1^{\pr}}},{\bf {1^{\pr\pr}}}\},~\rho_{F}={\bf 3},~\rho_{E}={\bf 3},~\rho_{S}={\bf 3}$.

In this scenario, the $F, E, S$ superfields transform as triplets while the ${\bar{f}_i}$ of the three generations transform as singlets. Different assignments of ${\bf 1},{\bf {1^{\pr}}},{\bf {1^{\pr\pr}}}$ representations to three generations and choices of modular weights will lead to large number of sub-scenarios. In our numerical study, we show two representative sample sub-scenarios, which can lead to better fittings
\bit
\item ${\bf IX}$:~~$\rho_{\bar{f}_{1,2,3}}=({\bf {1^{\pr}}},{\bf {1^{\pr\pr}}},{\bf {1^{\pr\pr}}}$) with modular weights $k_{\bar{f}_{1,2,3}} =(2,0,2)$; $k_F=4$ and $k_E=k_S=2$.
\item ${\bf X}$:~~$\rho_{\bar{f}_{1,2,3}}=({\bf 1},{\bf {1^{\pr}}},{\bf {1^{\pr\pr}}})$ with modular weights $k_{\bar{f}_{1,2,3}}=(4,2,4)$; $k_F=4$ and $k_E=k_S=2$.
\eit
     The parameters of the best-fit points for sub-scenario ${\bf IX}$ and ${\bf X}$ are listed in Table~\ref{table:f1FES3-IX} and Table~\ref{table:f1FES3-X}, respectively. We can see that, when the quark sector and lepton sector share the same $\tau$, the best-fit point of sub-scenario ${\bf IX}$ predicts $\chi_q^2=16.408$ and $\chi_l^2=486.036$ while that of the sub-scenario ${\bf X}$ predicts $\chi_q^2=69.216$ and $\chi_l^2=208.261$. Obviously, although the fitting of the quark sector can be fairly good, the fitting of the lepton part is still unsatisfactory. So, again we can adopt multiple moduli fields to introduce an independent $\tau_l$ parameter for the lepton sector in addition to the $\tau_q$ for the quark sector. By re-scanning the parameter spaces, we find that, with multiple moduli values, we can get much better fitting. The best-fit point of sub-scenario ${\bf IX}$ now gives $\chi_l^2=30.215$ and that of sub-scenario ${\bf X}$ gives a much lower value $\chi_l^2=0.878$. We should note again that, when we carry out best-fitting with a single modulus value, we carry out fitting for the quark sector first before we carry out fitting for the lepton sector with the best-fitting value of modulus $\tau_q$ obtained for the quark sector.

    In principle, to obtain the best-fit point in the scenarios with multiple (two) moduli values, we need to repeatedly best fit the quark sector with the original $\tau_q$ each time we choose a new $\tau_l$ value to fit the lepton sector. However, we anticipate that the changing in the lepton sector will feed back into the quark sector by only affecting slightly their RGE evolutions. As the small changes of flavor parameters in the lepton sector (when varying $\tau_l$) can only slightly alter the beta functions for the flavor parameters in the quark sector, the best fit point for the quark sector is almost insensitive to the changes in the lepton sector. So, in our best fitting, we keep fixed the best fit point for the quark sector while we further carry out best fitting for the lepton sector in the scenarios with two different moduli values for quark and lepton sectors. To show the effects of such approximation, we list the predictions for the quark in Table 4 with (red colored) and without (uncolored) taking into account the RGE feeding back effects. It can be seen that, taking into account the feeding back effects from the lepton sector when varying $\tau_l$ away from $\tau_q$, the predictions of flavor parameters in the quark sector with the best fit points will increases the $\chi_q^2$ by $4.423$. Such an increase in $\chi_q^2$ can indeed be acceptable and can be neglected in most cases, which justify the use of this approximation.

\item {\bf III:} $\rho_{\bar{f}}={\bf 3},~\rho_{F}\in\{{\bf 1},{\bf {1^{\pr}}},{\bf {1^{\pr\pr}}}\},~\rho_{E}={\bf 3},~\rho_{S}={\bf 3}$.

In this scenario, the ${\bar{f}}, E, S$ superfields transform as triplets while the $F_i$ of the three generations transform as singlets. Similar to previous scenarios, we just show two representative sample sub-scenarios that can lead to better fitting
\bit
\item ${\bf IX^\pr}$:~~$\rho_{F_{1,2,3}}=({\bf {1^{\pr}}},{\bf {1^{\pr\pr}}},{\bf {1^{\pr\pr}}}$) with modular weights $k_{F_{1,2,3}} =(2,4,2)$;

    ~~~~~$k_{\bar{f}}=k_E=k_S=2$.
\item ${\bf X^\pr}$:~~$\rho_{F_{1,2,3}}=({\bf 1},{\bf {1^{\pr}}},{\bf {1^{\pr\pr}}})$ with modular weights $k_{F_{1,2,3}}=(0,2,4)$;

    ~~~~~ $k_{\bar{f}}=k_E=k_S=2$.
\eit

The parameters of the best-fit points for sub-scenario ${\bf IX^\pr}$ and ${\bf X^\pr}$ are listed in Table~\ref{table:fES3F1-IX} and Table~\ref{table:fES3F1-X}, respectively.  We can see that, when the quark sector and lepton sector share the same $\tau$, the best-fit point of sub-scenario ${\bf IX^\pr}$ predicts $\chi_q^2=1.221$, $\chi_l^2=0.358$ while that of the sub-scenario ${\bf X^\pr}$ predicts $\chi_q^2=50.511$, $\chi_l^2=232.993$. So, it can be seen that the best-fit point for sub-scenario ${\bf IX^\pr}$ can lead to excellent fitting even if only a single common modulus value for both quark sector and lepton sector are adopted.

However, from the $\chi_{q,l}^2$ values in the ${\bf X^\pr}$ model, it seems that we still need to introduce additional $\tau_l$ to obtain better fitting for the lepton sector.
 Indeed, after introducing additional $\tau_l$, the $\chi_l^2$ for the lepton sector can be reduced to $58.908$. Similar to the discussions in scenario {\bf II}, improved best-fitting for the lepton sector will feed back into the quark sector. In this scenario, it will further optimize the original fitting in the quark sector 
 (see the predictions of quark sector in Table~\ref{table:fES3F1-IX}).

 \item {\bf IV:} $\rho_{\bar{f}}={\bf 3},~\rho_{F}={\bf 3},~\rho_{E}={\bf 1},{\bf {1^{\pr}}},{\bf {1^{\pr\pr}}},~\rho_{S}={\bf 3}$.

In this scenario, the ${\bar{f}}, F, S$ superfields transform as triplets while the $E_i$ of the three generations transform as singlets. As the ${\bar{f}}$ and $F$ are both triplets, the best-fit point for the quark sector is almost the same as in scenario {\bf I}. Motivated by the best-fitting results in scenario {\bf I}, we adopt the following sample sub-scenario that can lead to better fitting
\bit
\item $\rho_{E_{1,2,3}}=({\bf 1},{\bf {1^{\pr}}},{\bf {1^{\pr\pr}}})$ with the corresponding modular weights $k_{E_{1,2,3}}=(2,0,2)$; $k_S=2$ and $k_{{\bar{f}}}=k_F=4$.
\eit
 The values of the best-fit points are shown in Table~\ref{table:E1fFS3-X}. The values of the best-fit points for the quark part are in agreement with that in scenario {\bf I}, which gives $\chi^2_q=47.396$ and is acceptable. The fitting of the lepton sector gives $\chi_l^2=53.711$, which is also acceptable and no longer needs the introduction of additional $\tau_l$.

 \item {\bf V:} $\rho_{\bar{f}}={\bf 3},~\rho_{F}={\bf 3},~\rho_{E}={\bf 3},~\rho_{S}={\bf 1},{\bf {1^{\pr}}},{\bf {1^{\pr\pr}}}$.

In this scenario, the ${\bar{f}}, F, E$ superfields transform as triplets while the $S_i$ of the three generations transform as singlets. Such a scenario with singlets $S_i$ is also similar to that in the scenario {\bf I}. Again, motivated by the best-fitting results in scenario {\bf I}, we adopt the following sample sub-scenario that can lead to better fitting
\bit
\item $\rho_{S_{1,2,3}}=({\bf 1},{\bf {1^{\pr}}},{\bf {1^{\pr\pr}}})$ with the corresponding modular weights $k_{S_{1,2,3}}=(2,4,2)$; $k_E=2$ and $k_{{\bar{f}}}=k_F=4$.
\eit

 The values of the best-fit point are shown in table \ref{table:S1fFE3-X}. The values of the best-fit point for the quark sector are also in agreement with that in scenario {\bf I} and {\bf IV}, which gives $\chi^2_q=46.121$. However, the $\chi_l^2$ for the lepton sector reaches $206.527$. Therefore, we still need to introduce an additional $\tau_l$ for lepton sector only to obtain better fitting. With an additional $\tau_l$ for the lepton sector, the $\chi_l^2$ can be reduced to $37.067$.

 \eit

\section{\label{sec:con}Conclusions}
 It is very interesting to check if the low energy flavor structures can be successfully predicted by the GUT models. In this paper, we try to explain the flavor structures of the Standard Model plus neutrinos in the framework of flipped SU(5) GUT with $A_4$ modular flavor symmetry.
 We classify all possible scenarios in this scheme according to the assignments of the modular $A_4$ representations for matter superfields and give the expressions of the quark and lepton mass matrices predicted by our scenarios at the GUT scale. After the RGE evolutions of the GUT scale parameters to the $M_Z$ scale, we can check whether our predictions can be consistent with the experimental data. By properly selecting the modular weights for various superfields that can lead to better fitting, the best-fit points can be found numerically with their corresponding $\chi^2$ values for the sample subscenarios.

  Our numerical results indicate that predictions of many scenarios can fit nicely to the experimental data when a common modulus value for quark sector and lepton sector is adopted. Especially, the $\chi^2_{total}$ of our fitting can be as low as $1.558$ for sample ${\bf IX^\pr}$ of scenario ${\bf III}$. However, the fitting with a single modulus value do not work very well for some scenarios, which then prefer the introduction of multiple moduli values for different sectors. In this paper, we concentrate on the possibility with two different moduli fields responsible for quark sector and lepton sector separately.

   We know that some representation of the GUT group contains both quarks and leptons. So, it seems that the unification of matter contents will be spoiled if different values of moduli fields are assigned separately for quarks and leptons. We propose that, such an inconsistency can be solved in orbifold GUT. In previous studies,  bi-triplet type Higgs fields with non-renormalizable interactions are always needed to break the multiple modular symmetries to a single modular symmetry so as that the UV theory with multiple modular symmetries and multiple moduli fields can be reduced to a IR theory with a single modular symmetry and multiple moduli fields. We propose a new approach to realize such reductions by adopting proper boundary conditions for the breaking of multiple modular symmetries into the surviving one.

 The most predictive scenario ${\bf III}$, in which all superfields transform as triplets of $A_4$, can be well fitted with two independent moduli values $\tau_q,\tau_l$ for quark sector and lepton sector, which gives $\chi^2_{total}\approx 95$ for the fitting. With one common modulus value for both quark sector and lepton sector, the value of $\chi^2_{total}$ for the best-fit point will be increased to $282.4$.

Note added: While we are preparing this draft, we notice the work in~\cite{Charalampous:2021gmf}, which also discuss the generations of flavor structures in flipped SU(5) GUT with $A_4$ modular symmetry. Although there are small overlaps, this work contains many new ingredients not covered  in~\cite{Charalampous:2021gmf}.  For example, we classify all the possible scenarios in flipped SU(5) GUT with $A_4$ modular symmetry according to the assignments of the modular $A_4$ representations for matter superfields. Besides, we do not adopt the Froggatt-Nielsen mechanism to generate the hierarchies in the flavor structures. Scenarios with multiple moduli fields in modular GUT and the breaking of modular flavor symmetries by BCs are also new, which also had not been discussed in~\cite{Charalampous:2021gmf}.
\begin{acknowledgments}
This work was supported by the National Natural Science Foundation of China (NNSFC) under grant Nos. 12075213, by the Key Research Project of Henan Education Department for colleges and universities under grant number 21A140025.
\end{acknowledgments}

\appendix
\section{\label{modular-weight} Modular form $Y_{\bf r}^{(k)}$ with weight $k$ and the level 3 under $A_4$}
We collect the modular forms $Y_{\bf r}^{(k)}$ with weight $k$ and the level 3 under $A_4$~\cite{feruglio}:
\bit
\item Modular weight $k=2$ and the representation ${\bf r}={\bf 3}$:

 $Y^{(2)}_{\mathbf{3}}=\left(Y_1, Y_2, Y_3\right)^{T}$
 with
\begin{eqnarray}
Y_1(\tau) &=& \frac{i}{2\pi}\left[ \frac{\eta'(\tau/3)}{\eta(\tau/3)}  +\frac{\eta'((\tau +1)/3)}{\eta((\tau+1)/3)}
+\frac{\eta'((\tau +2)/3)}{\eta((\tau+2)/3)} - \frac{27\eta'(3\tau)}{\eta(3\tau)}  \right], \nonumber \\
Y_2(\tau) &=& \frac{-i}{\pi}\left[ \frac{\eta'(\tau/3)}{\eta(\tau/3)}  +\omega^2\frac{\eta'((\tau +1)/3)}{\eta((\tau+1)/3)}
+\omega \frac{\eta'((\tau +2)/3)}{\eta((\tau+2)/3)}  \right] ,\nonumber \\
Y_3(\tau) &=& \frac{-i}{\pi}\left[ \frac{\eta'(\tau/3)}{\eta(\tau/3)}  +\omega\frac{\eta'((\tau +1)/3)}{\eta((\tau+1)/3)}
+\omega^2 \frac{\eta'((\tau +2)/3)}{\eta((\tau+2)/3)} \right]\,,
\end{eqnarray}
where $\eta(\tau)$ is the Dedekind eta-function,
\begin{equation}
\eta(\tau)=q^{1/24} \prod_{n =1}^\infty (1-q^n), ~~~  q=e^{2\pi i\tau}\,.
\end{equation}

The $q-$expansions of $Y_{1,2,3}(\tau)$ are given as
\begin{eqnarray}
\nonumber&&Y_1(\tau)=1 + 12q + 36q^2 + 12q^3 + 84q^4 + 72q^5 +\dots\,,\\
\nonumber&&Y_2(\tau)=-6q^{1/3}(1 + 7q + 8q^2 + 18q^3 + 14q^4 +\dots)\,,\\
&&Y_2(\tau)=-18q^{2/3}(1 + 2q + 5q^2 + 4q^3 + 8q^4 +\dots)\,.
\end{eqnarray}
\item  Modular weight $k=4$ and the representation ${\bf r}={\bf 3}, {\bf 1}, {\bf 1^\pr}$:
\begin{eqnarray}
\nonumber Y^{(4)}_{\mathbf{3}}&=&\frac{1}{2}(Y^{(2)}_{\mathbf{3}}Y^{(2)}_{\mathbf{3}})_{\mathbf{3}}=
\left(\bea{c}
Y_1^2-Y_2 Y_3\\
Y_3^2-Y_1 Y_2\\
Y_2^2-Y_1 Y_3
\eea\right)\,, \\
\nonumber Y^{(4)}_{\mathbf{1}}&=&(Y^{(2)}_{\mathbf{3}}Y^{(2)}_{\mathbf{3}})_{\mathbf{1}}=Y_1^2+2 Y_2 Y_3\,, \\
Y^{(4)}_{\mathbf{1}'}&=&(Y^{(2)}_{\mathbf{3}}Y^{(2)}_{\mathbf{3}})_{\mathbf{1}'}=Y_3^2+2 Y_1 Y_2\,.
\end{eqnarray}
\item  Modular weight $k=6$ and  the representation ${\bf r}={{\bf 3}I}, {{\bf 3}II}, {\bf 1}$:
\begin{eqnarray}
Y^{(6)}_{\mathbf{1}}&=&(Y^{(2)}_{\mathbf{3}}Y^{(4)}_{\mathbf{3}})_{\mathbf{1}}=Y_1^3+Y_2^3+Y_3^3-3 Y_1 Y_2 Y_3\,,\nonumber\\
Y^{(6)}_{\mathbf{3}I}&=&Y^{(2)}_{\mathbf{3}}Y^{(4)}_{\mathbf{1}}=(Y_1^2+2Y_2Y_3)\left(\bea{c}
Y_1\\
Y_2\\
Y_3
\eea\right)\,,\nonumber\\
Y^{(6)}_{\mathbf{3}II}&=&Y^{(2)}_{\mathbf{3}}Y^{(4)}_{\mathbf{1}'}=
(Y_3^2+2 Y_1Y_2)\left(\bea{c}
Y_3\\
Y_1\\
Y_2
\eea\right)\,.
\end{eqnarray}
\item  Modular weight $k=8$ and  the representation ${\bf r}={{\bf 3}I}, {{\bf 3}II}, {\bf 1},{\bf 1^\pr},{\bf 1^{\pr\pr}}$:
\begin{eqnarray}
\nonumber Y^{(8)}_{\mathbf{1}}&=&(Y^{(2)}_{\mathbf{3}}Y^{(6)}_{\mathbf{3}I})_{\mathbf{1}}=(Y_1^2+2 Y_2 Y_3)^2\,,\\
\nonumber Y^{(8)}_{\mathbf{1'}}&=&(Y^{(2)}_{\mathbf{3}}Y^{(6)}_{\mathbf{3}I})_{\mathbf{1'}}=(Y_1^2+2 Y_2 Y_3)(Y_3^2+2 Y_1 Y_2)\,, \\
\nonumber Y^{(8)}_{\mathbf{1''}}&=&(Y^{(2)}_{\mathbf{3}}Y^{(6)}_{\mathbf{3}II})_{\mathbf{1''}}=(Y_3^2+2 Y_1 Y_2)^2\,,\\
\nonumber Y^{(8)}_{\mathbf{3}I}&=&Y^{(2)}_{\mathbf{3}}Y^{(6)}_{\mathbf{1}}=(Y_1^3+Y_2^3+Y_3^3-3 Y_1 Y_2 Y_3)\left(\bea{c}
Y_1 \\
Y_2\\
Y_3
\eea\right)\,,\\
Y^{(8)}_{\mathbf{3}II}&=&(Y^{(2)}_{\mathbf{3}}Y^{(6)}_{\mathbf{3}II})_{\mathbf{3}_A}=(Y_3^2+2 Y_1Y_2)\left(\bea{c}
Y^2_2-Y_1Y_3\\
Y^2_1-Y_2Y_3\\
Y^2_3-Y_1Y_2
\eea\right)\,.
\end{eqnarray}
\eit
\begin{minipage}{\textwidth}
\renewcommand\arraystretch{0.9}
\makeatletter\def\@captype{table}
\begin{minipage}[H]{0.58\textwidth}
\begin{tabular}{|c|c|}

    \hline
    Parameter     & Value1  \\
    \hline
    $\beta_1/10^{-2}$ & 5.292            \\
    $\beta_2/10^{-4}$ & 1588.315-6.410i  \\
    $\beta_3/10^{-2}$ & -3.704+10.670i   \\
    $\beta_4/10^{-3}$ & -5.817+2.049i    \\
    $\beta_5/10^{-3}$ & -1058.838-1.620i \\
    $\beta_6/10^{-2}$ & 2.055+9.933i     \\
    $\beta_7/10^{-1}$ & 1.710-1.460i     \\
    $\al_1/10^{-4}$ & 3.774            \\
    $\al_2/10^{-3}$ & -2.821-273.122i  \\
    $\al_3/10^{-5}$ & -70.375+1.104i   \\
    $\al_4/10^{-4}$ & -3.691+2707.307i \\
    $\al_5/10^{-4}$ & -411.085-1.680i \\
    $\tau$          & 1.198+2.830i    \\
    \hline
    \hline
    $\ga_1/10^{-2}$ & 1.203           \\
    $\ga_2/10^{-4}$ & 3.536-9.922i    \\
    $\ga_3/10^{-3}$ & 1.713-3.086i    \\
    $\ga_4/10^{-4}$ & -6.732+105.230i \\
    $\ga_5/10^{-4}$ & 123.915-9.913i  \\
    $\Lambda_1/(10^{9}$ GeV) & 1.207              \\
    $\Lambda_2/(10^{2}$ GeV) & 2.006             \\
    $\la_1/10^{-3}$ & 8.678           \\
    $\la_2/10^{-3}$ & 3.249-81.540i   \\
    $\la_3/10^{-4}$ & 9.383-154.099i  \\
    $\la_4/10^{-2}$ & 8.095+18.176i   \\
    $\la_5/10^{-2}$ & 1.193-8.156i    \\
    $\ka_1/10^{-2}$ & -4.752          \\
    $\ka_2/10^{-2}$ & -2.809+2.805i   \\
    $\ka_3/10^{-2}$ & -1.245-1.635i   \\
    \hline
    \hline
    $\ga_1/10^{-2}$ & 1.204            \\
    $\ga_2/10^{-4}$ & 3.522-9.722i     \\
    $\ga_3/10^{-3}$ & 1.710-3.078i     \\
    $\ga_4/10^{-4}$ & --6.723+105.127i \\
    $\ga_5/10^{-4}$ & 123.760-9.751i   \\
    $\Lambda_1/(10^{9}$ GeV) & 1.298               \\
    $\Lambda_2/(10^{2}$ GeV) & 1.833              \\
    $\la_1/10^{-3}$ & 8.674            \\
    $\la_2/10^{-3}$ & 3.246-81.548i    \\
    $\la_3/10^{-4}$ & 9.385-154.042i   \\
    $\la_4/10^{-2}$ & 8.095+18.164i    \\
    $\la_5/10^{-2}$ & 1.194-8.155i     \\
    $\ka_1/10^{-2}$ & -4.753           \\
    $\ka_2/10^{-2}$ & -2.808+2.804i    \\
    $\ka_3/10^{-2}$ & -1.244-1.635i    \\
    $\tau_l$        & 1.180+2.711i     \\
    \hline
\end{tabular}
\end{minipage}
\begin{minipage}[t]{0.45\textwidth}
\begin{tabular}{|c|c|}
    \hline
    observable     & Value2                \\
    \hline
     $y_u/10^{-6}$  & 6.644      \\
     $y_c/10^{-3}$  & 3.445      \\
     $y_t$          & 0.868      \\
     $y_d/10^{-5}$     & 1.323      \\
     $y_s/10^{-4}$     & 1.841      \\
     $y_b/10^{-2}$     & 1.395      \\
     $\theta_{12}^q$             & 0.22737     \\
     $\theta_{13}^q/10^{-3}$     & 3.716       \\
     $\theta_{23}^q/10^{-2}$     & 4.296       \\
     $\delta_{CP}^q$                 & 1.194       \\
    \hline
     $\chi_{q}^2$                    & 46.122      \\
    \hline
    \hline
     $y_e/10^{-6}$      & 2.848      \\
     $y_\mu/10^{-4}$    & 6.104      \\
     $y_\tau/10^{-2}$   & 1.022      \\
     $\Delta m_{21}^2/10^{-5}/$eV$^2$$^2$         & 7.419      \\
     $\Delta m_{31}^2/10^{-3}/$eV$^2$         & 2.516      \\
     ${\sin}^2 \theta_{12}^l  $           & 0.457      \\
     ${\sin}^2 \theta_{13}^l/10^{-2}$     & 2.304      \\
     ${\sin}^2 \theta_{23}^l$             & 0.806      \\
    \hline
     $\chi_l^2$         & 236.259       \\
    \hline
    \hline
     $y_e/10^{-6}$      & 2.847      \\
     $y_\mu/10^{-4}$    & 6.109      \\
     $y_\tau/10^{-2}$   & 1.022      \\
     $\Delta m_{21}^2/10^{-5}/$eV$^2$         & 7.405      \\
     $\Delta m_{31}^2/10^{-3}/$eV$^2$         & 2.511      \\
     ${\sin}^2 \theta_{12}^l  $           & 0.384      \\
     ${\sin}^2 \theta_{13}^l/10^{-2}$     & 2.260      \\
     ${\sin}^2 \theta_{23}^l$             & 0.642      \\
    \hline
     $\chi_l^2$         & 48.806       \\
    \hline
\end{tabular}
\end{minipage}
\caption{The input parameters and low energy predictions of the best-fit point for the sample sub-scenario of scenario ${\bf I}$, which is given by $\rho_{\bar{f}}=\rho_{F}=\rho_{E}=\rho_{S}={\bf 3}$ and $k_{\bar{f}}=k_F=4,k_E=k_S=2$. The upper and middle mini-tables (in the left and right columns) for the case with one modulus value $\tau$, and the lower mini-table for the case with two moduli values $\tau_q,\tau_l$. }
\label{table:fFES3}
\end{minipage}

\begin{minipage}{\textwidth}
\makeatletter\def\@captype{table}
\begin{minipage}[t!]{0.58\textwidth}
\begin{tabular}{|c|c|}
    \hline
    Parameter     & Value1  \\
    \hline
    $\tilde{\beta}_1/10^{-6}$ & -5.602         \\
    $\tilde{\beta}_2/10^{-2}$ & -42.016-2.423i \\
    $\beta_1/10^{-2}$ & -62.245+1.201i \\
    $\tilde{\beta}_3/10^{-2}$ & -1.228+7.162i  \\
    $\tilde{\beta}_4/10^{-1}$ & -1.614+1.022i  \\
    $\al_1/10^{-3}$ & -1.152           \\
    $\al_2/10^{-2}$ & -1.244-148.050i  \\
    $\al_3/10^{-5}$ & 232.706+3.884i   \\
    $\al_4/10^{-4}$ & 3.011-14663.865i \\
    $\al_5/10^{-4}$ & -5.366-25.651i   \\
    $\tau/10^{-2}$  & 3.310+359.899i   \\
    \hline
    \hline
    $\ga_1/10^{-3}$ & -2.173           \\
    $\ga_2/10^{-4}$ & -14.990+7.062i    \\
    $\ga_3/10^{-2}$ & -2.669+2.470i    \\
    $\Lambda_1/(10^{6}$ GeV) & 4.838              \\
    $\Lambda_2/(10^{2}$ GeV) & 4.920             \\
    $\la_1/10^{-1}$ & 1.202           \\
    $\la_2/10^{-2}$ & 11.344-1.374i   \\
    $\la_3/10^{-3}$ & -2.706+35.392i  \\
    $\la_4/10^{-2}$ & -5.099+6.058i   \\
    $\la_5/10^{-3}$ & -234.073-6.489i    \\
    $\ka_1/10^{-2}$ & 8.870          \\
    $\ka_2/10^{-2}$ & 9.847-1.585i   \\
    $\ka_3/10^{-2}$ & 2.516+1.240i   \\
    \hline
    \hline
    $\ga_1/10^{-5}$ & 1.096            \\
    $\ga_2/10^{-2}$ & -2.892-2.143i    \\
    $\ga_3/10^{-3}$ & 2.073-2.232i     \\
    $\Lambda_1/(10^{8}$ GeV) & 3.204               \\
    $\Lambda_2/(10^{2}$ GeV) & 2.504              \\
    $\la_1/10^{-1}$ & 1.546            \\
    $\la_2/10^{-2}$ & -12.002-8.792i   \\
    $\la_3/10^{-2}$ & -6.180-1.039i    \\
    $\la_4/10^{-2}$ & -2.485+4.358i    \\
    $\la_5/10^{-2}$ & 1.017+6.638i     \\
    $\ka_1/10^{-2}$ & -8.731           \\
    $\ka_2/10^{-3}$ & -7.501-61.795i   \\
    $\ka_3/10^{-2}$ & 3.557+1.782i     \\
    $\tau_l/10^{-1}$ & 1.744+13.565i   \\
    \hline
\end{tabular}

\end{minipage}
\begin{minipage}[t]{0.45\textwidth}

\begin{tabular}{|c|c|}
    \hline
    observable     & Value2                \\
    \hline
     $y_u/10^{-6}$  & 10.688      \\
     $y_c/10^{-3}$  & 3.044      \\
     $y_t$          & 0.869      \\
     $y_d/10^{-5}$     & 1.626      \\
     $y_s/10^{-4}$     & 2.953      \\
     $y_b/10^{-2}$     & 1.388      \\
     $\theta_{12}^q$             & 0.22741     \\
     $\theta_{13}^q/10^{-3}$     & 3.721       \\
     $\theta_{23}^q/10^{-2}$     & 4.296       \\
     $\delta_{CP}^q$                 & 1.216       \\
    \hline
     $\chi_{q}^2$                    & 16.408      \\
    \hline
    \hline
     $y_e/10^{-6}$      & 2.888      \\
     $y_\mu/10^{-4}$    & 6.103      \\
     $y_\tau/10^{-2}$   & 1.022      \\
     $\Delta m_{21}^2/10^{-5}/$eV$^2$         & 7.286      \\
     $\Delta m_{31}^2/10^{-3}/$eV$^2$         & 2.521      \\
     ${\sin}^2 \theta_{12}^l  $           & 0.159      \\
     ${\sin}^2 \theta_{13}^l/10^{-2}$     & 2.466      \\
     ${\sin}^2 \theta_{23}^l$             & 0.124     \\
    \hline
     $\chi_l^2$         & 486.036       \\
    \hline
    \hline
     $y_e/10^{-6}$      & 2.781      \\
     $y_\mu/10^{-4}$    & 6.100      \\
     $y_\tau/10^{-2}$   & 1.022      \\
     $\Delta m_{21}^2/10^{-5}/$eV$^2$         & 7.601      \\
     $\Delta m_{31}^2/10^{-3}/$eV$^2$         & 2.504      \\
     ${\sin}^2 \theta_{12}^l  $           & 0.354      \\
     ${\sin}^2 \theta_{13}^l/10^{-2}$     & 2.353      \\
     ${\sin}^2 \theta_{23}^l$             & 0.598     \\
    \hline
     $\chi_l^2$         & 30.215       \\
    \hline
\end{tabular}
\end{minipage}
\caption{The input parameters and low energy predictions of the best-fit point for the sample sub-scenario  ${\bf {IX}}$ of scenario ${\bf {II}}$, which is $\rho_{F}=\rho_{E}=\rho_{S}={\bf 3}$, $\rho_{\bar{f}}={\bf {1^{\pr}}},{\bf {1^{\pr\pr}}},{\bf {1^{\pr\pr}}}$ and $k_{\bar{f}_{1,2,3}}=2,0,2$,$k_F=4$,$k_E=k_S=2$. The overall structure of this table is the same as Table~\ref{table:fFES3}. }
\label{table:f1FES3-IX}
\end{minipage}

\begin{minipage}{\textwidth}
\renewcommand\arraystretch{0.9}
\makeatletter\def\@captype{table}
\begin{minipage}[t!]{0.58\textwidth}
\begin{tabular}{|c|c|}
    \hline
    Parameter     & Value1  \\
    \hline
    $\hat{\beta}_1/10^{-4}$ & 2.021         \\
    $\hat{\beta}_2/10^{-2}$ & -3.445-6.671i  \\
    $\tilde{\beta}_1/10^{-2}$ & 1.950+31.706i  \\
    $\tilde{\beta}_2/10^{-2}$ & 2.515+15.885i  \\
    $\hat{\beta}_3/10^{-4}$ & 4.817-1.422i   \\
    $\hat{\beta}_4/10^{-2}$ & 2.343+1.423i   \\
    $\al_1/10^{-4}$ & 3.614             \\
    $\al_2/10^{-3}$ & -2.705-274.976i   \\
    $\al_3/10^{-5}$ & -71.175+1.010i    \\
    $\al_4/10^{-4}$ & 3.673-2746.837i   \\
    $\al_5/10^{-4}$ & -411.476-1.676i   \\
    $\tau$          & 1.198+2.835i      \\
    \hline
    \hline
    $\tilde{\ga}_1/10^{-5}$ & -1.347           \\
    $\tilde{\ga}_2/10^{-4}$ & -1.709+1.341i    \\
    $\ga_1/10^{-3}$ & -1.705+1.348i    \\
    $\tilde{\ga}_3/10^{-3}$ & -27.601+1.343i    \\
    $\tilde{\ga}_4/10^{-3}$ & -8.541-748.529i    \\
    $\Lambda_1/(10^{8}$ GeV) & 2.930              \\
    $\Lambda_2/(10^{3}$ GeV) & 1.186             \\
    $\la_1/10^{-3}$ & -6.147           \\
    $\la_2/10^{-2}$ & -1.456+2.594i   \\
    $\la_3/10^{-3}$ & 22.618+3.510i  \\
    $\la_4/10^{-3}$ & 17.130+3.472i   \\
    $\la_5/10^{-4}$ & 1.680+3.592i    \\
    $\ka_1/10^{-3}$ & 8.146          \\
    $\ka_2/10^{-2}$ & -6.236-3.579i   \\
    $\ka_3/10^{-3}$ & -9.665+47.141i   \\
    \hline
    \hline
    $\tilde{\ga}_1/10^{-5}$ & -1.321           \\
    $\tilde{\ga}_2/10^{-2}$ & -10.421-9.577i   \\
    $\ga_1/10^{-3}$ & 1.464+1.726i     \\
    $\tilde{\ga}_3/10^{-3}$ & -2.979+1.974i    \\
    $\tilde{\ga}_4/10^{-3}$ & 18.076-6.972i    \\
    $\Lambda_1/(10^{9}$ GeV) & 1.086               \\
    $\Lambda_2/(10^{2}$ GeV) & 7.729              \\
    $\la_1/10^{-2}$ & 1.092            \\
    $\la_2/10^{-2}$ & -5.706-1.620i    \\
    $\la_3/10^{-3}$ & -4.109+4.722i    \\
    $\la_4/10^{-3}$ & 8.417+29.966i    \\
    $\la_5/10^{-6}$ & -10.407-9.570i   \\
    $\ka_1/10^{-2}$ & -1.378           \\
    $\ka_2/10^{-3}$ & -85.215-6.487i   \\
    $\ka_3/10^{-2}$ & -2.525+1.181i    \\
    $\tau_l$        & 1.326+1.317i     \\
    \hline
\end{tabular}

\end{minipage}
\begin{minipage}[t]{0.45\textwidth}

\begin{tabular}{|c|c|c|}
    \hline
    observable     & \multicolumn{2}{|c|}{Value2}                \\
    \hline
     $y_u/10^{-6}$  & 6.445  & \textcolor{red}{6.354}    \\
     $y_c/10^{-3}$  & 2.836  & \textcolor{red}{2.787}    \\
     $y_t$          & 0.869  & \textcolor{red}{0.856}    \\
     $y_d/10^{-5}$     & 2.138 & \textcolor{red}{2.138}     \\
     $y_s/10^{-4}$     & 1.697 & \textcolor{red}{1.697}    \\
     $y_b/10^{-2}$     & 1.395 & \textcolor{red}{1.395}     \\
     $\theta_{12}^q$             & 0.22732  & \textcolor{red}{0.22732}    \\
     $\theta_{13}^q/10^{-3}$     & 3.726  & \textcolor{red}{3.726}    \\
     $\theta_{23}^q/10^{-2}$     & 4.296  & \textcolor{red}{4.296}     \\
     $\delta_{CP}^q$                 & 1.242  & \textcolor{red}{1.242}      \\
    \hline
     $\chi_{q}^2$                    & 69.216  & \textcolor{red}{73.639}     \\
    \hline
    \hline
     $y_e/10^{-6}$      & \multicolumn{2}{|c|}{2.848}      \\
     $y_\mu/10^{-4}$    & \multicolumn{2}{|c|}{6.103}      \\
     $y_\tau/10^{-2}$   & \multicolumn{2}{|c|}{1.022}      \\
     $\Delta m_{21}^2/10^{-5}/$eV$^2$         & \multicolumn{2}{|c|}{7.274}      \\
     $\Delta m_{31}^2/10^{-3}/$eV$^2$         & \multicolumn{2}{|c|}{2.521}      \\
     ${\sin}^2 \theta_{12}^l  $           & \multicolumn{2}{|c|}{0.212}      \\
     ${\sin}^2 \theta_{13}^l/10^{-2}$     & \multicolumn{2}{|c|}{2.393}      \\
     ${\sin}^2 \theta_{23}^l$             & \multicolumn{2}{|c|}{0.865}     \\
    \hline
     $\chi_l^2$         & \multicolumn{2}{|c|}{208.261}       \\
    \hline
    \hline
     $y_e/10^{-6}$      & \multicolumn{2}{|c|}{2.848}      \\
     $y_\mu/10^{-4}$    & \multicolumn{2}{|c|}{6.103}      \\
     $y_\tau/10^{-2}$   & \multicolumn{2}{|c|}{1.022}      \\
     $\Delta m_{21}^2/10^{-5}/$eV$^2$         & \multicolumn{2}{|c|}{7.492}      \\
     $\Delta m_{31}^2/10^{-3}/$eV$^2$         & \multicolumn{2}{|c|}{2.510}      \\
     ${\sin}^2 \theta_{12}^l  $           & \multicolumn{2}{|c|}{0.313}      \\
     ${\sin}^2 \theta_{13}^l/10^{-2}$     & \multicolumn{2}{|c|}{2.195}      \\
     ${\sin}^2 \theta_{23}^l$             & \multicolumn{2}{|c|}{0.560}     \\
    \hline
     $\chi_l^2$         & \multicolumn{2}{|c|}{0.878}       \\
    \hline
\end{tabular}
\end{minipage}
\caption{The input parameters and low energy predictions of the best-fit point for the sample sub-scenario ${\bf {X}}$ of scenario ${\bf {II}}$, whic is $\rho_{F}=\rho_{E}=\rho_{S}={\bf 3}$, $\rho_{\bar{f}}={\bf 1},{\bf {1^{\pr}}},{\bf {1^{\pr\pr}}}$, $k_{\bar{f}_{1,2,3}}=4,2,4,~k_F=4,k_E=k_S=2$. The overall structure of this table is similar to Table~\ref{table:fFES3}. The values marked with (without) red color denote the predictions of quark sectors with (without) taking into account the feeding back RGE effects from the changing of the lepton parameters when varying $\tau_l$.}
\label{table:f1FES3-X}
\end{minipage}

\begin{minipage}{\textwidth}
\makeatletter\def\@captype{table}
\begin{minipage}[t!]{0.58\textwidth}
\begin{tabular}{|c|c|}
    \hline
    Parameter     & Value1  \\
    \hline
    $\beta_1/10^{-1}$ & -1.011        \\
    $\tilde{\beta}_1/10^{-5}$ & 3.031-12.327i \\
    $\tilde{\beta}_2/10^{-5}$ &-3.761-12.431i \\
    $\beta_2/10^{-7}$ & 6.423+6.285i  \\
    $\al_1/10^{-3}$ & 6.676           \\
    $\al_2/10^{-5}$ & 190.308-6.370i  \\
    $\al_3/10^{-5}$ & -2.466-3.369i   \\
    $\al_4/10^{-5}$ & -182.524-4.239i \\
    $\tau/10^{-3}$  & 3.040+719.434i  \\
    \hline
    \hline
    $\ga_1/10^{-3}$ & 1.387           \\
    $\ga_2/10^{-4}$ & 46.093-4.339i   \\
    $\ga_3/10^{-5}$ & -279.543+5.181i \\
    $\ga_4/10^{-4}$ & -28.062+2.663i \\
    $\Lambda_1/(10^{8}$ GeV) & 2.685              \\
    $\Lambda_2/(10^{3}$ GeV) & 1.202             \\
    $\la_1/10^{-2}$ & 1.239           \\
    $\tilde{\la}_1/10^{-2}$ & -2.379-58.255i  \\
    $\tilde{\la}_2/10^{-3}$ & -8.173-215.115i \\
    $\la_2/10^{-3}$ & -37.241+2.455i  \\
     $\ka_1/10^{-2}$ &-4.784          \\
    $\ka_2/10^{-2}$ & 24.526+4.093i   \\
    $\ka_3/10^{-2}$ & -3.330+1.181i   \\
    \hline

\end{tabular}

\end{minipage}
\begin{minipage}[t]{0.45\textwidth}

\begin{tabular}{|c|c|}
    \hline
    observable     & Value2      \\
    \hline
     $y_u/10^{-6}$  & 6.403      \\
     $y_c/10^{-3}$  & 3.103      \\
     $y_t$          & 0.868      \\
     $y_d/10^{-5}$     & 1.189   \\
     $y_s/10^{-4}$     & 2.611   \\
     $y_b/10^{-2}$     & 1.390    \\
     $\theta_{12}^q$             & 0.22735    \\
     $\theta_{13}^q/10^{-3}$     & 3.725      \\
     $\theta_{23}^q/10^{-2}$     & 4.296      \\
     $\delta_{CP}^q$                 & 1.221  \\
    \hline
     $\chi_{q}^2$                    & 1.221  \\
    \hline
    \hline
     $y_e/10^{-6}$      & 2.848    \\
     $y_\mu/10^{-4}$    & 6.103    \\
     $y_\tau/10^{-2}$   & 1.022    \\
     $\Delta m_{21}^2/10^{-5}/$eV$^2$         & 7.405    \\
     $\Delta m_{31}^2/10^{-3}/$eV$^2$         & 2.507    \\
     ${\sin}^2 \theta_{12}^l  $           & 0.309    \\
     ${\sin}^2 \theta_{13}^l/10^{-2}$     & 2.214    \\
     ${\sin}^2 \theta_{23}^l$             & 0.563    \\
    \hline
     $\chi_l^2$         & 0.358     \\
     \hline
\end{tabular}
\end{minipage}
\caption{The input parameters and low energy predictions of the best-fit point for the sample sub-scenario ${\bf {IX}}$ of scenario ${\bf {III}}$, which is $\rho_{{\bar{f}}}=\rho_{E}=\rho_{S}={\bf 3}$, $\rho_F={\bf {1^{\pr}}},{\bf {1^{\pr}}},{\bf {1^{\pr\pr}}}$, $k_{F_{1,2,3}}=2,4,2$ and $k_{\bar{f}}=k_E=k_S=2$. The sub-scenario adopts a common $\tau$ for both quark sector and lepton sector.}
\label{table:fES3F1-IX}
\end{minipage}

\begin{minipage}{\textwidth}
\makeatletter\def\@captype{table}
\begin{minipage}[t!]{0.58\textwidth}
\begin{tabular}{|c|c|}
    \hline
    Parameter     & Value1  \\
    \hline
    $\beta_1/10^{-1}$ & -6.267        \\
    $\beta_2/10^{-4}$ & -9.660-13.517i\\
    $\tilde{\beta}_1/10^{-6}$ & -3.706+3.607i \\
    $\tilde{\beta}_2/10^{-6}$ & -2.620-5.096i \\
    $\al_1/10^{-2}$ & -2.851          \\
    $\al_2/10^{-4}$ & -127.076+3.644i \\
    $\al_3/10^{-4}$ & 1.659+1.722i \\
    $\al_4/10^{-5}$ & 5.443-5.549i \\
    $\al_5/10^{-4}$ & 3.540+9.221i  \\
    $\tau/10^{-3}$  & 1.017+1.913i \\
    \hline
    \hline
    $\ga_1/10^{-2}$ & -1.254         \\
    $\ga_2/10^{-4}$ & 4.974+7.795i   \\
    $\ga_3/10^{-4}$ & -126.546-1.708i\\
    $\ga_4/10^{-3}$ & -18.843+4.274i  \\
    $\Lambda_1/(10^{8}$ GeV) & 4.056             \\
    $\Lambda_2/(10^{2}$ GeV) & 1.000            \\
    $\la_1/10^{-1}$ & -2.279         \\
    $\la_2/10^{-4}$ & -27379.394-3.830i \\
    $\tilde{\la}_1/10^{-2}$ & -7.307-1.826i  \\
    $\tilde{\la}_2/10^{-2}$ & -1.452-6.729i  \\
     $\ka_1/10^{-1}$ &1.773          \\
    $\ka_2/10^{-2}$ & 1.230+1.915i   \\
    $\ka_3/10^{-2}$ & -1.711+2.243i  \\
    \hline
    \hline
    $\ga_1/10^{-3}$ & 3.325          \\
    $\ga_2/10^{-4}$ & 98.069+7.731i  \\
    $\ga_3/10^{-4}$ & -3.702-10.643i \\
    $\ga_3/10^{-4}$ & -66.573-4.858i \\
    $\Lambda_1/(10^{9}$ GeV)   & 5.301           \\
    $\Lambda_2/(10^{3}$ GeV) & 4.672            \\
    $\la_1/10^{-2}$ & 1.770          \\
    $\la_2/10^{-2}$ & -3.087-8.997i  \\
    $\tilde{\la}_1/10^{-2}$ & -7.622-1.218i  \\
    $\tilde{\la}_2/10^{-2}$ & 10.711-2.220i  \\
     $\ka_1/10^{-2}$ & 7.139         \\
    $\ka_2/10^{-2}$ & -2.402+2.631i  \\
    $\ka_3/10^{-2}$ & -11.173+2.045i \\
    \hline
    $\tau/10^{-1}$   & 5.435+9.0761i \\
    \hline
\end{tabular}

\end{minipage}
\begin{minipage}[t]{0.45\textwidth}

\begin{tabular}{|c|c|c|}
    \hline
    observable     & \multicolumn{2}{|c|}{Value2}      \\
    \hline
     $y_u/10^{-6}$  & 9.660   & \textcolor{red}{7.076} \\
     $y_c/10^{-3}$  & 3.104   & \textcolor{red}{3.104} \\
     $y_t$          & 0.869   & \textcolor{red}{0.869} \\
     $y_d/10^{-5}$  & 1.884   & \textcolor{red}{1.357} \\
     $y_s/10^{-4}$  & 1.790   & \textcolor{red}{1.832} \\
     $y_b/10^{-2}$  & 1.388   & \textcolor{red}{1.388} \\
     $\theta_{12}^q$          & 0.22743 & \textcolor{red}{0.22737}\\
     $\theta_{13}^q/10^{-3}$  & 3.724   & \textcolor{red}{3.735}  \\
     $\theta_{23}^q/10^{-2}$  & 4.296   & \textcolor{red}{4.296}  \\
     $\delta_{CP}^q$          & 1.223   & \textcolor{red}{1.207}  \\
    \hline
     $\chi_{q}^2$             & 50.511  & \textcolor{red}{33.856} \\
    \hline
    \hline
     $y_e/10^{-6}$      & \multicolumn{2}{|c|}{2.834}    \\
     $y_\mu/10^{-4}$    & \multicolumn{2}{|c|}{6.103}      \\
     $y_\tau/10^{-2}$   & \multicolumn{2}{|c|}{1.022}      \\
     $\Delta m_{21}^2/10^{-5}/$eV$^2$         & \multicolumn{2}{|c|}{7.329}      \\
     $\Delta m_{31}^2/10^{-3}/$eV$^2$         & \multicolumn{2}{|c|}{2.519}      \\
     ${\sin}^2 \theta_{12}^l  $           & \multicolumn{2}{|c|}{0.152}      \\
     ${\sin}^2 \theta_{13}^l/10^{-2}$     & \multicolumn{2}{|c|}{2.272}      \\
     ${\sin}^2 \theta_{23}^l$             & \multicolumn{2}{|c|}{0.336}      \\
    \hline
     $\chi_l^2$         & \multicolumn{2}{|c|}{232.993}    \\
    \hline
    \hline
     $y_e/10^{-6}$      & \multicolumn{2}{|c|}{2.849}      \\
     $y_\mu/10^{-4}$    & \multicolumn{2}{|c|}{6.105}      \\
     $y_\tau/10^{-2}$   & \multicolumn{2}{|c|}{1.022}      \\
     $\Delta m_{21}^2/10^{-5}/$eV$^2$         & \multicolumn{2}{|c|}{7.303}      \\
     $\Delta m_{31}^2/10^{-3}/$eV$^2$         & \multicolumn{2}{|c|}{2.520}      \\
     ${\sin}^2 \theta_{12}^l  $           & \multicolumn{2}{|c|}{0.287}     \\
     ${\sin}^2 \theta_{13}^l/10^{-2}$     & \multicolumn{2}{|c|}{2.268}     \\
     ${\sin}^2 \theta_{23}^l$             & \multicolumn{2}{|c|}{0.390}     \\
    \hline
     $\chi_l^2$         & \multicolumn{2}{|c|}{58.908}      \\
    \hline
\end{tabular}
\end{minipage}
\caption{ The input parameters and low energy predictions of the best-fit point for the sample sub-scenario ${\bf {X}}$ of scenario ${\bf {III}}$, which is $\rho_{{\bar{f}}}=\rho_{E}=\rho_{S}={\bf 3}$, $\rho_F={\bf 1},{\bf {1^{\pr}}},{\bf {1^{\pr\pr}}}$, $k_{F_{1,2,3}}=0,2,4$ and $k_{\bar{f}}=k_E=k_S=2$. The overall structure of the table is the same as Table~\ref{table:f1FES3-X}.}
\label{table:fES3F1-X}
\end{minipage}

\begin{minipage}{\textwidth}
\makeatletter\def\@captype{table}
\begin{minipage}[t!]{0.58\textwidth}
\begin{tabular}{|c|c|}
    \hline
    Parameter     & Value1  \\
    \hline
    $\beta_1/10^{-2}$ & 5.292            \\
    $\beta_2/10^{-4}$ & 1588.315-6.410i  \\
    $\beta_3/10^{-2}$ & -3.704+10.670i   \\
    $\beta_4/10^{-3}$ & -5.817+2.049i    \\
    $\beta_5/10^{-3}$ & -1058.838-1.620i \\
    $\beta_6/10^{-2}$ & 2.055+9.933i     \\
    $\beta_7/10^{-1}$ & 1.710-1.460i     \\
    $\al_1/10^{-4}$ & 3.774              \\
    $\al_2/10^{-3}$ & -2.821-273.122i    \\
    $\al_3/10^{-5}$ & -70.375+1.104i     \\
    $\al_4/10^{-4}$ & -3.691+2707.307i   \\
    $\al_5/10^{-4}$ & -411.085-1.680i    \\
    $\tau$          & 1.198+2.830i       \\
    \hline
    \hline
    $\tilde{\ga}_1/10^{-5}$ & 8.417           \\
    $\tilde{\ga}_2/10^{-2}$ & 53.132-1.239i   \\
    $\ga_1/10^{-3}$ & 31.013+8.621i   \\
    $\tilde{\ga}_3/10^{-4}$ & 4.794-20.290i   \\
    $\tilde{\ga}_4/10^{-3}$ &   8.497+8.861i  \\
    $\Lambda_1/(10^{9}$ GeV) & 1.363              \\
    $\Lambda_2/(10^{2}$ GeV) & 1.788             \\
    $\la_1/10^{-3}$ & 8.676           \\
    $\la_2/10^{-3}$ & 3.250-81.511i   \\
    $\la_3/10^{-4}$ & 9.385-154.120i  \\
    $\la_4/10^{-2}$ & 8.091+18.199i   \\
   $\la_5/10^{-2}$  & 1.193-8.152i    \\
     $\ka_1/10^{-2}$ &-4.752          \\
    $\ka_2/10^{-2}$ &-2.809+2.805i    \\
    $\ka_3     $ & 1.198+2.835i       \\
    $\ka_4/10^{-6}$ &6.945+3891.526i  \\
    \hline

\end{tabular}

\end{minipage}
\begin{minipage}[t]{0.45\textwidth}

\begin{tabular}{|c|c|}
    \hline
    observable     & Value2    \\
    \hline
     $y_u/10^{-6}$  & 6.683    \\
     $y_c/10^{-3}$  & 3.459    \\
     $y_t$          & 0.868    \\
     $y_d/10^{-5}$     & 1.338 \\
     $y_s/10^{-4}$     & 1.839 \\
     $y_b/10^{-2}$     & 1.396 \\
     $\theta_{12}^q$             & 0.22737  \\
     $\theta_{13}^q/10^{-3}$     & 3.714    \\
     $\theta_{23}^q/10^{-2}$     & 4.296    \\
     $\delta_{CP}^q$             & 1.191    \\
    \hline
     $\chi_{q}^2$                & 47.396   \\
    \hline
    \hline
     $y_e/10^{-6}$      & 2.847   \\
     $y_\mu/10^{-4}$    & 6.102   \\
     $y_\tau/10^{-2}$   & 1.022   \\
     $\Delta m_{21}^2/10^{-5}/$eV$^2$         & 7.436      \\
     $\Delta m_{31}^2/10^{-3}/$eV$^2$         & 2.514      \\
     ${\sin}^2 \theta_{12}^l  $           &0.217    \\
     ${\sin}^2 \theta_{13}^l/10^{-2}$     & 2.241   \\
     ${\sin}^2 \theta_{23}^l$             &0.642    \\
    \hline
     $\chi_l^2$         &53.771   \\
     \hline
\end{tabular}
\end{minipage}
\caption{The input parameters and low energy predictions of the best-fit point for the sample sub-scenario of scenario ${\bf {IV}}$, which is $\rho_{\bar{f}}=\rho_{F}=\rho_{S}={\bf 3}$, $\rho_E={\bf 1},{\bf {1^{\pr}}},{\bf {1^{\pr\pr}}}$, $k_{E_{1,2,3}}=2,0,2$ and $k_{\bar{f}}=k_F=4,~k_S=2$. The overall structure of this table is the same as Table~\ref{table:fES3F1-IX}.}
\label{table:E1fFS3-X}
\end{minipage}

\begin{minipage}{\textwidth}
\renewcommand\arraystretch{0.92}
\makeatletter\def\@captype{table}
\begin{minipage}[t!]{0.58\textwidth}
\begin{tabular}{|c|c|}
    \hline
    Parameter     & Value1  \\
    \hline
    $\beta_1/10^{-2}$ & 5.292           \\
    $\beta_2/10^{-4}$ & 1588.315-6.410i \\
    $\beta_3/10^{-2}$ & -3.704+10.670i  \\
    $\beta_4/10^{-3}$ & -5.817+2.049i   \\
    $\beta_5/10^{-3}$ & -1058.838-1.620i\\
    $\beta_6/10^{-2}$ & 2.055+9.933i    \\
    $\beta_7/10^{-1}$ & 1.710-1.460i    \\
    $\al_1/10^{-4}$ & 3.774             \\
    $\al_2/10^{-3}$ & -2.821-273.122i   \\
    $\al_3/10^{-5}$ & -70.375+1.104i    \\
    $\al_4/10^{-4}$ & -3.691+2707.307i  \\
    $\al_5/10^{-4}$ & -411.085-1.680i   \\
    $\tau$          & 1.198+2.830i      \\
    \hline
    \hline
    $\ga_1/10^{-2}$ & -1.347          \\
    $\ga_2/10^{-5}$ & -40.960+6.261i  \\
    $\ga_3/10^{-3}$ & -5.940+1.341i   \\
    $\ga_4/10^{-3}$ & -9.701-271.749i \\
    $\ga_5/10^{-5}$ & 1152.424-7.288i \\
    $\Lambda_1/(10^{8}$ GeV) & 2.309              \\
    $\Lambda_2/(10^{3}$ GeV) & 1.370             \\
    $\tilde{\la}_1/10^{-1}$ & -1.287          \\
    $\tilde{\la}_2/10^{-2}$ & 42.415-2.024i   \\
    $\hat{\la}_1/10^{-2}$ & 1.172-18.705i   \\
    $\hat{\la}_2/10^{-3}$ & 39.161+8.635i   \\
    $\tilde{\la}_3/10^{-2}$ & -1.075-1.045i   \\
    $\tilde{\la}_4/10^{-4}$ & 2317.508+7.364i \\
    $\ka_1/10^{-2}$ & -7.844          \\
    $\ka_2/10^{-2}$ & -6.140+3.706i   \\
    $\ka_3/10^{-2}$ & -7.963-3.454i   \\
    $\ka_4/10^{-2}$ & -1.385-10.892i  \\
    \hline
    \hline
    $\ga_1/10^{-5}$ & 1.130           \\
    $\ga_2/10^{-2}$ & 29.237-3.487i   \\
    $\ga_3/10^{-3}$ & -8.728-3.466i   \\
    $\ga_4/10^{-3}$ & -14.891+3.175i  \\
    $\ga_5/10^{-3}$ & 1220.349+2.453i \\
    $\Lambda_1/(10^{8}$ GeV) & 2.316              \\
    $\Lambda_2/(10^{3}$ GeV) & 1.296             \\
    $\tilde{\la}_1/10^{-2}$ & -3.811          \\
    $\tilde{\la}_2/10^{-2}$ & 3.904+4.544i    \\
    $\hat{\la}_1/10^{-2}$ & -1.667+4.209i   \\
    $\hat{\la}_2/10^{-2}$ & 1.518+4.197i    \\
    $\tilde{\la}_3/10^{-2}$ & 4.172+3.284i    \\
    $\tilde{\la}_4/10^{-2}$ & -7.439-1.976i   \\
    $\ka_1/10^{-2}$ & 5.123           \\
    $\ka_2/10^{-2}$ & -7.910+9.687i   \\
    $\ka_3/10^{-2}$ & -9.404+8.927i   \\
    $\ka_4/10^{-2}$ & 7.421+1.231i    \\
    $\tau_l$        & 1.937+2.607i    \\
    \hline
\end{tabular}

\end{minipage}
\begin{minipage}[t]{0.45\textwidth}

\begin{tabular}{|c|c|}
    \hline
    observable     & Value2    \\
    \hline
     $y_u/10^{-6}$  & 6.683    \\
     $y_c/10^{-3}$  & 3.459    \\
     $y_t$          & 0.868    \\
     $y_d/10^{-5}$     & 1.338 \\
     $y_s/10^{-4}$     & 1.839 \\
     $y_b/10^{-2}$     & 1.396 \\
     $\theta_{12}^q$             & 0.22737    \\
     $\theta_{13}^q/10^{-3}$     & 3.714      \\
     $\theta_{23}^q/10^{-2}$     & 4.296      \\
     $\delta_{CP}^q$                 & 1.191  \\
    \hline
     $\chi_{q}^2$                    & 46.121 \\
    \hline
    \hline
     $y_e/10^{-6}$      & 2.849      \\
     $y_\mu/10^{-4}$    & 6.103      \\
     $y_\tau/10^{-2}$   & 1.022      \\
     $\Delta m_{21}^2/10^{-5}/$eV$^2$         & 7.230      \\
     $\Delta m_{31}^2/10^{-3}/$eV$^2$         & 2.524      \\
     ${\sin}^2 \theta_{12}^l  $           & 0.445      \\
     ${\sin}^2 \theta_{13}^l/10^{-2}$     & 2.377      \\
     ${\sin}^2 \theta_{23}^l$             & 0.790      \\
    \hline
     $\chi_l^2$         & 206.527    \\
    \hline
    \hline
     $y_e/10^{-6}$      & 2.848      \\
     $y_\mu/10^{-4}$    & 6.103      \\
     $y_\tau/10^{-2}$   & 1.022      \\
     $\Delta m_{21}^2/10^{-5}/$eV$^2$         & 7.494      \\
     $\Delta m_{31}^2/10^{-3}/$eV$^2$         & 2.510      \\
     ${\sin}^2 \theta_{12}^l  $           & 0.320      \\
     ${\sin}^2 \theta_{13}^l/10^{-2}$     & 2.222      \\
     ${\sin}^2 \theta_{23}^l$             & 0.713      \\
    \hline
     $\chi_l^2$         & 37.067     \\
    \hline
\end{tabular}
\end{minipage}
\caption{The input parameters and low energy predictions of the best-fit point for the sample sub-scenario of scenario ${\bf {V}}$, which is $\rho_{\bar{f},F,E}={\bf 3}$, $\rho_S={\bf 1},{\bf {1^{\pr}}},{\bf {1^{\pr\pr}}}$, $k_{S_{1,2,3}}=2,4,2$, $k_{\bar{f},F}=4$ and $k_E=2$. The overall structure of this table is the same as Table~\ref{table:fFES3}. }
\label{table:S1fFE3-X}
\end{minipage}



\end{document}